\documentclass[aip,amsmath,amssymb,reprint,]{revtex4-1}

\usepackage{graphicx}
\usepackage{dcolumn}
\usepackage{bm}

\usepackage[utf8]{inputenc}
\usepackage[T1]{fontenc}
\usepackage{mathptmx}
\usepackage{etoolbox}
\usepackage{xcolor}

\makeatletter
\def\@email#1#2{
 \endgroup
 \patchcmd{\titleblock@produce}
  {\frontmatter@RRAPformat}
  {\frontmatter@RRAPformat{\produce@RRAP{*#1\href{mailto:#2}{#2}}}\frontmatter@RRAPformat}
  {}{}
}
\makeatother

\begin{document}

\title[]{Modelling of saturated external MHD instabilities in tokamaks: a comparison of 3D free boundary equilibria and nonlinear stability calculations}

\author{R Ramasamy}
\affiliation{Max-Planck Institut f\"ur Plasmaphysik, Boltzmannstraße 2, 85748 Garching bei München, Germany}
\affiliation{Max-Planck Princeton Center for Plasma Physics, Princeton, New Jersey 08544 USA}
\email{rohan.ramasamy@ipp.mpg.de}

\author{G Bustos Ramirez}
\affiliation{Ecole Polytechnique Fédérale de Lausanne (EPFL), Swiss Plasma Center (SPC), CH-1015 Lausanne, Switzerland}

\author{M Hoelzl}
\affiliation{Max-Planck Institut f\"ur Plasmaphysik, Boltzmannstraße 2, 85748 Garching bei München, Germany}

\author{J Graves} 
\affiliation{Ecole Polytechnique Fédérale de Lausanne (EPFL), Swiss Plasma Center (SPC), CH-1015 Lausanne, Switzerland}

\author{G Su\'arez L\'opez} 
\affiliation{Max-Planck Institut f\"ur Plasmaphysik, Boltzmannstraße 2, 85748 Garching bei München, Germany}

\author{K Lackner}
\affiliation{Max-Planck Institut f\"ur Plasmaphysik, Boltzmannstraße 2, 85748 Garching bei München, Germany}

\author{S G\"unter} 
\affiliation{Max-Planck Institut f\"ur Plasmaphysik, Boltzmannstraße 2, 85748 Garching bei München, Germany}

\author{the JOREK team}
\altaffiliation{See author list of Ref. \onlinecite{hoelzl2021jorek} for full list of team members.}

\date{\today}

\begin{abstract}
    3D free boundary equilibrium computations have recently been used to model external kinks and edge harmonic oscillations (EHOs), comparing with linear MHD stability codes, and nonlinear analytic theory [Kleiner et al, PPCF 61 084005 (2019)]. In this study, results of the VMEC equilibrium code are compared further with nonlinear reduced MHD simulations, using the JOREK code. {The purpose of this investigation is to understand the extent to which the modelling approaches agree, and identify the important physical effects which can modify the dynamics.} For the simulated external kink, which is dominated by a single toroidal harmonic, {good agreement is found when a large Lundquist number {is used} in the JOREK simulation, such that resistive effects are sub-dominant}. Modelling EHOs where multiple toroidal harmonics are linearly unstable, the saturated perturbation observed can differ in the dominant toroidal harmonic. On the ideal timescale, a $n=2$ EHO is observed in JOREK, while the saturated perturbation predicted by VMEC is a $n=1$ mode. Extending simulations into timescales where resistive effects can play a role, similar $n=1$ perturbations can be found. The coupling of different linearly unstable toroidal harmonics in the JOREK simulation broadens the magnetic energy spectrum and ergodises the plasma edge region, resulting in a more localised pressure perturbation. These effects are not observed in VMEC, because closed magnetic flux surfaces are enforced. {Despite the sensitivity of JOREK results on the assumed resistivity, saturated states can be found using both approaches that are in reasonable agreement, even for this more advanced case}.
\end{abstract}

\maketitle

\section{Introduction}

Over the past 10 years, there have been considerable efforts to model nonlinearly perturbed tokamak equilibria \cite{chapman2014three, liu2017nonlinear}. Such configurations start out as axisymmetric equilibria, that are then driven to a non-axisymmetric state either by MHD instabilities, which are typically radially localised near the plasma edge, or by resonant magnetic perturbations. The perturbed equilibria are attractive, because the modest increase in particle and thermal transport limits the pedestal build up, and the onset of type I ELMs in H-mode discharges, while maintaining good confinement across the edge region.

Numerical simulations of weakly 3D equilibria can provide insights into how to reach attractive perturbed tokamak states. There are several approaches to modelling this problem, making use of equilibrium, linear and nonlinear MHD codes \cite{chapman2014three, turnbull2013comparisons}. The most physically accurate description should be the use of nonlinear MHD codes, but the uncertainty in the prescribed diffusive parameters applied in these codes can make it difficult to find a physically meaningful perturbed equilibrium state. This is because these parameters modify the trajectory of the simulated plasma towards a nonlinearly saturated state. Linear MHD codes can characterise the initial response of the plasma to MHD instabilities, or {Resonant Magnetic Perturbation (RMP)} coils, but the final displacement of the nonlinear state remains somewhat arbitrary. Ideal MHD equilibrium codes like VMEC{\cite{hirshman1983steepest}} enforce the preservation of nested flux surfaces, which can be a severe limitation to their physical validity in this context. {Extensive cross-code validation studies studies of RMPs\cite{turnbull2013comparisons, reiman2015tokamak}  on the DIII-D tokamak suggest that the differences found between VMEC and other models is mainly due to poor resolution of localised currents at low order rational surfaces \cite{lazerson2016verification, lopez2020validation}}.

{On the other hand, VMEC is commonly used in RMP studies as a first order approximation \cite{lopez2019icrf}, and the expected ideal screening on resonant surfaces has been observed in VMEC solutions, despite resolution constraints\cite{willensdorfer2016plasma}}. With respect to MHD instabilities, which are not externally driven by RMPs, previous studies using the code have been successful in identifying saturated core (1, 1) ideal MHD instabilities. {Similar saturated core instabilities have been observed on the NSTX and MAST tokamaks \cite{menard2006observation, chapman2010saturated}}. The helical core instabilities found using VMEC have been linked to the experimentally observed `snakes' in JET \cite{weller1987persistent}, and stationary states near to the MHD stability threshold of sawtooth crashes \cite{reimerdes2006sawtooth}. The equilibrium approach has been compared with results of nonlinear simulations using XTOR \cite{lutjens2010xtor}, showing reasonable agreement between the different approaches, when ideal MHD assumptions are used in the nonlinear code \cite{brunetti2014ideal}. Following this study, the different modelling approaches were linked to experimental results, where it was shown that the imposed ideal MHD assumption was valid for experimentally observed modes in TCV \cite{graves2012magnetohydrodynamic}. This implies that the equilibrium approach can be used for more advanced, experimentally relevant studies, considering confinement in the presence of saturated MHD instabilities.

More recently external instabilities such as {Edge Harmonic Oscillations} (EHOs) have been investigated using the free boundary 3D VMEC equilibrium approach \cite{cooper2016saturated}. {EHOs are typically interpreted as benign low-n kink/peeling modes that are localised in the plasma edge region \cite{snyder2007stability}. They are generally observed during quiescent H-mode (QH-mode) operation, first demonstrated on the DIII-D tokamak \cite{burrell2001quiescent}. The QH-mode successfully avoids type-I ELMs, while maintaining good thermal confinement, and impurity control, offering an attractive operational scenario for future devices. For this reason, understanding the mechanisms behind EHOs is important for optimising the QH-mode and reaching improved confinement scenarios.} 

{Modelling of EHOs using VMEC} was progressed further in Ref. \onlinecite{kleiner2018free} where the amplitude of the VMEC perturbations were compared with an analytical description of non-linear external kink modes. Key to this work was the conversion of the VMEC spectra into straight field line coordinates, and the construction of associated Fourier spectra.  It was found in Ref. \onlinecite{kleiner2019current} that these current driven modes disappear for more realistic equilibria which include a separatrix, but it was confirmed using the KINX code \cite{degtyarev1997kinx} that the pressure driven {exfernal (external-infernal \cite{brunetti2019excitation})} modes robustly continue to exist. {It has been shown in previous studies that exfernal modes can be distinguished from current driven external kinks by their pressure drive in regions with low magnetic shear. This drives instabilities of the infernal kind, and thus generates strong poloidal mode coupling between neighbouring rational surfaces \cite{kleiner2019current}. As such, even when the edge safety factor is only slightly above a given rational surface $q=q_\mathrm{0}$, a $(nq_\mathrm{0}+1,n)$ external perturbation can still be observed.} The application of straight field line coordinates in Ref. \onlinecite{kleiner2019current} demonstrated that exfernal modes have a dominant poloidal mode number, and this in turn enabled the comparison of mode structures and existence conditions across VMEC-3D equilibria, KINX linear code with separatrix and linear analytical exfernal modes.

While these results are encouraging, a comparison with a nonlinear MHD code for these external modes is desirable to clarify whether nonlinear effects do not influence the results significantly. {This comparison is important, because, unlike RMPs, where the nonlinear state is driven externally, and internal modes, which are dominated by the (1, 1) internal kink instability, external modes are often global, and governed by multiple linearly independent modes. In such a way, external MHD instabilities are expected to be more challenging to model using VMEC, because the global nature of the mode increases the threat of nonlinearly triggering additional MHD activity, and there are likely to be multiple competing solutions for the nonlinearly saturated state, each governed by a different dominant toroidal harmonic.} {Nonlinear triggering has been observed in previous simulation studies of ELMs \cite{holzl2012reduced, krebs2013nonlinear}, and has been shown to be important in determining the dominant toroidal harmonics contributing to the ELM crash. As these dynamic effects are not captured explicitly by the VMEC model, confirming that a saturated state is also obtained using a nonlinear code, and that the nonlinear structure is similar to the VMEC result, is considered instructive.} 

The nonlinear MHD code JOREK \cite{hoelzl2021jorek} is suitable for a comparison with VMEC results. {By coupling JOREK to the STARWALL code \cite{hoelzl2012coupling}, free boundary MHD activity can be modelled. Using this approach, experimental observations of EHOs have been reproduced in past studies \cite{garofalo2015quiescent, liu2015nonlinear}}. In this paper, we focus on the comparison of saturated external modes in tokamak equilibria observed using JOREK and VMEC, to further understand the region of validity of both approaches for modelling ideal MHD external modes {and the extent to which results of the two approaches agree. In particular, the influence of viscoresistive effects and diamagnetic flows are studied in the JOREK simulations to understand the influence they can have on the nonlinear dynamics. The results may be used to gain an understanding of how the MHD activity is modified as the assumptions of VMEC, namely ideal MHD and closed magnetic flux surfaces, are relaxed.}

The rest of the paper is outlined as follows. In Section \ref{sec:numerical_methods}, the numerical methods used in JOREK and VMEC are described, paying particular attention to the conditions for the VMEC computations to be physically valid. A (5, 1) external kink is then modelled in Section \ref{sec:epfl_kink}. For this simple case, where the linear dynamics are governed by the $n=1$ toroidal harmonic alone, good agreement is found between the two codes {when the the assumed Lundquist number across the plasma region in JOREK is sufficiently large, such that the dynamics can be expected to produce a similar result to the ideal MHD limit over the simulated timescale}. In such a way, it can be said that the result observed in JOREK is consistent with the previous validation for this test case in the ideal MHD limit. It is then shown that the dynamics are modified when including a Spitzer resistivity, because higher toroidal harmonics which are unstable to ballooning mode instabilities interfere with the kink dynamics. As anticipated by theoretical studies \cite{brunetti2019excitation}, the inclusion of realistic diamagnetic flows can suppress this higher n mode, such that the kink structure observed in VMEC is similar even when these additional effects are included in the nonlinear MHD calculations. 

A more challenging EHO case is then modelled in Section \ref{sec:eho_comparison}. For this case, multiple low n toroidal harmonics are linearly unstable, such that mode coupling is expected to have a more significant effect. The $n=1$ instability, which dominates the VMEC solution, has a smaller linear growth rate than the other low-n toroidal harmonics. {As a result, the saturation of the $n=2$ and $n=3$ mode delays the onset of the $n=1$ instability in JOREK simulations, such that resistive effects can play a role in the dynamics. For simulations with higher Lundquist numbers across the plasma edge region, the $n=2$ perturbation was found to dominate on the ideal MHD timescale.} Extending simulations into hybrid timescales where resistive effects can play a role, $n=1$ dominant perturbations could be found. These resistive saturated states show a strong ergodisation of the plasma edge region, but this effect does not lead to a significant loss in confinement, such that the final state has the characteristics of an EHO. Comparing with VMEC, the magnetic energy spectra of the two approaches show a deviation in the energy of higher toroidal harmonics, and the expected toroidal dependence of the pressure is not observed using the equilibrium approach. Despite these limitations, the saturated states observed by the two codes are reasonably similar. {Linear studies of the influence of diamagnetic flows show that the higher n modes are stabilised by this effect, giving some indication that flows would improve the agreement between the approaches, by reducing the width of the toroidal magnetic energy spectrum}. The paper is concluded with an outlook for future work in Section \ref{sec:conclusion}.

\section{Numerical methods}  \label{sec:numerical_methods}

\subsection{Numerical model in VMEC free-boundary code}

The algorithm implemented in VMEC is well documented in Ref. \onlinecite{hirshman1983steepest}, and the methods applied in this paper follow the approach of previous studies \cite{kleiner2019current, strumberger2014mhd}. A brief review of this approach follows in this section. 

The ideal MHD potential energy, $W_{\mathrm{mhd}}$, can be written as

\begin{equation} \label{eq:mhd_energy}
    W_{\mathrm{mhd}} = \int \frac{B^2}{2 \mu_\mathrm{0}} + \frac{p}{\gamma - 1}\ dV,
\end{equation}

{where $\gamma$ is the ratio of specific heats.} VMEC minimises $W_{\mathrm{mhd}}$ of the plasma and vacuum region up to a prescribed level of accuracy in the ideal MHD force balance equation

\begin{equation} \label{eq:force_residual}
    \mathbf{j} \times \mathbf{B} - \nabla p = 0,
\end{equation}

where ${j}$, $\mathbf{B}$ and $p$ are the current, magnetic field and plasma pressure respectively. It is well known that ideal MHD equilibria are overconstrained by the equilibrium profiles for pressure, toroidal plasma current, and rotational transform, $\iota$, such that only two of these profiles need to be specified in order to define the last for a given plasma boundary. 

In the free boundary version of the VMEC code, the plasma boundary is allowed to evolve, keeping two of the three equilibrium profiles fixed, while minimising $W_{\mathrm{mhd}}$ \cite{hirshman1986three}. A representation of the vacuum magnetic field is necessary in order to carry out free boundary computations. For the computations presented here, the use of a coil set was avoided by using the EXTENDER code \cite{drevlak2005pies} to compute the vacuum magnetic field representation.

For MHD unstable equilibria, it is possible for the convergence algorithm to find a nonlinearly perturbed state. This happens because the MHD energy can be further minimised by physical MHD perturbations, if the initially targeted axisymmetric equilibrium is ideal MHD unstable. In this case, the equilibrium will converge to a new equilibrium, which is physically interpreted as a nonlinearly saturated MHD instability. Typically when searching for this state, a perturbation is added to the otherwise axisymmetric initial guess for the equilibrium. This can either be through a small RMP field, or by defining an initially non-axisymmetric axis, or plasma boundary. {With this perturbation to the initial guess of the equilibrium state}, the saturated non-axisymmetric equilibrium can be found more easily. For this study, a $n=1$ helical axis perturbation is used.

Before proceeding to the comparison of the two codes in Section \ref{sec:epfl_kink} and Section \ref{sec:eho_comparison}, it is important to justify the use of VMEC computations to model nonlinear MHD phenomena. In particular, during VMEC iterations, the equilibrium is modified by a minimisation of the ideal MHD equilibrium energy, without use of the full ideal MHD time evolution equations. In such a way, the dynamics associated with the momentum of the plasma, and the evolution of its kinetic energy are neglected. For this reason, it is difficult to justify that the trajectory of subsequent iterations follows a physical path. In order to use this approach for nonlinear MHD studies, a physical link needs to be enforced between the unperturbed axisymmetric equilibrium, and the final perturbed state of the free boundary computation. 

As VMEC assumes the ideal MHD force balance, a reasonable choice for this constraint is to assume the conservation of helicity during the computation, because this quantity is conserved during the evolution of ideal MHD instabilities. It is shown in Appendix \ref{app:helicity_conservation} that this can be achieved by fixing $\iota$ during the computations. As such, all VMEC computations in this study have fixed the rotational transform. {As two of the three profiles need to be defined in the VMEC algorithm, the pressure profile is then also artificially constrained, while allowing the toroidal current profile to change as the equilibrium is converged.}

The results reported in this paper have assumed up-down symmetry, and used 311 radial grid points, 15 poloidal and 6 toroidal harmonics. {The equilibria are considered converged when the force residuals in VMEC fall below $f_{\mathrm{tol}} = 5\times10^{-17}$.} These are similar resolution parameters to those used in previous studies of low-n external MHD perturbations \cite{kleiner2019current}. Increasing the toroidal resolution to $n=10$, similar nonlinear perturbed states were found as the results shown in this paper.

\subsection{Numerical model in JOREK}

The viscoresistive reduced MHD model used in this study is outlined in detail in Section 2.3.1 of Ref. \onlinecite{hoelzl2021jorek}. Unless stated otherwise, all variables in the equations of this section have the same definition as in this reference. This reduced MHD model has been validated against the full MHD version of JOREK for a variety of test cases \cite{pamela2020extended}. This exercise showed that the reduced model is sufficient for capturing the nonlinear dynamics of external modes in typical tokamaks. {The reduced model leads to the following strong form of the system of equations to be evolved in time.}

\begin{equation}
    \frac{\partial \rho}{\partial t} = -\nabla\cdot\left(\rho \mathbf{V}\right) + \nabla \cdot \left(D_\bot \nabla_\bot \rho \right) + S_\rho
\end{equation}

\begin{equation} \label{eq:pol_momentum}
\begin{split}
    \mathbf{e_\phi}\cdot\nabla \times \Bigg( \Bigg.\rho \frac{\partial \mathbf{V}}{\partial t} = & -\rho\left(\mathbf{V}\cdot \nabla \right)\mathbf{V} - \nabla \left(\rho T \right) + \mathbf{J} \times \mathbf{B} \\
    & + \mu \Delta \mathbf{V} + S_\mathrm{V} \Bigg. \Bigg)
\end{split}
\end{equation}

\begin{equation} \label{eq:par_momentum}
\begin{split}
    \mathbf{B}\cdot \Bigg( \Bigg. \rho \frac{\partial \mathbf{V}}{\partial t} = &-\rho\left(\mathbf{V}\cdot \nabla \right)\mathbf{V} - \nabla \left(\rho T \right) + \mathbf{J} \times \mathbf{B} \\
    & + \mu \Delta \mathbf{V} + S_\mathrm{V} \Bigg. \Bigg)
\end{split}
\end{equation}

\begin{equation}
\begin{split}
    \rho \frac{\partial T}{\partial t} = &-\rho \mathbf{V} \cdot \nabla T - \left(\gamma - 1\right) \rho T \nabla \cdot \mathbf{V} \\
    &+ \nabla \cdot \left(\kappa_\bot \nabla_\bot T + \kappa_\parallel T \nabla_\parallel T \right) + S_\mathrm{T}
\end{split}
\end{equation}

\begin{equation} \label{eq:psi_eq}
    \frac{1}{R^2} \frac{\partial \psi}{\partial t} = \frac{\eta}{R^2} j - \frac{1}{R} \left[u, \psi\right] - \frac{F_\mathrm{0}}{R^2} \frac{\partial u}{\partial \phi}
\end{equation}

The poisson bracket in equation \ref{eq:psi_eq} is defined as $\left[f, g\right]=\mathbf{e_{\phi}}\cdot \left(\nabla f \times \nabla g \right)$. In this model, the magnetic field is defined as 

\begin{equation} \label{eq:magnetic_field_ansatz}
    \mathbf{B} = \frac{F_\mathrm{0}}{R} \mathbf{e_{\phi}} + \frac{1}{R}\nabla \psi \times \mathbf{e_{\phi}}
\end{equation}

where $F_\mathrm{0}= R_\mathrm{0} B_{\phi 0}$ is a constant, $\psi$ is the poloidal flux, and $\mathbf{e_{\phi}}$ is the normalised toroidal basis vector in cylindrical coordinates. In such a way, the toroidal field is held constant through the simulation, while $\psi$ is evolved in time. In reduced MHD, the velocity is defined as 

\begin{equation} \label{eq:velocity_ansatz}
    \mathbf{V} = -R \nabla u \times \mathbf{e_{\phi}} + \frac{m_\mathrm{i} R}{e F_\mathrm{0} \rho} \nabla p_\mathrm{i} \times \mathbf{e_{\phi}} + v_\parallel \mathbf{B}
\end{equation}

where $u=\varphi / F_\mathrm{0}$ with $\varphi$ as the electrostatic potential, such that the first term is the $\mathbf{E} \times \mathbf{B}$ velocity. The second term is the source for diamagnetic flows, and the third governs the parallel velocity. Therefore, JOREK simulations only need to solve for a system of five unknowns ($\psi$, $u$, $v_\parallel$, $T$, and $\rho$) for experimentally relevant simulations of such modes. 

For the comparison of JOREK with VMEC, the system of equations has been simplified even further. {The VMEC computations in this study do not include equilibrium flows, such that the diamagnetic flow term in equation \ref{eq:velocity_ansatz} is neglected in the main comparison of the two approaches}. In addition, the dynamics of the simulated external modes is strongly dominated by the perpendicular direction, such that the parallel velocity in equation \ref{eq:velocity_ansatz} can also be neglected. Unless stated otherwise, the simulations carried out in this paper are run without these two terms. They are only included in Section \ref{sec:jet_kink_experimental_model} and Section \ref{sec:eho_flows} to assess the influence of more experimentally relevant parameters on the dynamics.

{With respect to the boundary conditions used in JOREK, Dirichlet boundary conditions are applied to all variables other than the poloidal flux. A resistive wall boundary condition can be applied to the poloidal flux by coupling JOREK to STARWALL\cite{hoelzl2012coupling}.} In this study, the no wall limit is applied for the $n>0$ poloidal flux components. A Dirichlet boundary condition is used for the $n=0$ component in order to prevent a vertical displacement event being triggered, which both test cases in Section \ref{sec:epfl_kink} and Section \ref{sec:eho_comparison} would be susceptible to.

{The unperturbed equilibria from VMEC are initially reconstructed in JOREK using its built-in Grad-Shafranov solver, preserving the pressure profile, q profile and plasma shape. The artificially prescribed diffusion parameters and sources in the JOREK model can change the equilibrium condition during the time evolution, such that it is difficult to maintain similar equilibrium profiles in JOREK as in the original VMEC equilibrium. To avoid this problem, JOREK is run axisymmetrically, suppressing the anticipated MHD activity, until new equilibrium profiles are reached that remain approximately constant over the timescales of interest for this study. In order to ensure that the profiles in JOREK and VMEC are similar, the new equilibrium profiles, and plasma boundary are used to construct an updated VMEC equilibrium, which is then used in the comparison.}

\subsection{Simulation parameters used in JOREK} \label{sec:kink_jorek_params}
In Section \ref{sec:epfl_kink} and \ref{sec:eho_comparison}, the main simulations that are reported have tried to approximate the ideal MHD conditions assumed in VMEC, in order to get a more reasonable comparison between the two codes. From these starting simulations, scans with more physically meaningful parameters have been performed where necessary to understand the influence of resistivity and equilibrium flows on the nonlinear dynamics. For nonlinear MHD studies, the resistivity is normally set such that the Lundquist number is 10 to 100 times lower than the experimentally relevant value. This is because lower resistivities require a higher numerical resolution and computational cost. As this study does not attempt to compare results with an experiment, the core plasma resistivity is chosen to be $1.9382 \times 10^{-7} \Omega m$, which is considered a suitable compromise between computational cost, and physical accuracy.

To achieve similar vacuum conditions as in the VMEC computation in JOREK, where the computational domain extends beyond the plasma boundary, an artificially high resistivity needs to be set outside the plasma to approximate a perfect vacuum. {This prevents currents being induced in the vacuum region by the plasma deformation, which would stabilise the plasma motion. For the cases simulated in Section \ref{sec:epfl_kink} and Section \ref{sec:eho_comparison}, this stabilising effect is negligible for vacuum resistivities higher than $1.9382\times10^{-2}\ \Omega m$. For this reason, the ratio between the assumed core and vacuum resistivities used in simulations which approximate the conditions in VMEC is set to $10^5$.}

During the nonlinear evolution, the resistivity needs to be linked to the plasma temperature in order for the vacuum region alone to remain highly resistive. In order to do this, the initial resistivity profile is converted into a function of the $n=0$ temperature that is used in the time evolution through the nonlinear phase. This choice of resistivity is of course somewhat artificial, and only used to compare against VMEC using similar assumptions. For both test cases, the simulations are re-run with the resistivity profile modified to have a Spitzer dependence on the temperature in order to determine the effect this can have on the dynamics. The resistivity profiles used for the external kink and EHO case are shown in more detail in Section \ref{sec:jet_kink_experimental_model} and Section \ref{sec:eho_resistivity_scan}, respectively.

The simulations in the following sections used the parameters shown in Table \ref{tab:jet_kink_mhd_params}.  In JOREK, particularly small densities and temperatures are computationally challenging as the nonlinear dynamics can lead to the formation of negative values. A perfect vacuum region can therefore not be assumed. Using the lower bound for the density and temperature shown in Table \ref{tab:jet_kink_mhd_params}, these numerical issues could be avoided sufficiently to run into the nonlinear phase. {Hyper-viscosity and hyper-resistivity, $\mu_{\mathrm{num}}$ and $\eta_{\mathrm{num}}$, terms are added to equations \ref{eq:pol_momentum} and \ref{eq:psi_eq}, respectively, to improve numerical stability. These terms act to damp fourth order derivatives of the poloidal flux and fluid vorticity, $\omega=\nabla \cdot \nabla_{\mathrm{pol}} u$, respectively, in order to prevent the development of sub-grid resolution numerical structures. The prescribed values are sufficiently low so as not to influence the dynamics.}

The diffusive parameters given are those assumed in the core of the plasma. For the parallel thermal conductivity, a Spitzer-Haerm dependence is assumed. A small amount of parallel particle diffusion is also used in the simulations to approximate the influence that the neglected parallel flow would have on the particle transport. For the perpendicular coefficients, a pedestal transport barrier is approximated by reducing the diffusivity coefficients near the plasma edge. Outside the plasma, the perpendicular diffusive coefficients are artificially increased to 10 times the core value, in order to keep the vacuum density and temperature relatively low in the axisymmetric state. This is necessary because a limiter geometry is used without modelling the target, so that there are closed magnetic flux surfaces outside the plasma in the simulation domain. For this reason, the parallel transport alone would not ensure reasonable vacuum conditions are reached.

\begin{table}
    \caption{Parameters used in nonlinear MHD simulation of the external kink and edge harmonic oscillation. The diffusive parameters are the values defined at the plasma core.}
    \begin{ruledtabular}
    \begin{tabular}{@{}c|c|c|c}
    Parameter & \multicolumn{2}{c|}{External Kink} & EHO \\ \hline
            $T\ [keV]$                       & \multicolumn{2}{c|}{$0.03 - 3.52$}                      &  $0.14 - 8.12$  \\
            $n\ [\times 10^{20}]$                    & \multicolumn{2}{c|}{$0.05 - 1.02$}                                    &  $0.014 - 0.826$ \\   
            $\kappa_{\parallel}\ [kg\cdot m^{-1}\cdot s^{-1}]$   & \multicolumn{2}{c|}{$4936.0$ }                    &  $39386.0$ \\
            $\kappa_{\perp}\ [kg\cdot m^{-1}\cdot s^{-1}]$  & \multicolumn{2}{c|}{$1.755$}          &  $1.755$ \\
            $D_{\perp}\ [m^2\cdot s^{-1}]$     & \multicolumn{2}{c|}{$1.54$}                   &  $1.54$ \\
            $\eta\ [\Omega\cdot m]$             & \multicolumn{2}{c|}{$1.9382\times 10^{-7}$}          &  $1.9382\times 10^{-7}$ \\
            $\eta_{\mathrm{num}}\ [\Omega \cdot {m^3}]$              & \multicolumn{2}{c|}{$1.9382\times 10^{-12}$}                             &  $1.9382\times 10^{-12}$ \\
            $\mu\ [kg\cdot m^{-1}\cdot s^{-1}]$                  & \multicolumn{2}{c|}{$5.1594\times 10^{-7}$}         &  $5.1594\times 10^{-7}$  \\
            $\mu_{\mathrm{num}}\ [kg {m} \cdot s^{-1}]$               & \multicolumn{2}{c|}{$5.1594 \times 10^{-12}$}                           &  $5.1594 \times 10^{-12}$ \\
            $n_{\mathrm{rad}}$                & $201\ \ \ \ \ \ \ $  & $115$                                                     &  $121$                         \\
            $n_{\mathrm{pol}}$                 & $121\ \ \ \ \ \ $  & $151$                                                     &  $151$                         \\
            $n_{\mathrm{plane}}$               & \multicolumn{2}{c|}{$32$}                       &   $32$
    \label{tab:jet_kink_mhd_params}
    \end{tabular}
    \end{ruledtabular}
\end{table}

\begin{figure}
    \centering
    \includegraphics[width=0.49\textwidth]{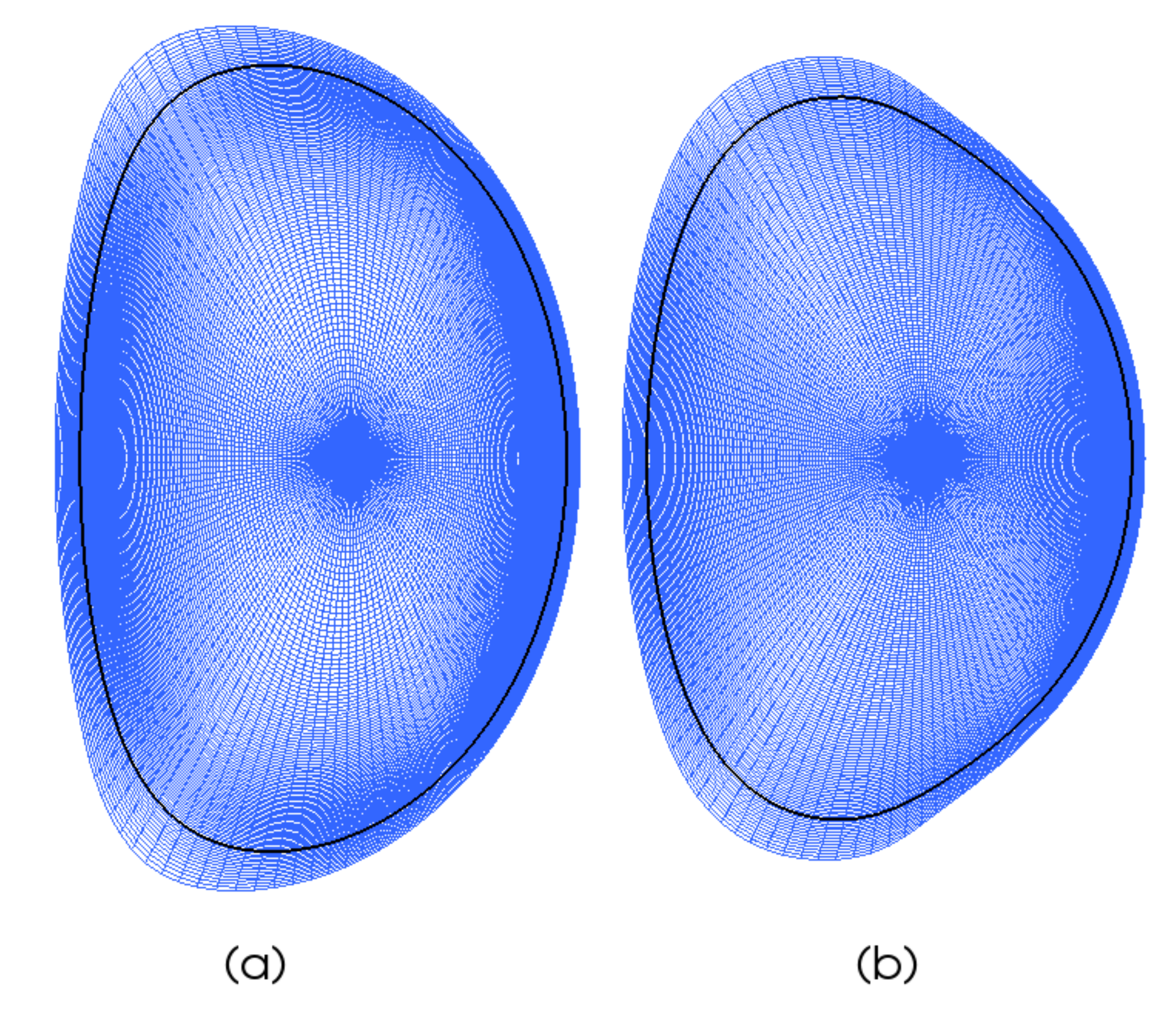}
    \caption{{Poloidal finite element meshes} used in {JOREK simulations for the} external kink (a) and EHO (b) test cases. The $\psi_\mathrm{N}=1$ surface (black) is shown to indicate the position of the plasma boundary.}
    \label{fig:jorek_meshes}
\end{figure}

Regarding resolution, the number of radial, $n_{\mathrm{rad}}$, and poloidal, $n_{\mathrm{pol}}$, grid elements were chosen to resolve the dynamics. For the external kink case shown in Section \ref{sec:kink_jorek_comparison}, a non-flux aligned polar grid with 201 radial and 121 poloidal elements was used. The dominant low-n modes which are simulated have large structures that are well resolved at this relatively high poloidal resolution. A flux aligned grid with 115 radial and 151 poloidal elements was used for the remaining simulations in Section \ref{sec:jet_kink_experimental_model}, in order to converge the linearly unstable $n=5$ mode that was found. For the EHO simulated in Section \ref{sec:eho_comparison}, the grid is aligned to the initial equilibrium flux surfaces. {The toroidal mode numbers $n=0-5$ are simulated.} 32 poloidal planes were used for both test cases, which is sufficient to resolve the representation of the toroidal harmonics. {The poloidal mesh used for the two cases is shown in Figure \ref{fig:jorek_meshes}.}

\section{Comparison of external kink} \label{sec:epfl_kink}

\subsection{VMEC Computation} \label{sec:epfl_kink_vmec}
\begin{figure}
    \centering
    \includegraphics[width=0.475\textwidth]{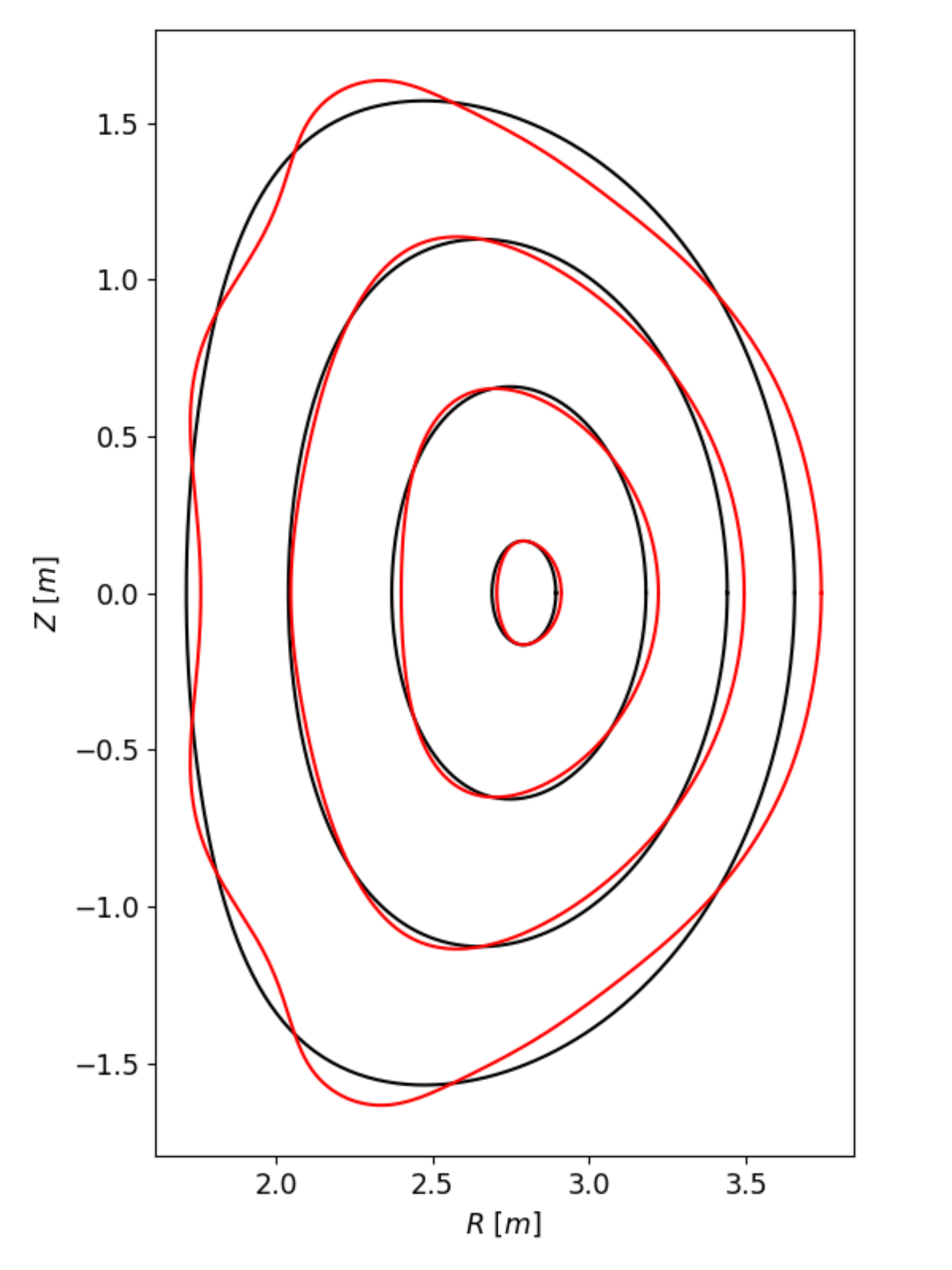}
    \small{(a)}
    \includegraphics[width=0.475\textwidth]{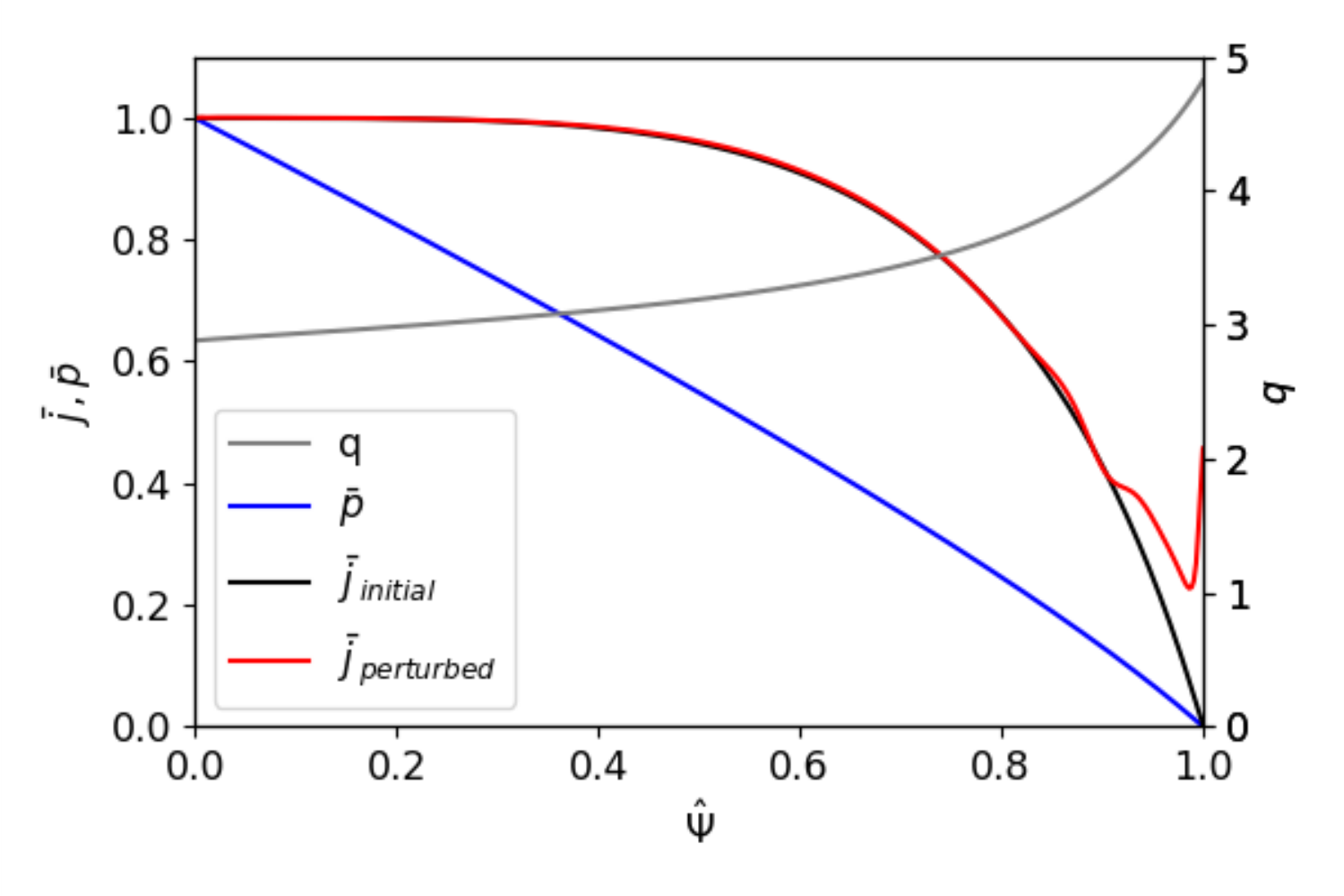}
    \small{(b)}
    \caption{Unperturbed (black) and perturbed (red) equilibrium {flux surfaces at $\sqrt{\hat \Phi} = 0.1$, 0.3, 0.7 and 1.0} (a), and radial profiles (b) {from VMEC} for the external kink case.}
    \label{fig:jet_kink_perturbation}
\end{figure}

The unperturbed equilibrium considered in this Section is a D-shaped tokamak, as shown in Figure \ref{fig:jet_kink_perturbation} (a). This test case was generated by modifying a previously studied JET-like equilibrium \cite{kleiner2018free}. {It has been shown previously that this case is a current driven external kink mode \cite{kleiner2019current}}. The equilibrium profiles of the unperturbed and perturbed equilibria are shown in Figure \ref{fig:jet_kink_perturbation} (b). A quartic current profile, and a linear pressure profile are assumed with respect to the normalised toroidal flux.

The edge safety factor is just below a value of 5, destabilising a (5, 1) external kink. As discussed in Appendix \ref{app:helicity_conservation}, the VMEC computations are carried out constraining the q profile, such that a physical link can be derived between the axisymmetric and final perturbed state.

It can be seen that the deformation leads to a sharp increase in the current density near the plasma edge. This effect is similar to the surface currents that were observed in past calculations of fast major radius compression in tokamaks, where flux conservation was also assumed \cite{albert1980adiabatic}. The reason for the increase in current density can be qualitatively understood in the following way; as the $q$ profile is fixed during the development of the 3D perturbed state, the ratio of the poloidal and toroidal pitch angles of magnetic field lines must remain effectively constant. With the major and minor radius, and the toroidal field approximately held constant, the poloidal field will vary according to

\begin{equation}
    \mu_\mathrm{0} \int \mathbf{j} \cdot \mathbf{dS} = \oint \mathbf{B} \cdot \mathbf{dl}.
\end{equation}

The total surface area is also approximately constant, while the line integral of the equilibrium flux surfaces has increased. In such a way, the total plasma current must increase to preserve the $q$ profile, leading to the spike in the current density observed in Figure \ref{fig:jet_kink_perturbation} (b). This can be interpreted as the current spike typically observed during fast MHD dynamics. In such a way, the perturbed equilibrium can be interpreted as the saturated state achieved after the fast phase, before diffusive effects or non-ideal MHD instabilities can modify the plasma state further.

\subsection{Comparison of VMEC perturbation with linear eigenfunctions} \label{sec:epfl_kink_linear}
\begin{figure}
    \centering
    \includegraphics[width=0.47\textwidth]{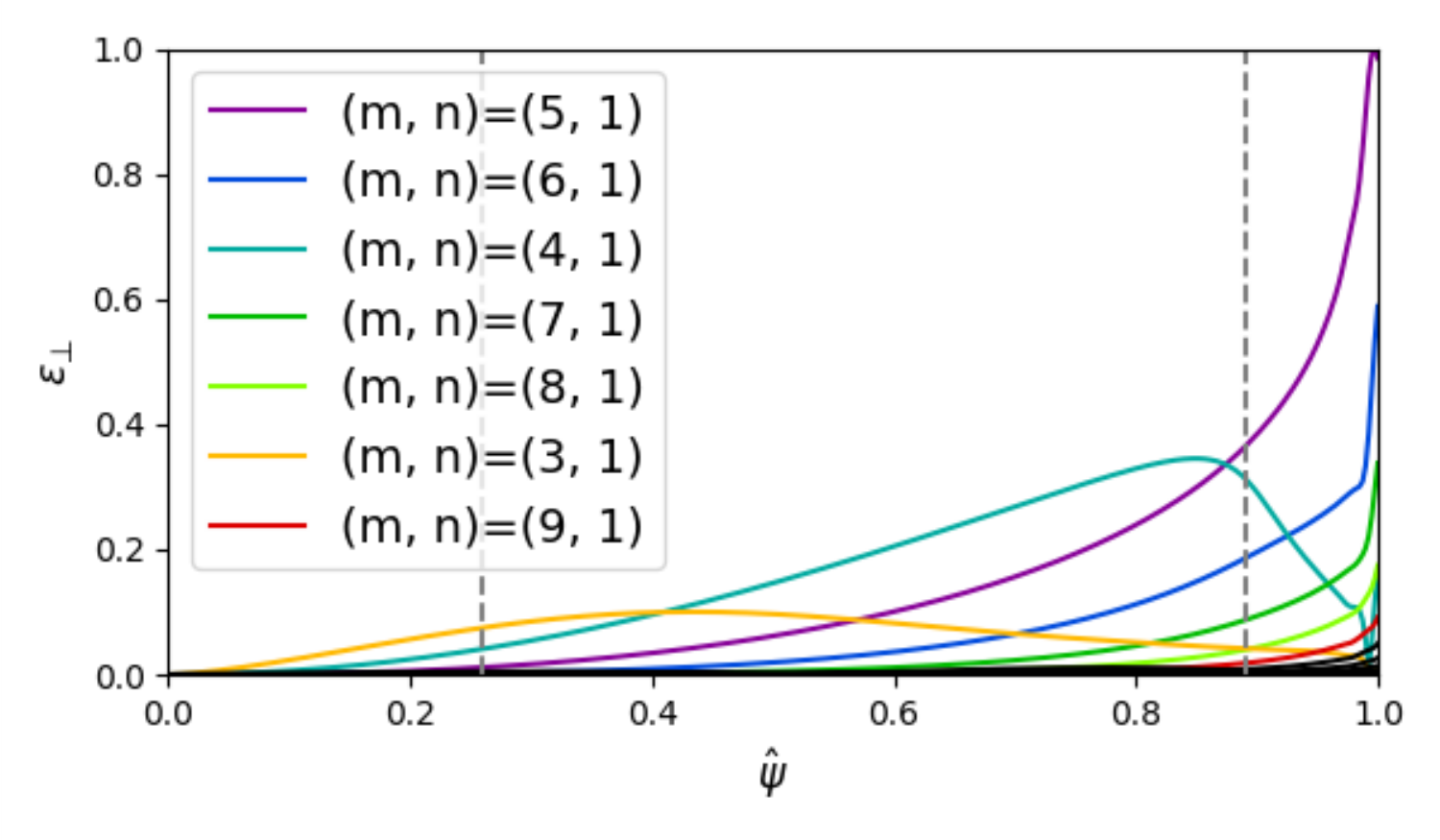}
    \small{(a)}
    \includegraphics[width=0.47\textwidth]{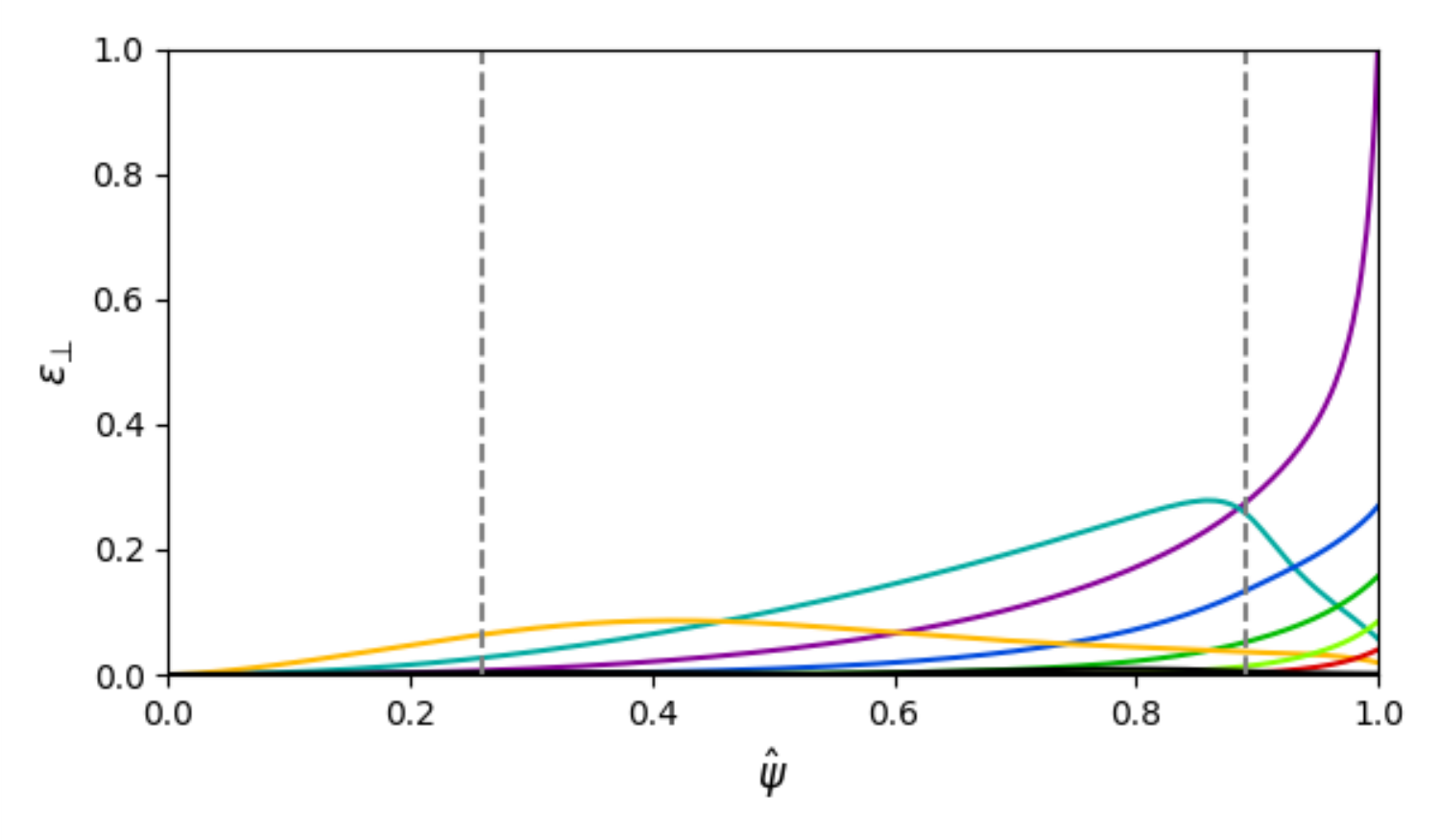}
    \small{(b)}
    \includegraphics[width=0.47\textwidth]{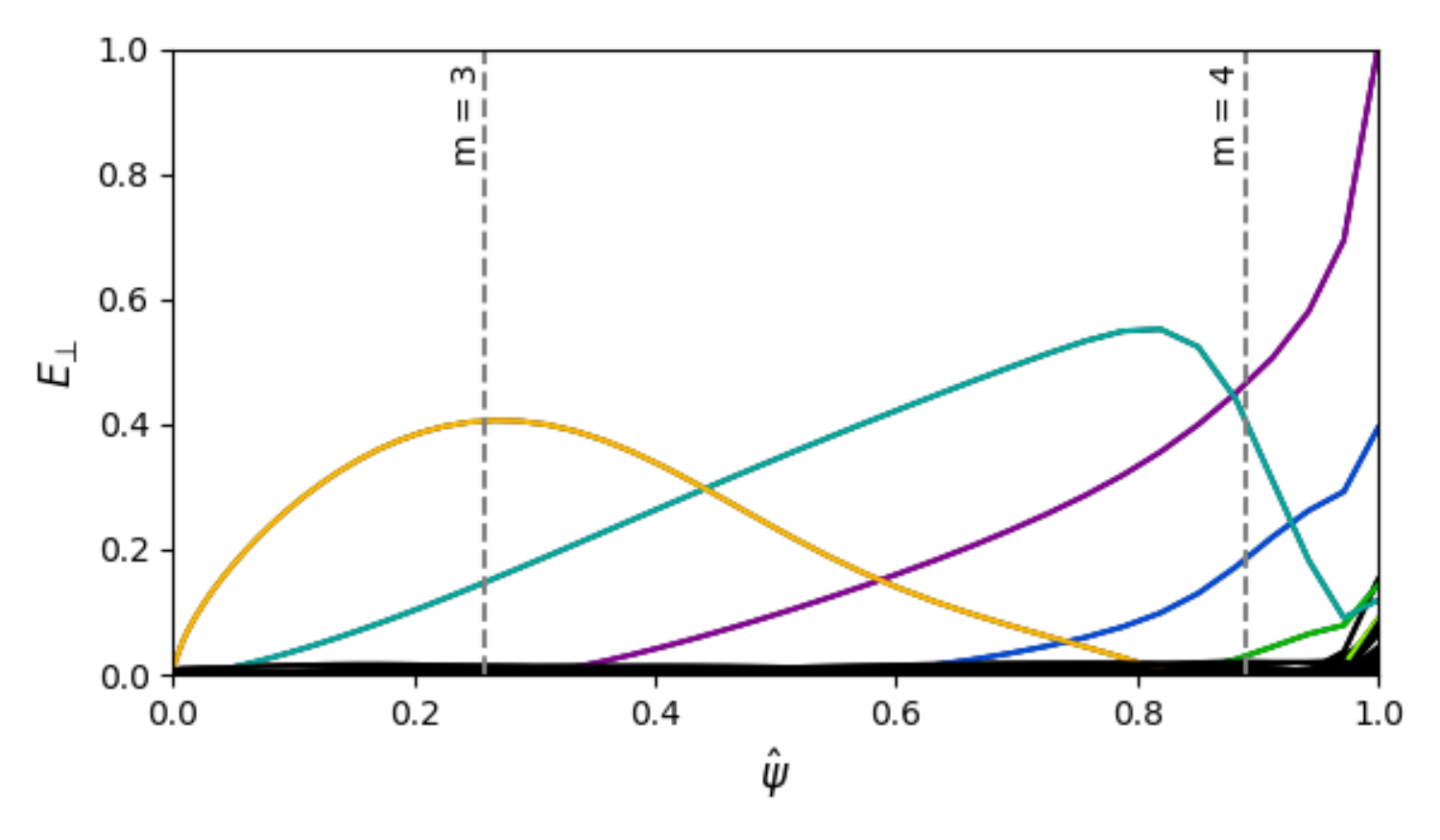}
    \small{(c)}
    \caption{Comparison of radial eigenfunctions observed in JOREK (a) during the linear phase, CASTOR3D (b), and the nonlinear perturbation observed in VMEC (c). The Fourier representation has been calculated using PEST coordinates. {The location of relevant rational surfaces with their corresponding poloidal mode number are marked by grey dashed lines.}}
    \label{fig:jet_kink_eigenfunctions}
\end{figure}

Before beginning the nonlinear comparison, the linear eigenfunction of the JOREK equilibrium is compared with other approaches to ensure that the expected instability is observed. The linear eigenfunction calculated by JOREK, and the viscoresistive linear full MHD code, CASTOR3D \cite{Strumberger2016}, are shown alongside the nonlinear Fourier representation of the VMEC perturbation in Figure \ref{fig:jet_kink_eigenfunctions}. 

{Previous studies have used similar comparisons to demonstrate the consistency of the VMEC approach with other methods \cite{kleiner2018free, kleiner2019current}. It should be noted that when comparing the codes in this way, quantitative agreement should not be expected between VMEC and the other approaches. This is because the perpendicular displacement between the stationary solutions, $E_\bot$, found in VMEC is a nonlinear result, and does not have the same meaning as the perpendicular displacement in linear theory, $\epsilon_\bot$. $\epsilon_\bot$ is dependent on the density of the plasma, while the nonlinearly saturated state found in VMEC is not. Because the density varies across the plasma region, the eigenfunctions in JOREK and CASTOR3D will have larger contributions towards the plasma boundary, where the density is lower. This leads to a larger relative displacement of the (5, 1) contribution close to the plasma edge, compared to the equivalent linear eigenfunction that would be obtained, if a constant density were assumed. While it has been verified that this effect does not significantly modify the overall structure of the linear eigenfunction, the nonlinear displacement from VMEC does not have the same dependency. It should therefore be made clear that such comparisons are only made to show that a similar mode structure is observed by the different approaches.}

{A global external kink mode is observed by all three codes, and it can be seen that there is qualitative agreement between the different approaches. While the perturbations are similar towards the plasma edge, a notable difference is that the VMEC solution appears to have a stronger internal contribution from the (3, 1) mode which is peaked closer to the $q=3$ rational surface, marked by a dashed grey line in the figure. This strong peak can also be seen in the inner flux surfaces of Figure \ref{fig:jet_kink_perturbation} (a). This is thought to be because the (3, 1) perturbation develops to a larger nonlinear amplitude in the saturated state.} 

\subsection{Comparison of flux surfaces and perturbed magnetic energies} \label{sec:kink_jorek_comparison}
\begin{figure*}
    \centering
    \begin{minipage}{0.32\textwidth}
    \includegraphics[width=\textwidth]{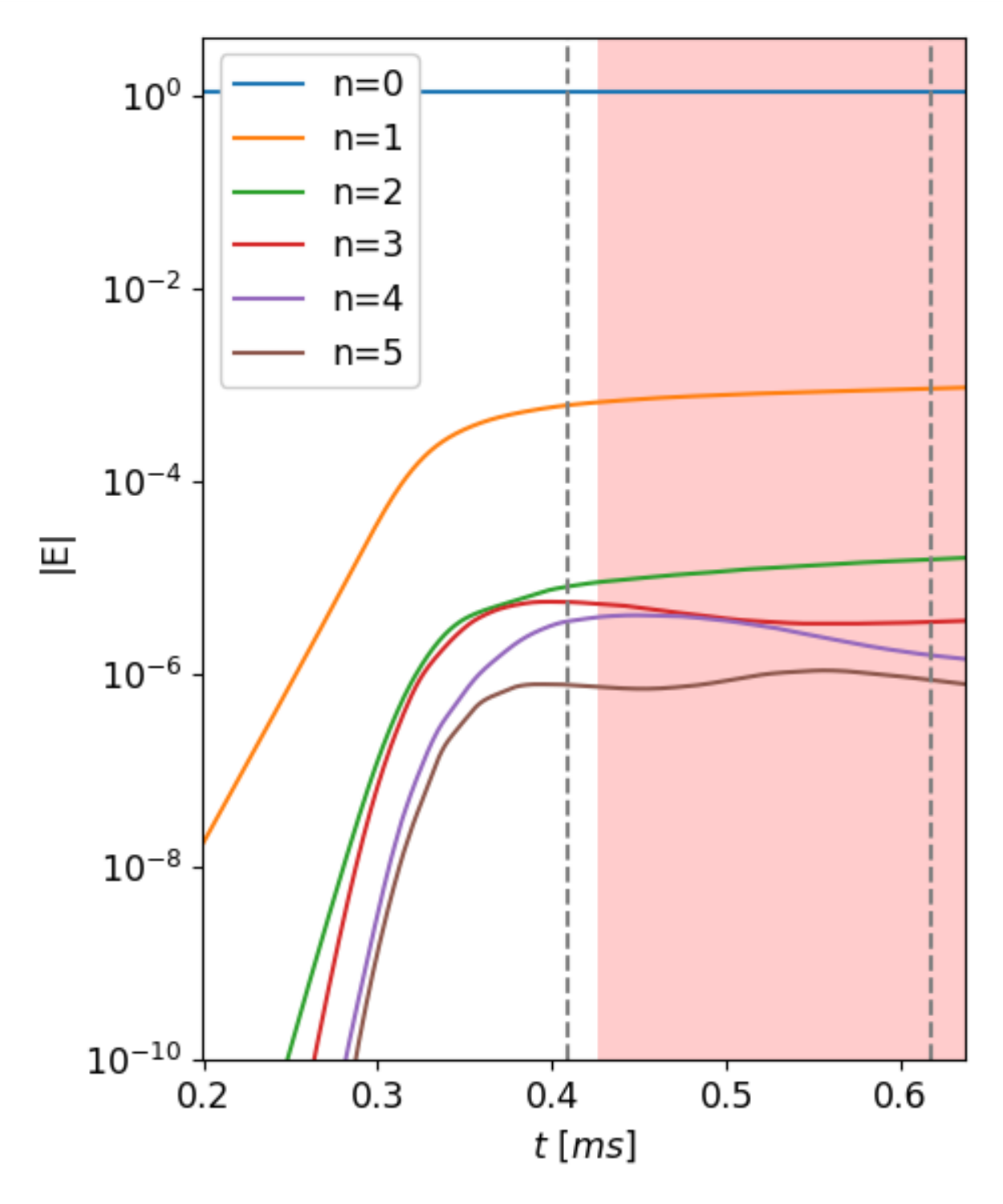}
    \centering
    \small{(a)}
    \end{minipage}
    \begin{minipage}{0.32\textwidth}
    \includegraphics[width=\textwidth]{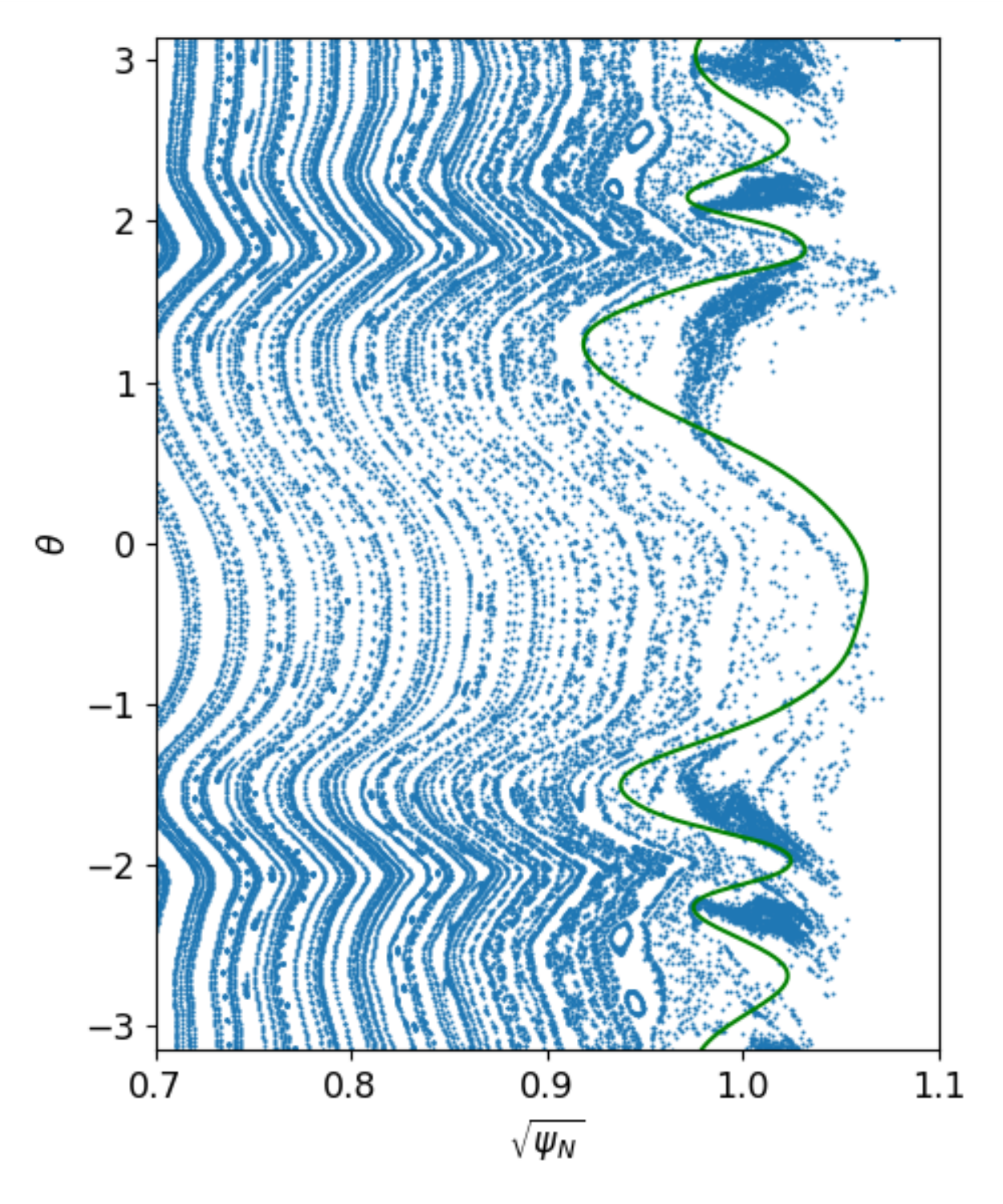}
    \centering
    \small{(b)}
    \end{minipage}
    \begin{minipage}{0.32\textwidth}
    \includegraphics[width=\textwidth]{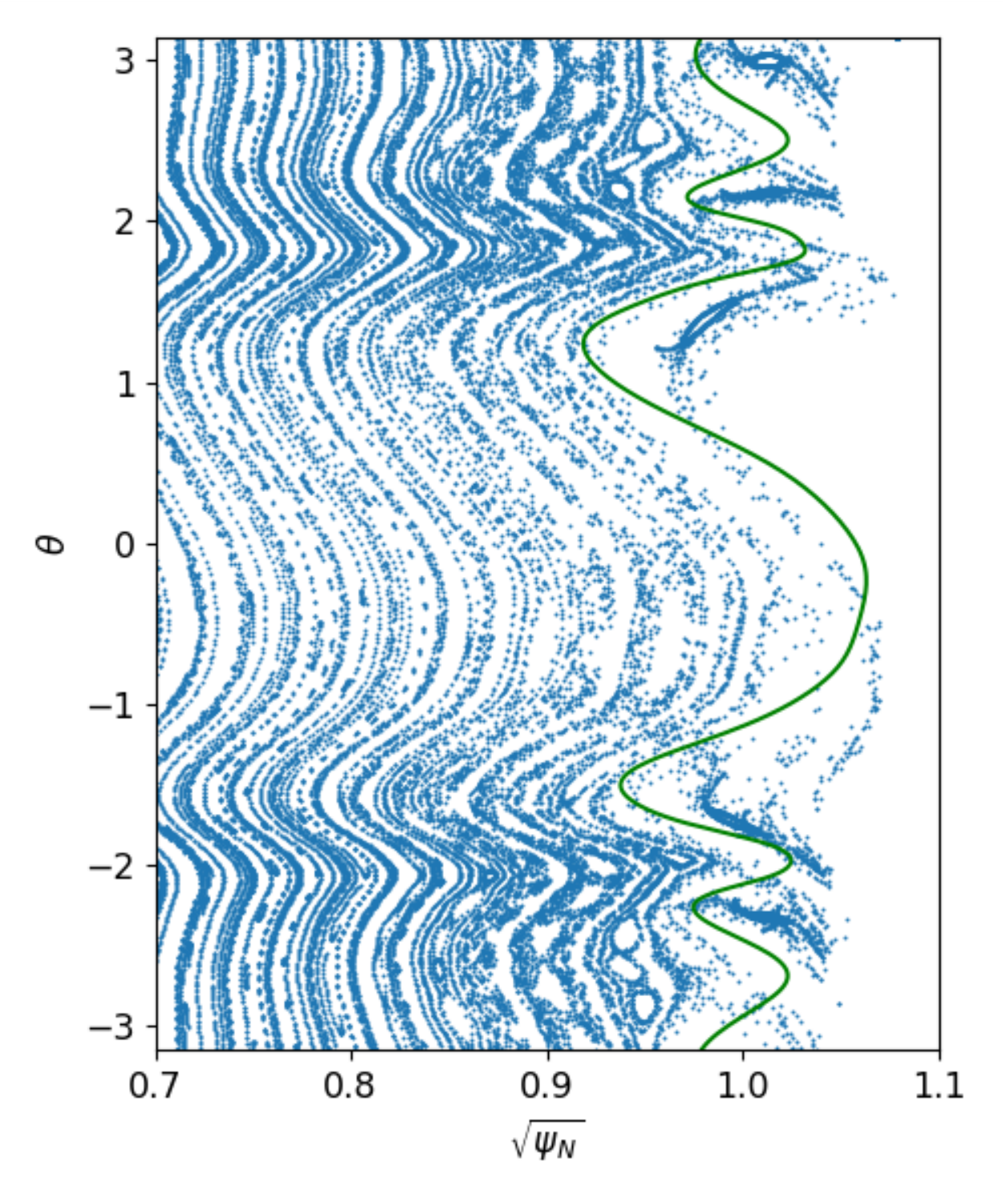}
    \centering
    \small{(c)}
    \end{minipage}
    \caption{Magnetic energy evolution of the JOREK simulation during the nonlinear phase (a). {Note that the energies are normalised by a factor $\mu_\mathrm{0}$.} The coloured regions correspond to the simulation time over which the energy spectrum is averaged over in Figure \ref{fig:jet_kink_perturbed_psi} (a). Poincar\'e comparisons are made with the perturbed VMEC equilibrium at the initial saturation of the $n=1$ mode (b) and near the end of the simulation run time (c). The times of the Poincar\'e plots are marked by grey dashed lines in Figure \ref{fig:jet_kink_poincare} (a). The last closed flux surface from VMEC (green) is overlaid on the Poincar\'e plots.}
    \label{fig:jet_kink_poincare}
\end{figure*}

To test the validity of the nonlinear perturbation further, the JOREK simulation is continued into the nonlinear phase, and the Poincar\'e surfaces are compared against the solution from VMEC. The results of this comparison are shown in Figure \ref{fig:jet_kink_poincare}. {There is good overall agreement between the flux surfaces observed in JOREK, and the predicted perturbation from VMEC}. The overall perturbation in JOREK is smaller, which is likely due to a combination of a small amount of plasma ergodisation near the plasma edge, and the pressure profile relaxing in the nonlinear evolution. In VMEC, the pressure profile is unchanged during the nonlinear evolution, which cannot be expected in JOREK due to diffusion and increased transport across the plasma boundary once the instability begins to saturate. 

\begin{figure*}
    \centering
    \begin{minipage}{0.42\textwidth}
    \includegraphics[width=\textwidth]{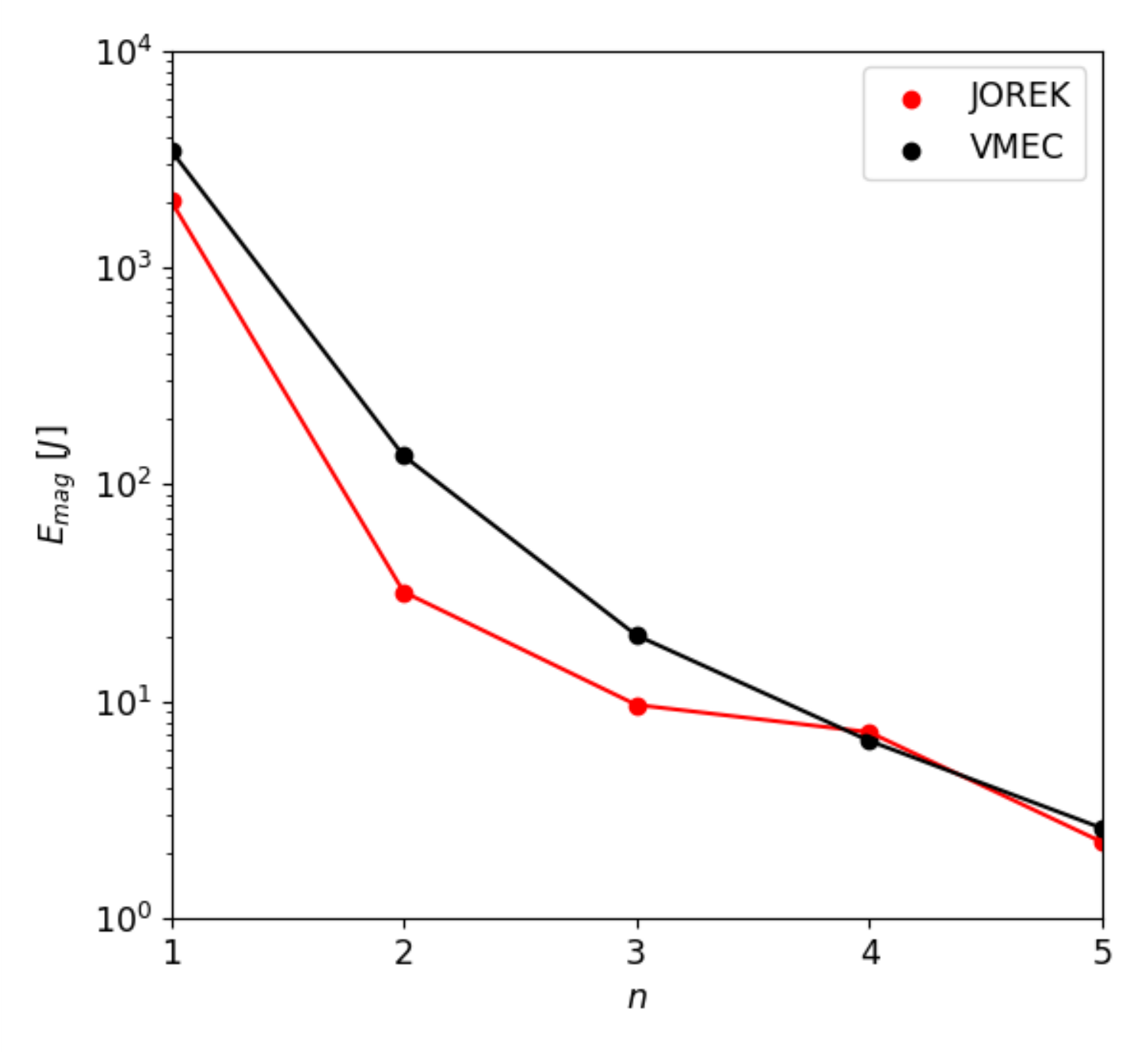}
    \centering
    \small{(a)}
    \end{minipage}
    \begin{minipage}{0.25\textwidth}
    \includegraphics[width=\textwidth]{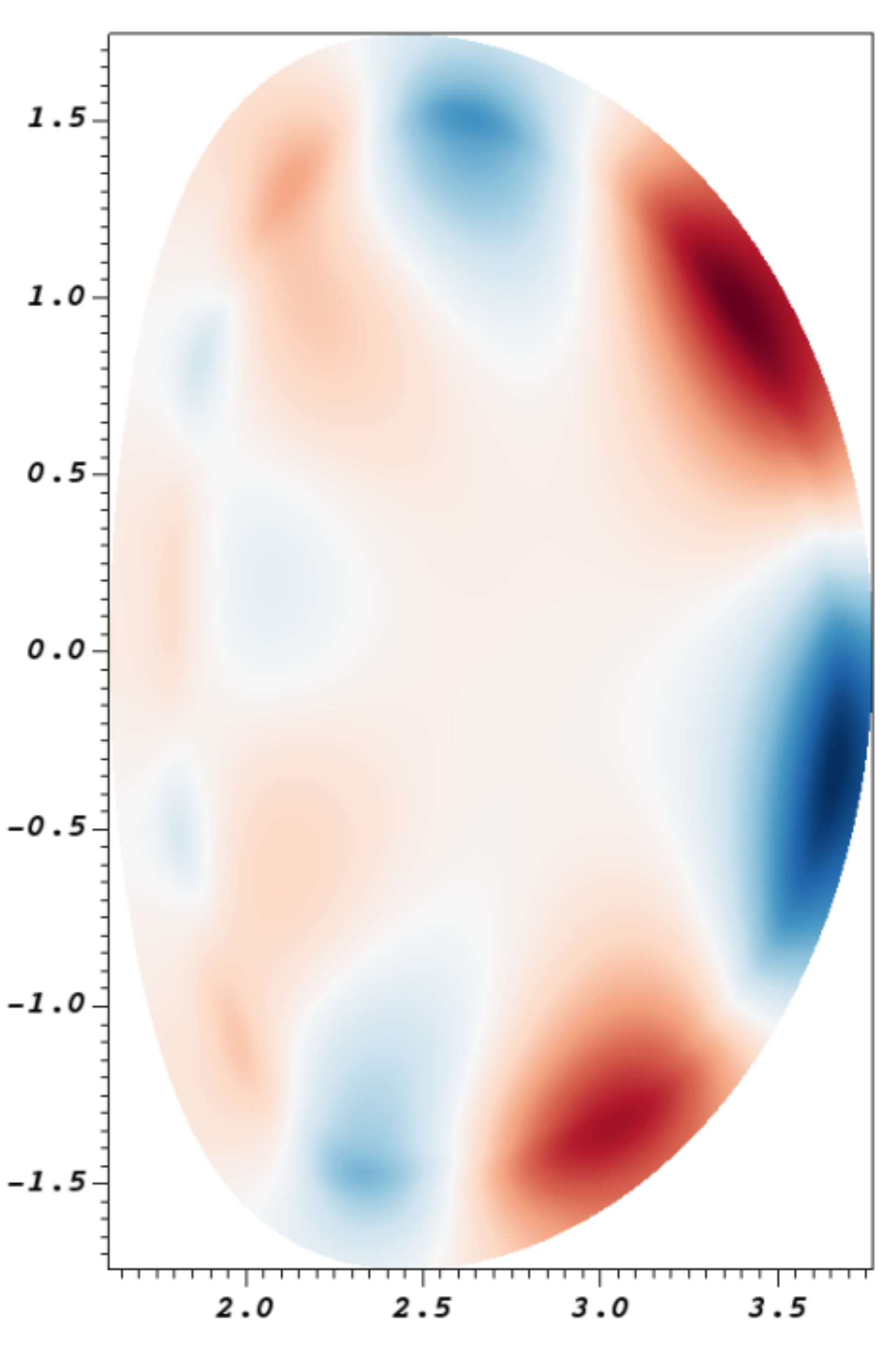}
    \centering
    \small{(b)}
    \end{minipage}
    \begin{minipage}{0.31\textwidth}
    \includegraphics[width=\textwidth]{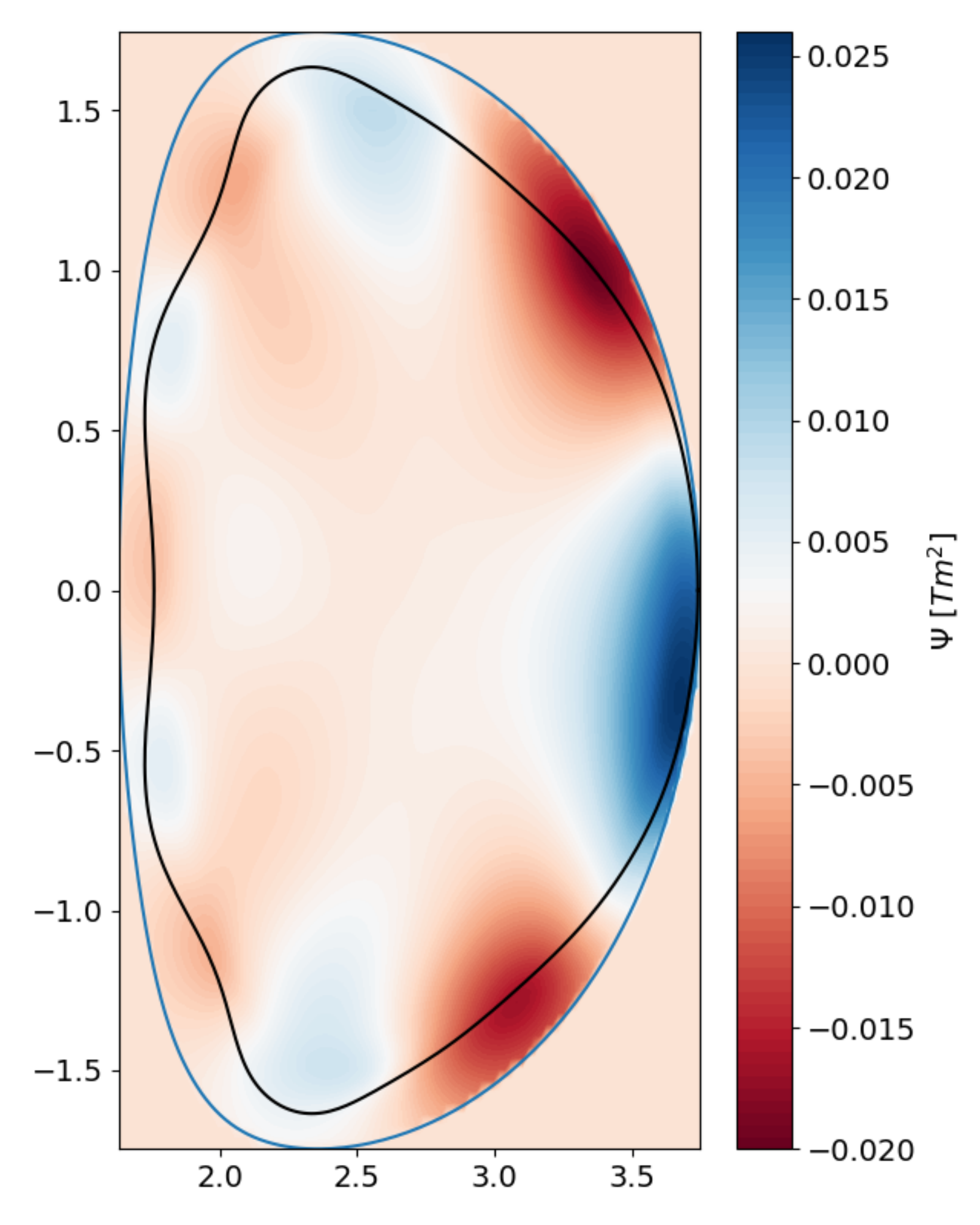}
    \centering
    \small{(c)}
    \end{minipage}
    \caption{Comparison of perturbed magnetic energy spectrum in JOREK and VMEC solutions (a). {The JOREK energy spectra is averaged over the corresponding red region shown in Figure \ref{fig:jet_kink_poincare} (a)}. The perturbed poloidal flux in JOREK at the end of the simulation time (b) is compared with the VMEC solution (c) for a comparable poloidal plane. The simulation boundary used in JOREK (blue), and the {plasma boundary} from the perturbed state {in VMEC} (black) are also shown. Note that the same colour bar is used in both pseudocolour plots.}
    \label{fig:jet_kink_perturbed_psi}
\end{figure*}

A final quantitative comparison is possible by comparing the magnetic energy spectra from the JOREK simulation with VMEC. In order to do this, the perturbed poloidal flux of the VMEC solution needs to be calculated in a similar representation as in JOREK. The magnetic field of the VMEC solution is therefore re-calculated on a $(R, Z, \phi)$ grid, using EXTENDER{, continuing the solution beyond the plasma boundary. The poloidal flux can then be solved for using}

\begin{equation} \label{eq:gs_sol}
    {\int \frac{1}{R} \nabla \psi \cdot \nabla w dA = \int w \left(\frac{\partial B_\mathrm{R}}{\partial Z} - \frac{\partial B_\mathrm{Z}}{\partial R} \right) dA.}
\end{equation} 

{Equation \ref{eq:gs_sol} is written in the weak form, where $w$ is the test function. The righthand side comes from substituting the magnetic field into the definition of $\Delta^*\psi = R\ \partial/\partial R(1/R\ \partial \psi / \partial R) + \partial^2 \psi/\partial Z^2$.} The solutions on each plane can then be Fourier transformed in the toroidal direction to get a comparable representation of the magnetic field as in JOREK. 

The magnetic energy spectrum of the two solutions is shown in Figure \ref{fig:jet_kink_perturbed_psi} (a). It can be seen that the two methods have relatively similar magnetic energies. It is expected that the JOREK simulation will have a slightly lower energy than the equilibrium approach, because of the inclusion of resistive magnetic energy dissipation near the plasma edge. The perturbed poloidal flux, omitting the $n=0$ mode, in JOREK and VMEC is compared in Figure \ref{fig:jet_kink_perturbed_psi} (b) and (c) at the poloidal plane where the mode structures observed in the two codes are approximately in phase. It can be seen that there is good agreement between the two codes, and that the perturbation is strongly dominated by the $n=1$ mode. $(5, 1)$ and $(4, 1)$ mode structures can be observed in both solutions.

\subsection{Modifications with realistic resistivity and equilibrium flows} \label{sec:jet_kink_experimental_model}
\begin{figure}
    \centering
    \includegraphics[width=0.475\textwidth]{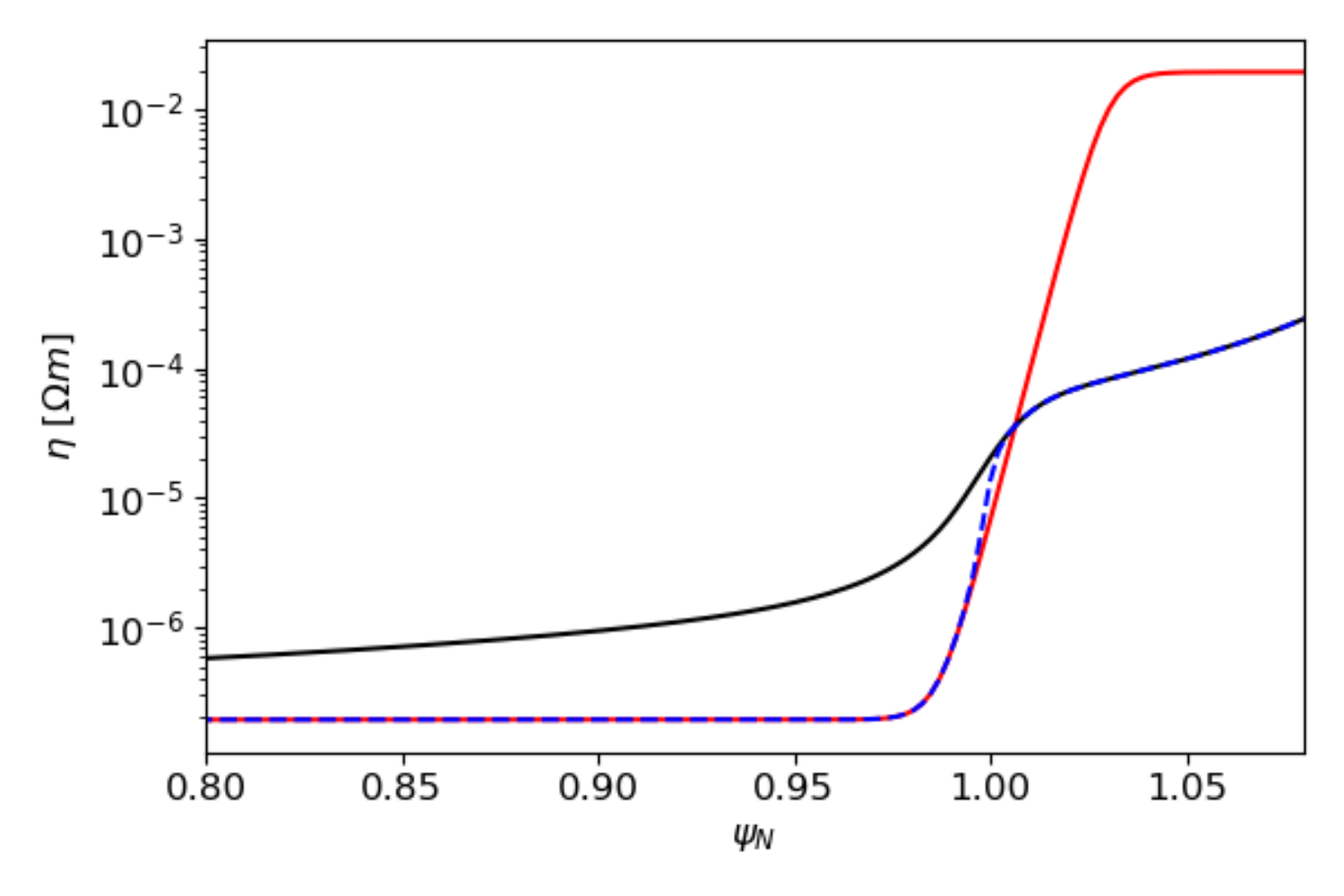}
    \small{(a)}
    \includegraphics[width=0.475\textwidth]{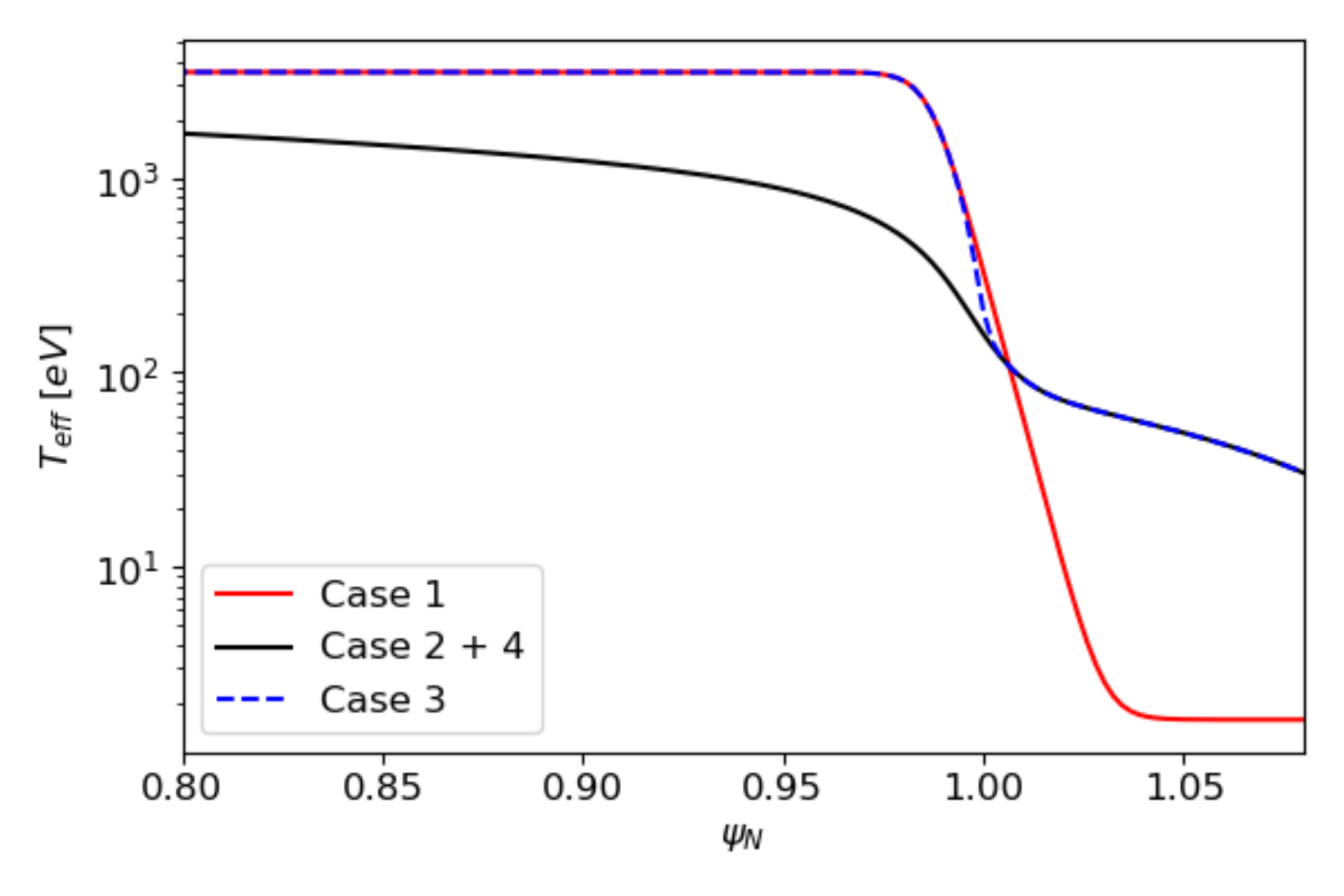}
    \small{(b)}
    \caption{Initial resistivity {(a)} and effective temperature {(b)} profiles for four simulated cases of the external kink. Case 1 is the simulation in Section \ref{sec:kink_jorek_comparison}. The resistivity profile has been artificially increased in this case, such that the effective temperature is just above $1\ eV$, which is comparable to ideal vacuum conditions. Case 2 and 4 assume a realistic Spitzer{-like} resistivity profile, such that the effective temperature this corresponds to is the prescribed plasma temperature in Table \ref{tab:jet_kink_mhd_params}. In case 3, the resistive effects inside the plasma are removed from case 2, by using a flat temperature profile up to $\psi_\mathrm{N} = 1$. This can be used to determine whether differences between case 1 and 2 are due to internal or external dynamics.}
    \label{fig:kink_resistivity_profiles}
\end{figure}

As discussed in Section \ref{sec:kink_jorek_params}, the resistivity profile was set up somewhat artificially to more accurately approximate the ideal MHD conditions in the VMEC computations, and therefore allow for a better comparison of the codes. In this section, a few additional permutations of the resistivity profile are simulated, to see how returning to a Spitzer-like profile can modify the dynamics. The simulated profiles of the resistivity are shown in Figure \ref{fig:kink_resistivity_profiles}. 

The first case has already been shown in detail in Section \ref{sec:kink_jorek_comparison}. Case 2 uses the expected Spitzer resistivity dependence inside and outside the plasma. In this case, the effective temperature $T_{\mathrm{eff}}$ that the simulated resistivity corresponds to, assuming a Spitzer dependence, is the same as the evolved temperature, $T$. Case 3 has a low resistivity inside the plasma, which transitions to a Spitzer-like resistivity outside. In such a way, internal, and external resistive effects can be differentiated using the three runs. Finally, case 4 is run with a Spitzer-like resistivity and diamagnetic flows to assess how equilibrium poloidal flows can modify the perturbed state. {As should be expected for these additional cases, which have a lower vacuum resistivity, the $n=1$ mode grows more slowly, but has a similar linear mode structure to Figure \ref{fig:jet_kink_eigenfunctions} (a).} {Note that while these profiles have the correct Spitzer-like dependence on resistivity, the absolute value of the resistivity is still significantly higher than the physical value, due to numerical constraints mentioned in Section \ref{sec:kink_jorek_params}.}

\begin{figure*}
    \centering
    \begin{minipage}{0.325\textwidth}
    \includegraphics[width=\textwidth]{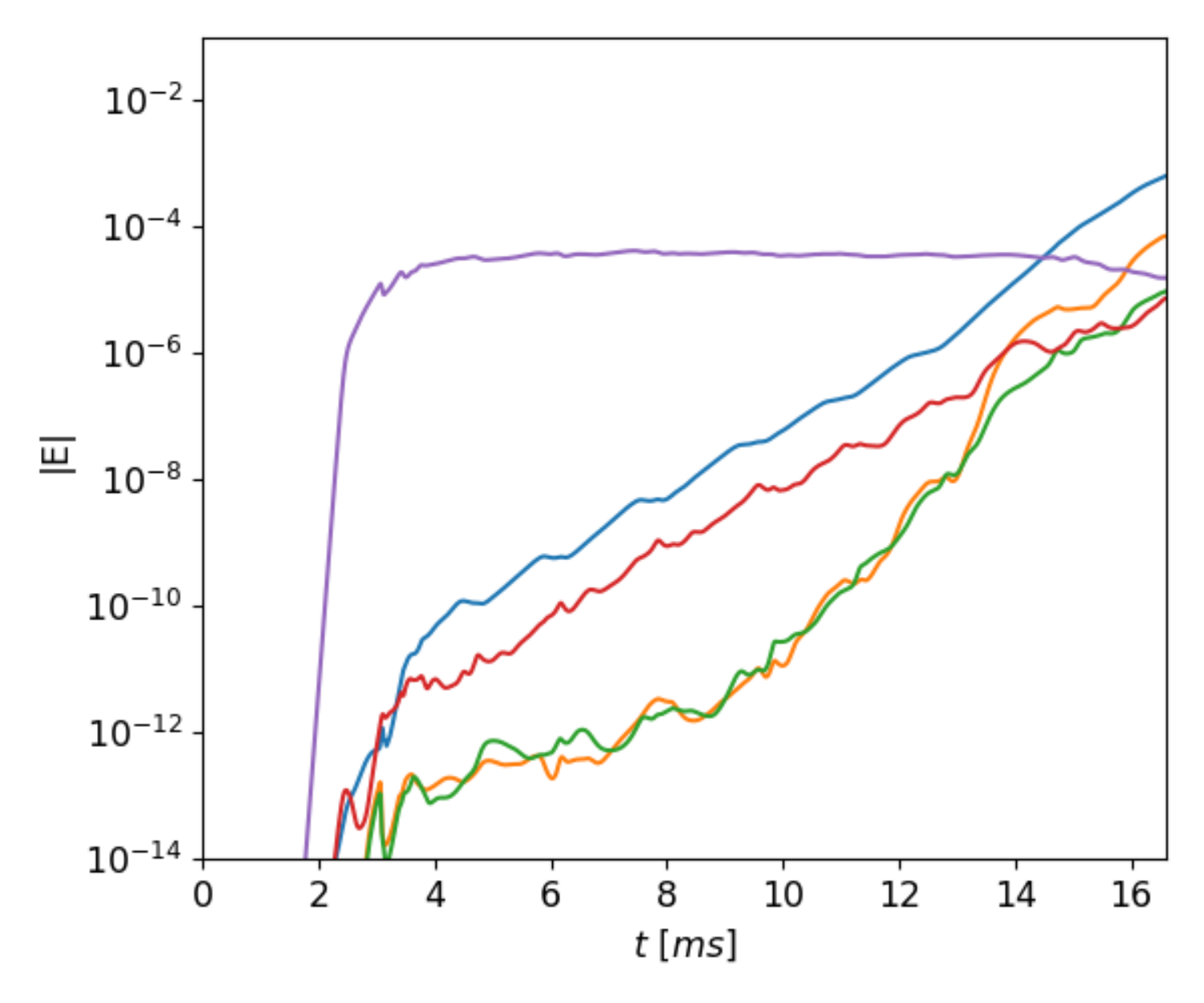}
    \centering
    \small{(a)}
    \end{minipage}
    \begin{minipage}{0.325\textwidth}
    \includegraphics[width=\textwidth]{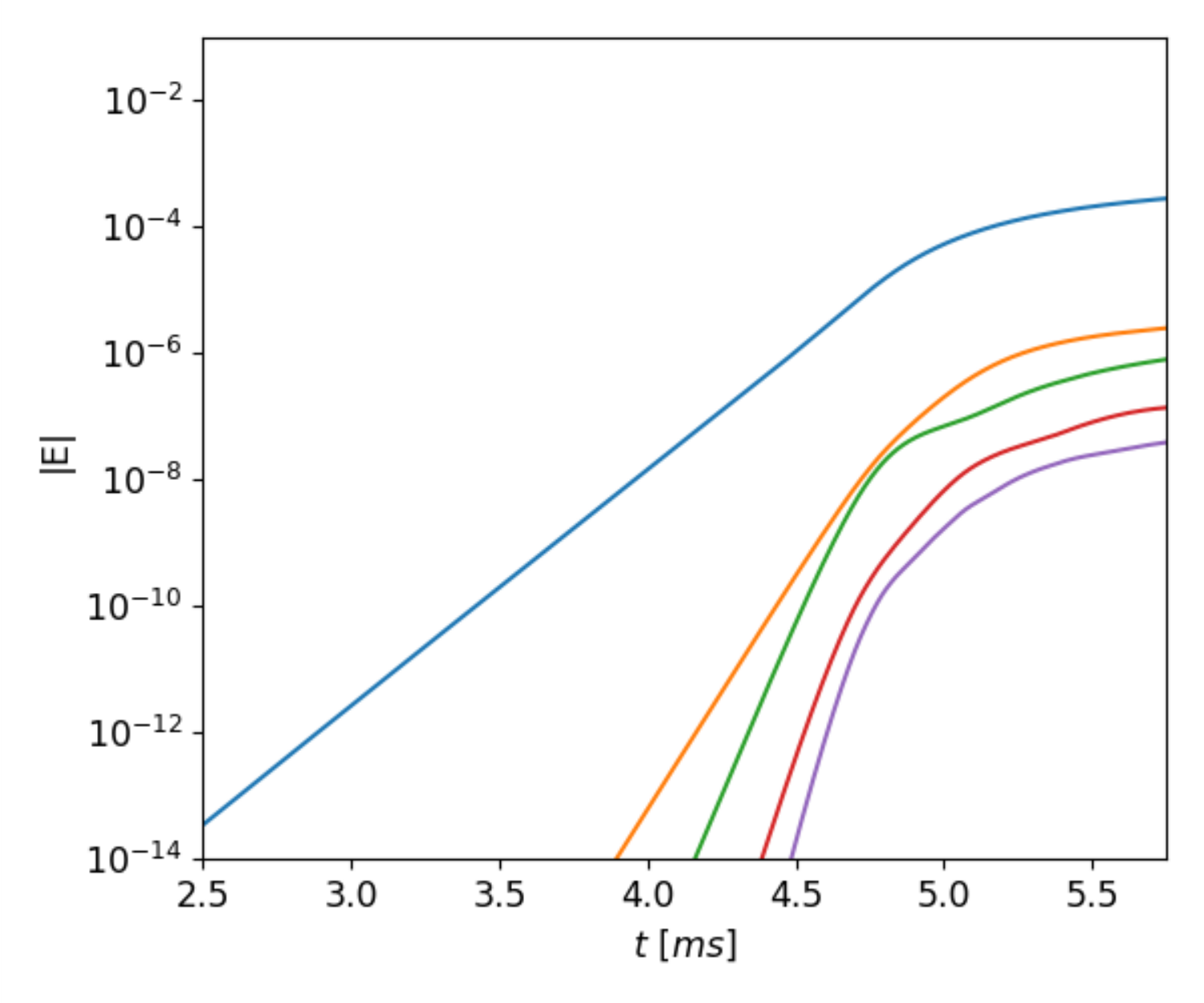}
    \centering
    \small{(b)}
    \end{minipage}
    \begin{minipage}{0.325\textwidth}
    \includegraphics[width=\textwidth]{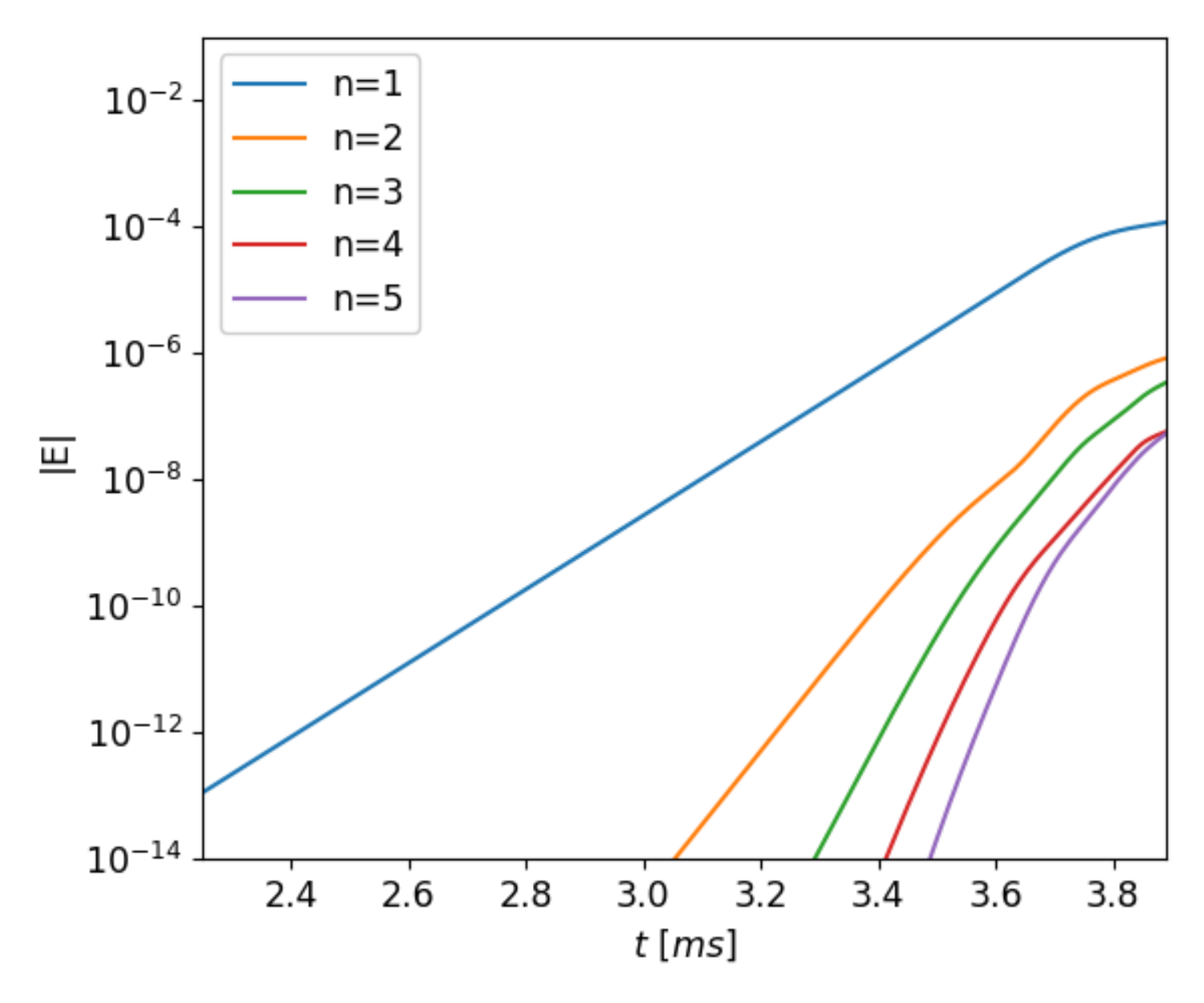}
    \centering
    \small{(c)}
    \end{minipage}
    \caption{Evolution of the toroidal magnetic energies {from JOREK} for case 2, 3, and 4 (a, b, and c) from Figure \ref{fig:kink_resistivity_profiles}.  {Note that the energies are normalised by a factor $\mu_\mathrm{0}$.} Case 1 is shown in Figure \ref{fig:jet_kink_poincare}.}
    \label{fig:jet_kink_res_E_comp}
\end{figure*}

\begin{figure}
    \centering
    \includegraphics[width=0.49\textwidth]{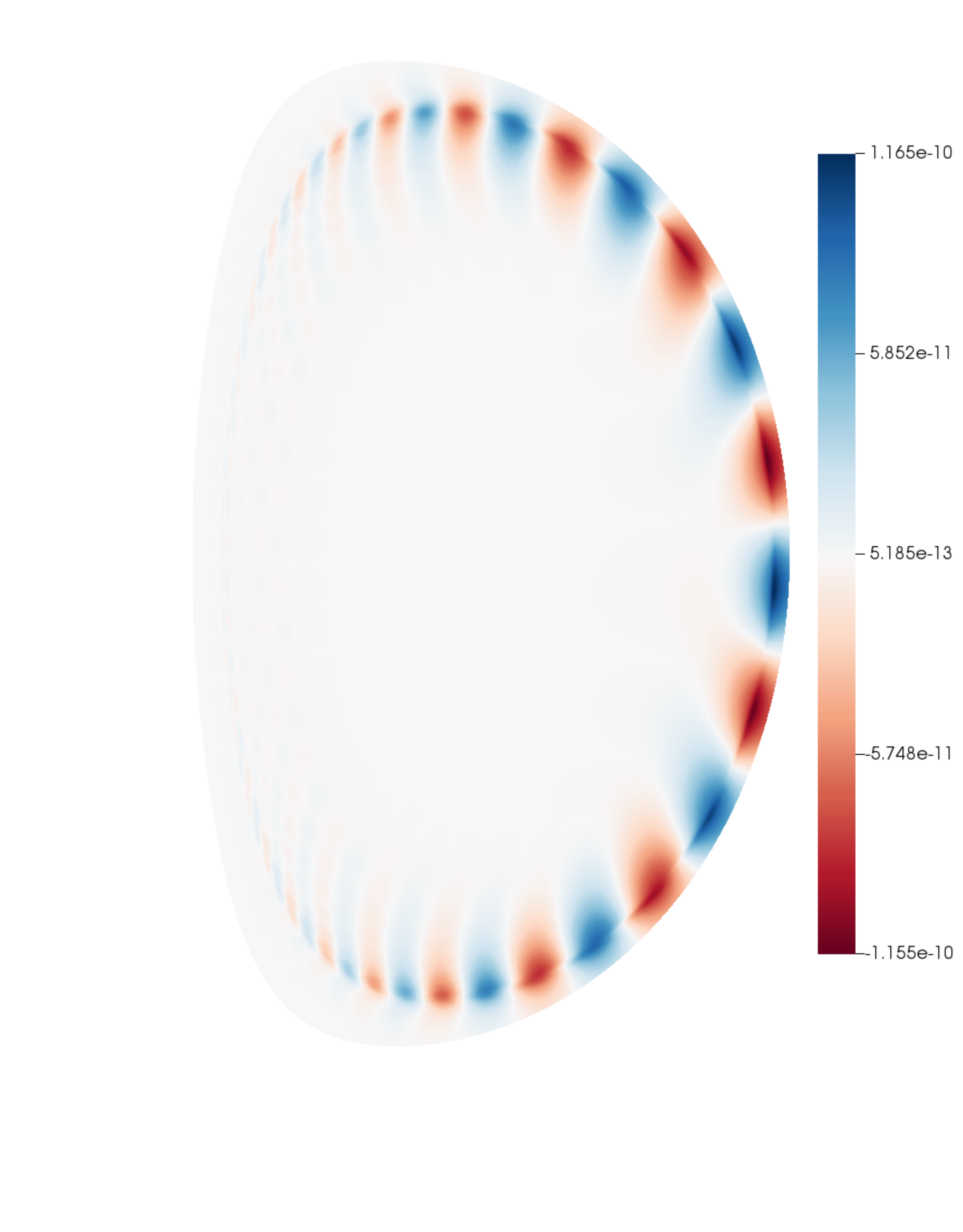}
    \caption{$n=5$ poloidal flux perturbation {from JOREK} during the linear phase of case 2.}
    \label{fig:n5_poloidal_flux}
\end{figure}

The evolution of the magnetic energies for the three additional cases is shown in Figure \ref{fig:jet_kink_res_E_comp}. It can be seen that, with the Spitzer-like profile in case 2, the dynamics are now led by the $n=5$ mode. Considering the $n=5$ poloidal flux perturbation associated with this instability, shown in Figure \ref{fig:n5_poloidal_flux}, the mode has a strong poloidal localisation on the low field side, typical of a ballooning mode.  The saturation of the $n=5$ instability leads to a nonlinear reduction in the growth rate of the $n=1$ kink mode, such that this instability takes longer to saturate. At the end of the simulation time, the $n=1$ harmonic overtakes the $n=5$ harmonic, and leads the dynamics. The dominant poloidal structures observed for the poloidal flux are the same as in case 1 at the end of the simulation run time.

When the resistivity in the plasma region is removed, as in case 3, it can be seen that the $n=5$ mode is stabilised, such that the $n=1$ mode leads the dynamics from the beginning once again. This indicates that the higher resistivity inside the plasma region in case 2 leads to the destabilisation of the $n=5$ mode, such that the mode must have some internal, resistive component. As shown in Figure \ref{fig:n5_poloidal_flux}, the observed mode is still localised near the plasma edge in the linear phase, where the edge current gradient is steepest. As such, the observed mode is thought to be a resistive peeling-ballooning mode.

The observed $n=5$ dynamics present a potential limitation of the VMEC approach, which could make the results less experimentally relevant. It is normally argued that such pressure driven modes would be stabilised in an experiment by flows, allowing the kink mode to still dominate \cite{brunetti2019excitation}. Case 4 was run with background $\mathbf{E} \times \mathbf{B}$ and diamagnetic flows, using the source in equation \ref{eq:velocity_ansatz} with similar parameters to what might be expected for the JET-like equilibrium being modeled. The simulation is re-run once again, with the Spitzer profile used as in case 2. In this instance, it can be seen that the $n=5$ mode is stabilised sufficiently by the presence of flows for the $n=1$ mode to dominate the dynamics. The $n=1$ mode has a faster linear growth rate than without poloidal flows, which is consistent with results from previous studies \cite{liu2017nonlinear}. The mode saturates at a smaller amplitude indicating the flows have affected the $n=1$ mode as well, but once again it has been confirmed that the poloidal mode structure of the $n=1$ mode is similar to that observed in Figure \ref{fig:jet_kink_perturbed_psi}. 

\section{Comparison of edge harmonic oscillation} \label{sec:eho_comparison}

\subsection{VMEC Computation}
\begin{figure}
    \centering
    \includegraphics[width=0.475\textwidth]{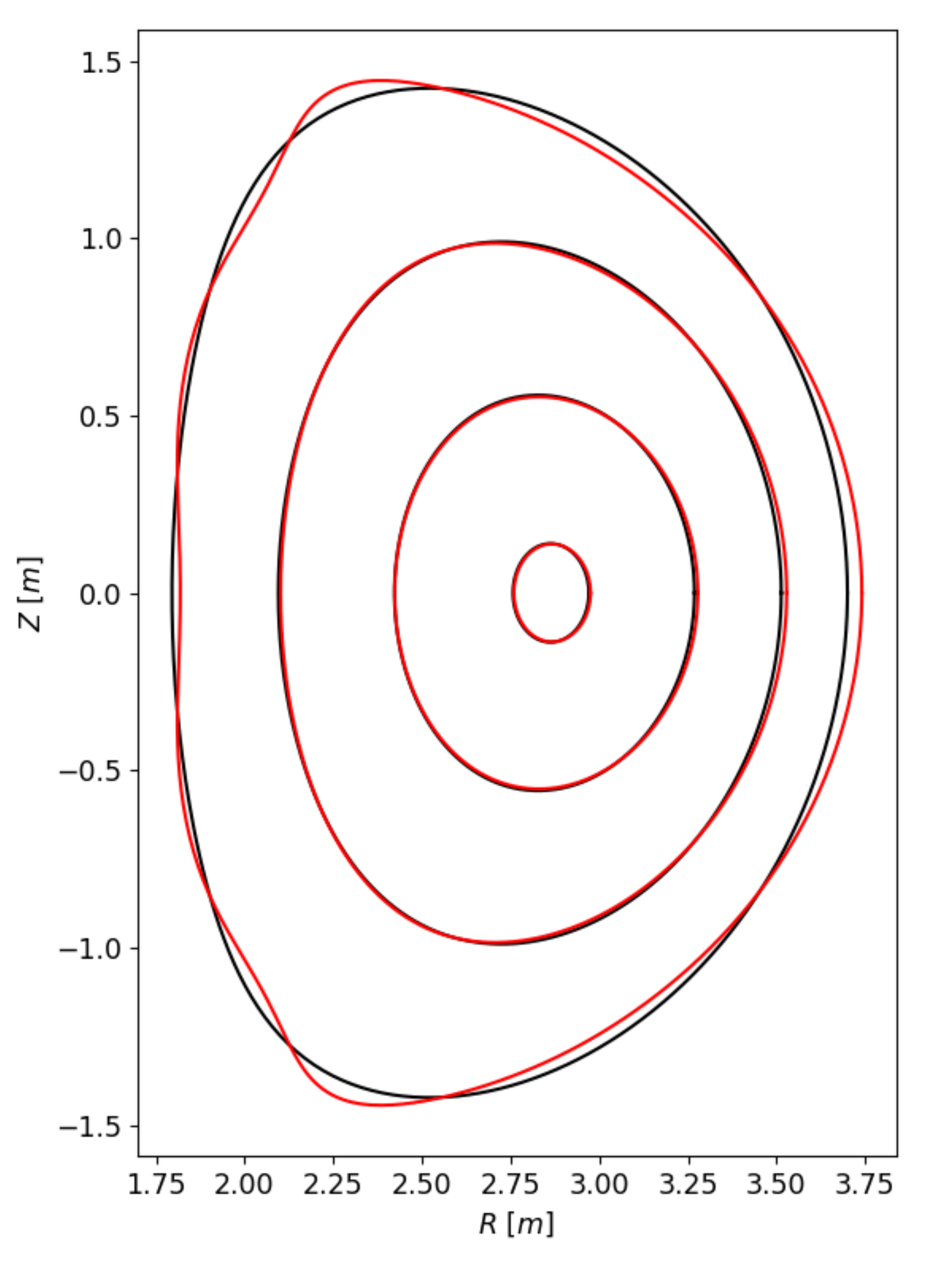}
    \small{(a)}
    \includegraphics[width=0.475\textwidth]{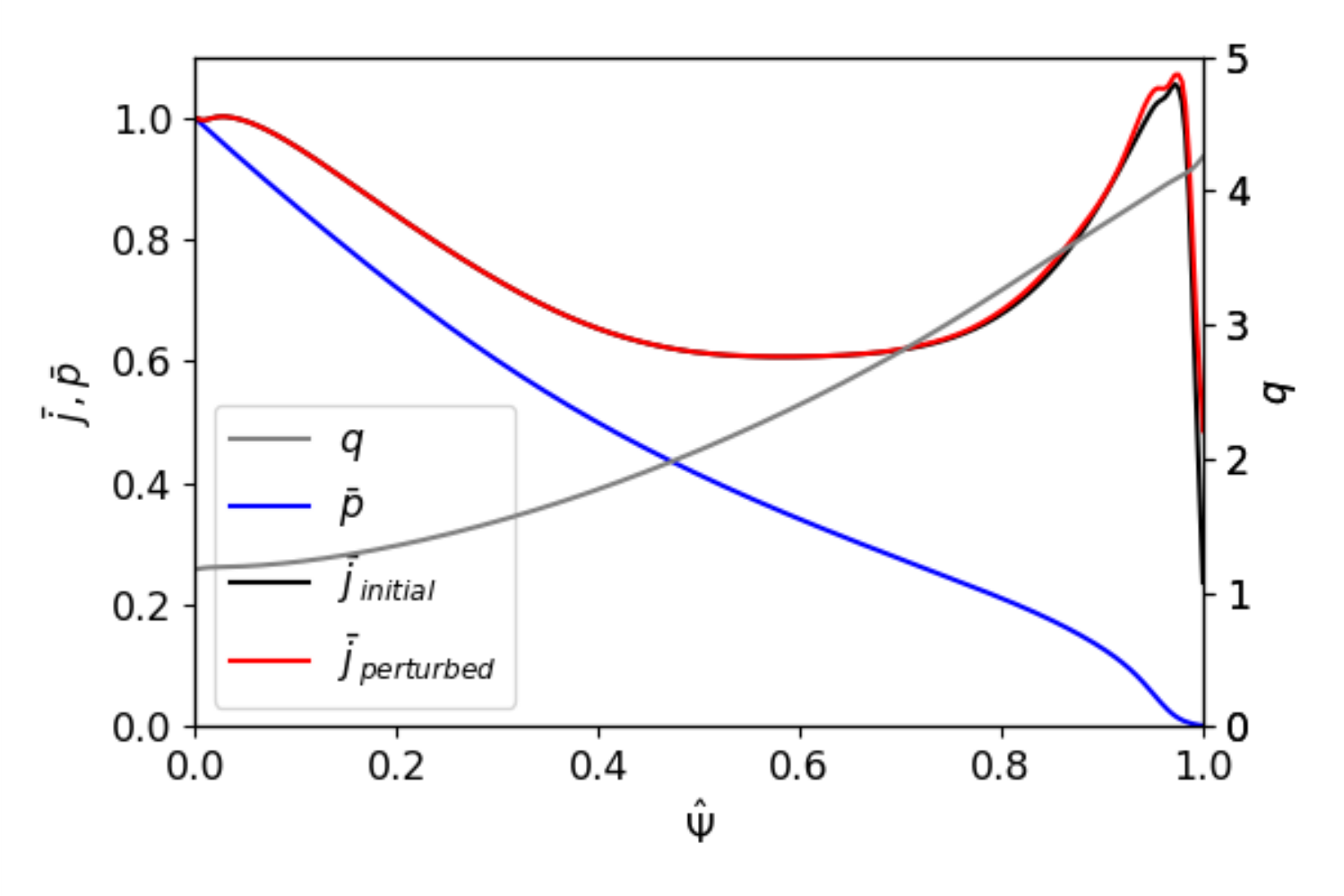}
    \small{(b)}
    \caption{Unperturbed and perturbed Equilibrium {flux surfaces at $\sqrt{\hat \Phi} = 0.1$, 0.3, 0.7 and 1.0} (a), and radial profiles (b) {from VMEC} for the edge harmonic oscillation case.}
    \label{fig:jet_eho_perturbation}
\end{figure}

The EHO test case considered is based on the previously presented JET-like equilibrium studied in Ref. \onlinecite{kleiner2019current}, where external infernal (exfernal) modes are calculated using VMEC. The nonlinearly saturated state computed in VMEC is shown in Figure \ref{fig:jet_eho_perturbation}. Comparing Figure \ref{fig:jet_eho_perturbation} to Figure \ref{fig:jet_kink_perturbation}, the deformation of the last closed flux surface, and corresponding current spike is smaller. This implies that the instability is more radially localised, as expected for an EHO, when compared to a global external kink mode. This type of instability is selected as a more challenging case for modelling with VMEC, because previous studies have shown that these instabilities involve strong toroidal mode coupling \cite{liu2015nonlinear}.

The simulated instability has been previously described as comparable to low-n kink-peeling modes \cite{brunetti2018analytic}, which have been observed in past JOREK simulations in the context of EHOs \cite{garofalo2015quiescent, liu2015nonlinear}. Such MHD instabilities typically rely on a large bootstrap current near the plasma edge generated by strong pressure gradients. The large bootstrap current can be seen in the non-monotonic profile shown in Figure \ref{fig:jet_eho_perturbation} (b). The current flattens the q profile near the plasma edge, which in turn allows the large pressure drive to strongly couple internal and external MHD perturbations. The case was chosen such that the flattening of the q profile is only partial, and remains realistic. It has been shown that partial weakening in the edge magnetic shear also leads to the excitation of exfernal modes \cite{ramirez2021edge}.

{In Section \ref{sec:epfl_eho_linear} and Section \ref{sec:eho_resistivity_scan}, it is shown that the simulated test case is linearly unstable for multiple toroidal harmonics, such that it is unclear which instability will lead the nonlinear dynamics. It should be noted that during the parameter scans carried out in this study, only the $n=1$ dominant saturated mode was found in VMEC for this test case.}

\subsection{Comparison of VMEC perturbation with linear eigenfunctions} \label{sec:epfl_eho_linear}
\begin{figure}
    \centering
    \includegraphics[width=0.44\textwidth]{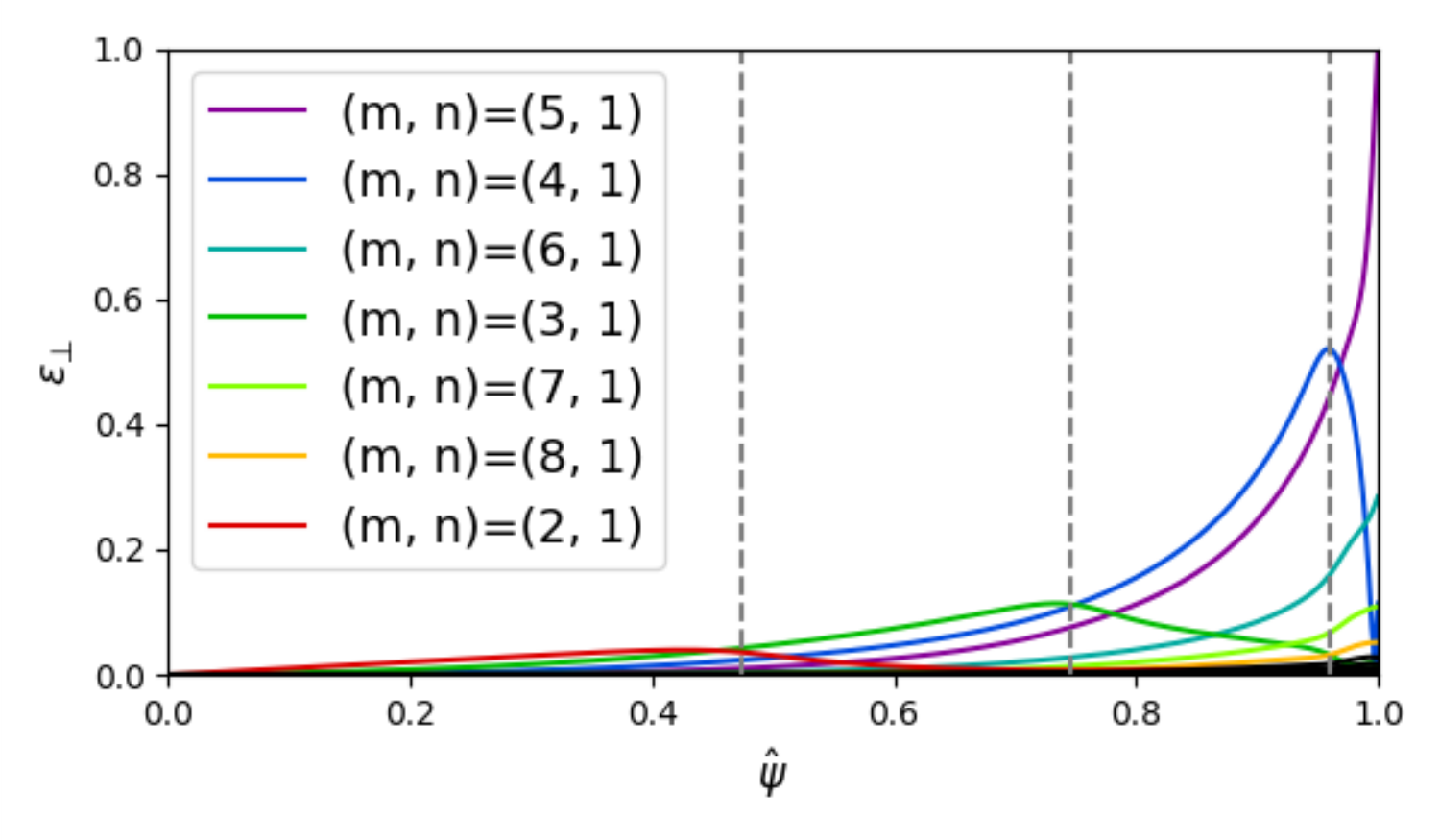}
    \small{(a)}
    \includegraphics[width=0.44\textwidth]{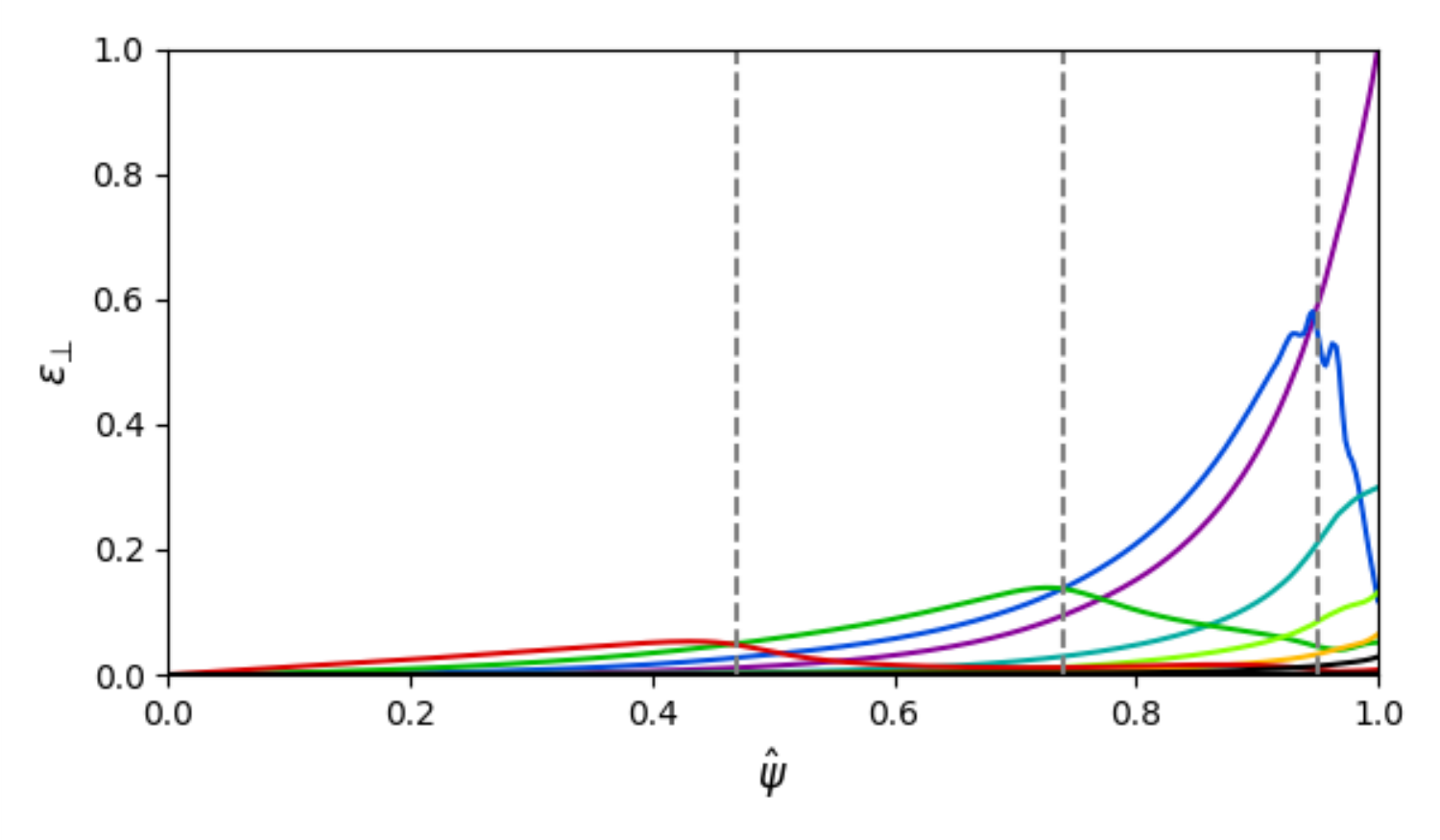}
    \small{(b)}
    \includegraphics[width=0.44\textwidth]{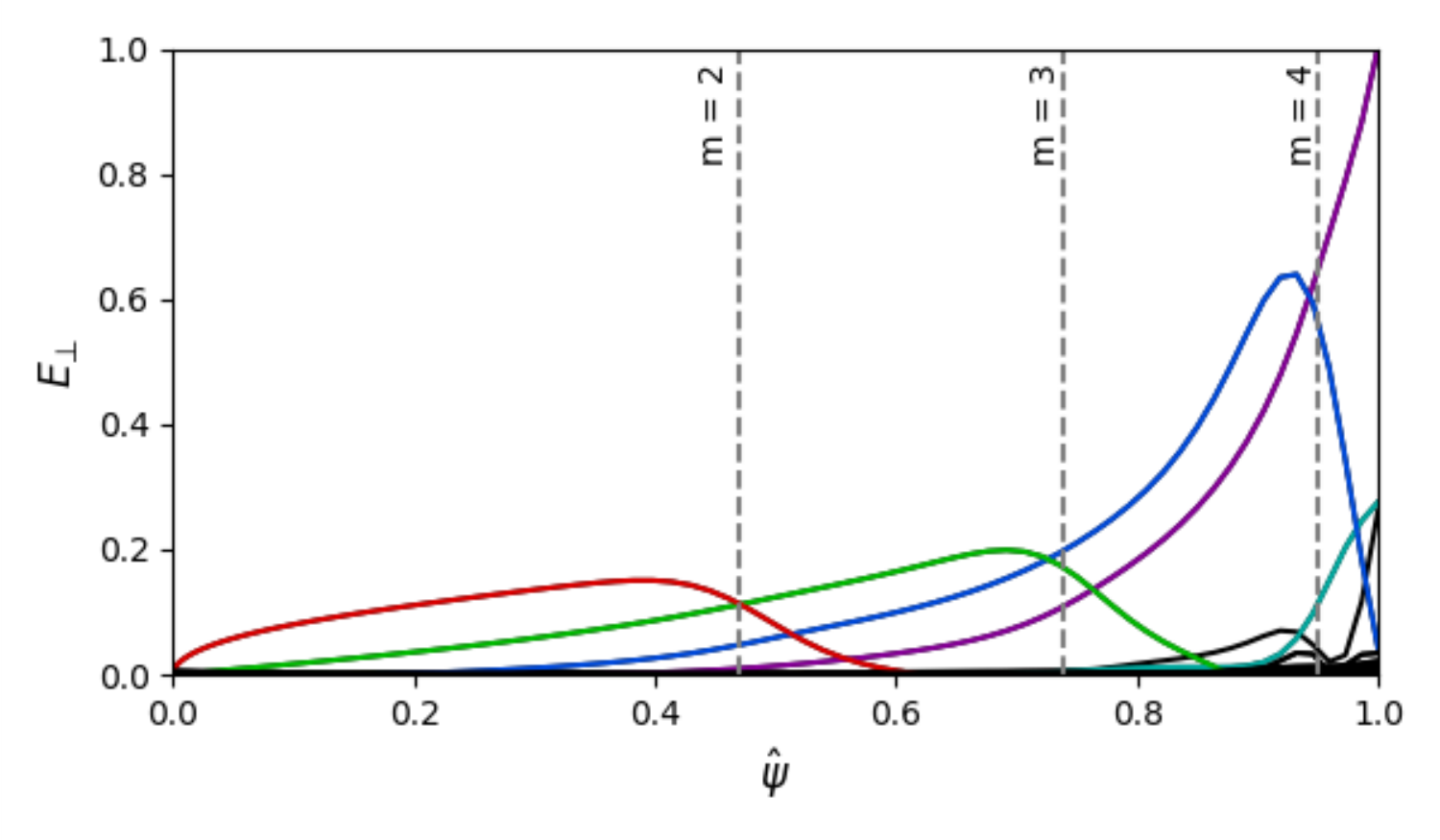}
    \small{(c)}
    \includegraphics[width=0.44\textwidth]{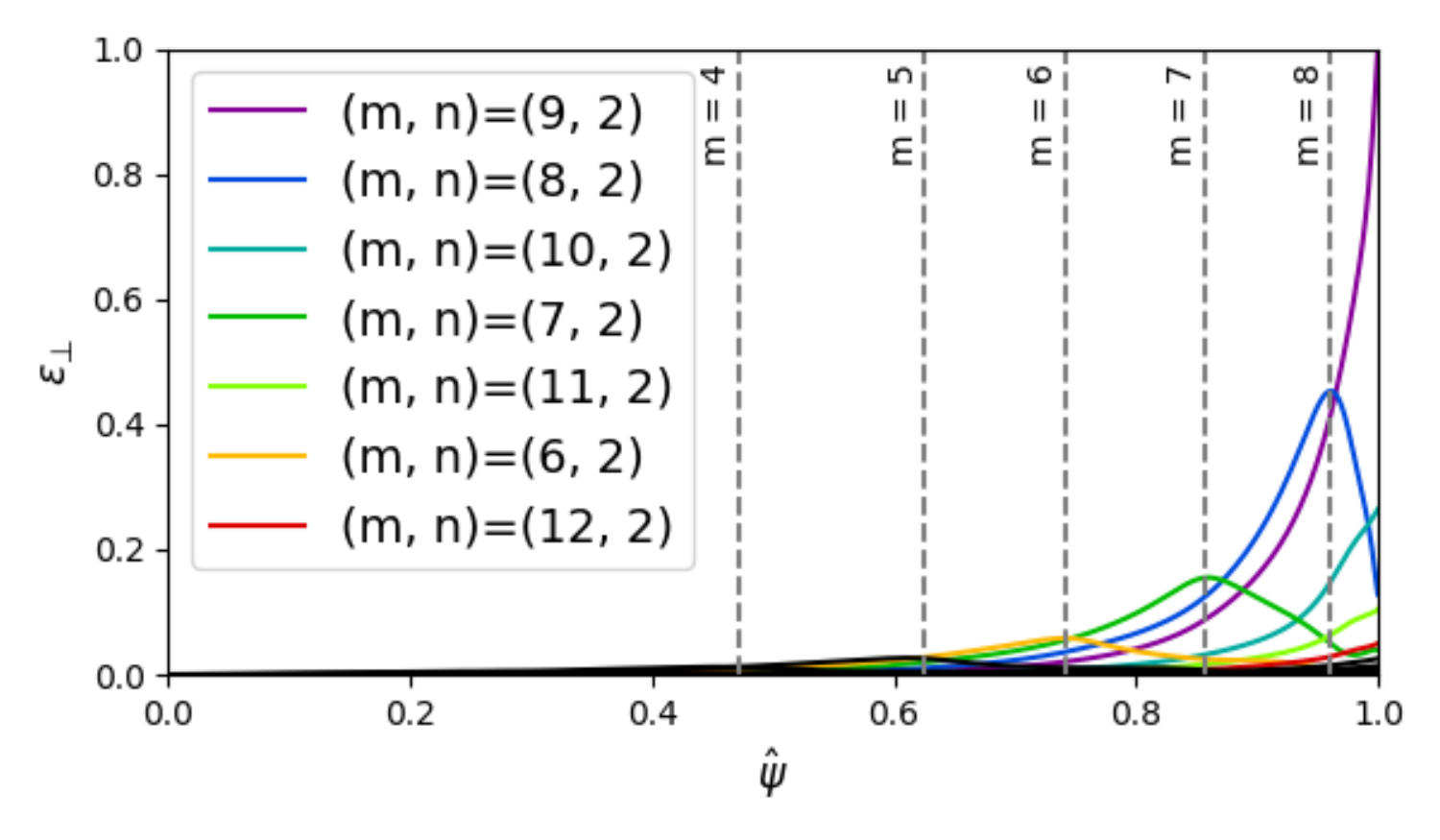}
    \small{(d)}
    \caption{Comparison of radial eigenfunctions observed in JOREK (a) during the linear phase of the EHO test case, with the radial eigenfunction found in CASTOR3D (b), and the nonlinear perturbation observed in VMEC (c) for the $n=1$ mode. {The linear $n=2$ mode observed in JOREK is also shown in (d)}. The Fourier representation has been calculated using PEST coordinates. {The location of relevant rational surfaces with their corresponding poloidal mode number are marked by grey dashed lines.}}
    \label{fig:jet_eho_eigenfunctions}
\end{figure}

Similar to Section \ref{sec:epfl_kink}, an equivalent equilibrium to the unperturbed state in VMEC is generated in JOREK, and the linear perturbation observed in JOREK is compared with CASTOR3D, and the nonlinearly perturbed VMEC equilibrium. The result of the comparison for the $n=1$ instability is shown in Figure \ref{fig:jet_eho_eigenfunctions}. It can be seen that there is {qualitative} agreement once again between the different approaches. The typical strong (4, 1) internal contribution to the edge harmonic oscillation is observed in all cases.

It should be noted that the linear dynamics are immediately different to the case considered in Section \ref{sec:epfl_kink}, because the $n=2-5$ toroidal harmonics are also unstable to linearly independent modes. The radial eigenfunction for the $n=2$ mode is shown in Figure \ref{fig:jet_eho_eigenfunctions} (d). It can be seen that this mode has a similar structure to the $n=1$ mode, with a large internal component localised at the $q=4$ rational surface, and a dominant external component corresponding to the nearest rational surface to the plasma edge, $(nq+1)/n$. For this case, the $n=2$ and $n=3$ toroidal harmonics have the largest linear growth rates. As shown in the next section, they will therefore lead the initial nonlinear dynamics of the instability in JOREK simulations.

\begin{figure}
    \centering
    \begin{minipage}{0.45\textwidth}
    \includegraphics[width=\textwidth]{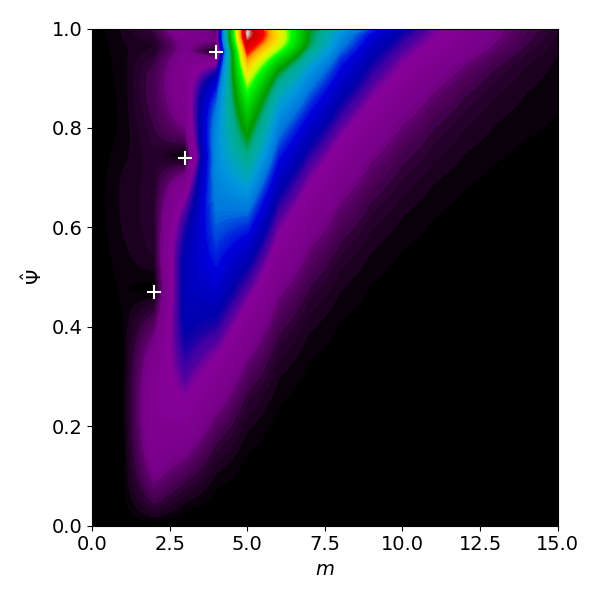}
    \centering
    \small{(a)}
    \end{minipage}
    \begin{minipage}{0.45\textwidth}
    \includegraphics[width=\textwidth]{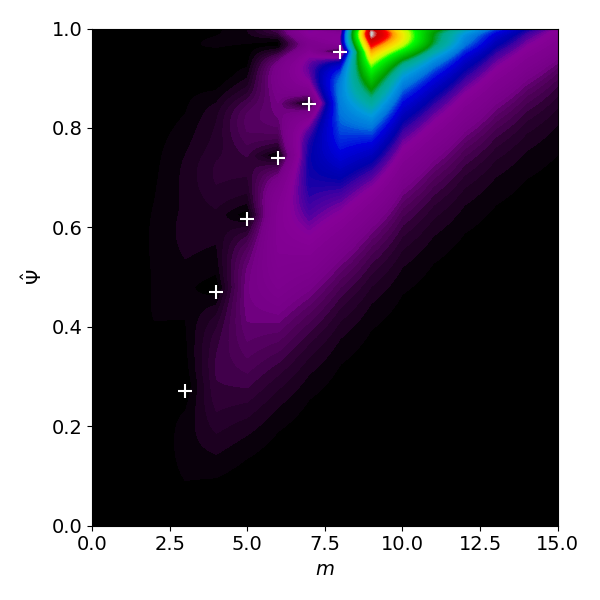}
    \centering
    \small{(b)}
    \end{minipage}
    \caption{{Linear mode spectrum of the $n=1$ (a) and $n=2$ (b) radial magnetic field perturbation in JOREK. The spectra are given over the poloidal components $m$ and the radial coordinate, $\hat \Psi$. The rational surfaces are marked by white crosses. It can be seen that the expected ideal screening occurs at these rational surfaces, even close to the plasma boundary.}}
    \label{fig:linear_psi_perturbation}
\end{figure}

{Lastly, for this case, the radial magnetic field perturbation for the $n=1$ and $n=2$ mode has been calculated in JOREK during the linear phase, as shown in Figure \ref{fig:linear_psi_perturbation}. The perturbations show the expected ideal screening of the mode at the rational surfaces marked by the white crosses. This indicates that resistive effects are sub-dominant in the linear dynamics.}

\subsection{Dependence of MHD dynamics on resistivity} \label{sec:eho_resistivity_scan}

Based on the intuition of previous nonlinear studies \cite{liu2017nonlinear, liu2015nonlinear, pankin2020towards}, the lowest toroidal harmonics are expected to control the nonlinearly saturated state of an EHO, but it is unclear whether the $n=1$ or $n=2$ toroidal harmonic will be the dominant perturbation. The eigenfunction from VMEC in Figure \ref{fig:jet_eho_eigenfunctions} implies that the $n=1$ component should be the principal mode structure.

\begin{figure}
    \centering
    \includegraphics[width=0.41\textwidth]{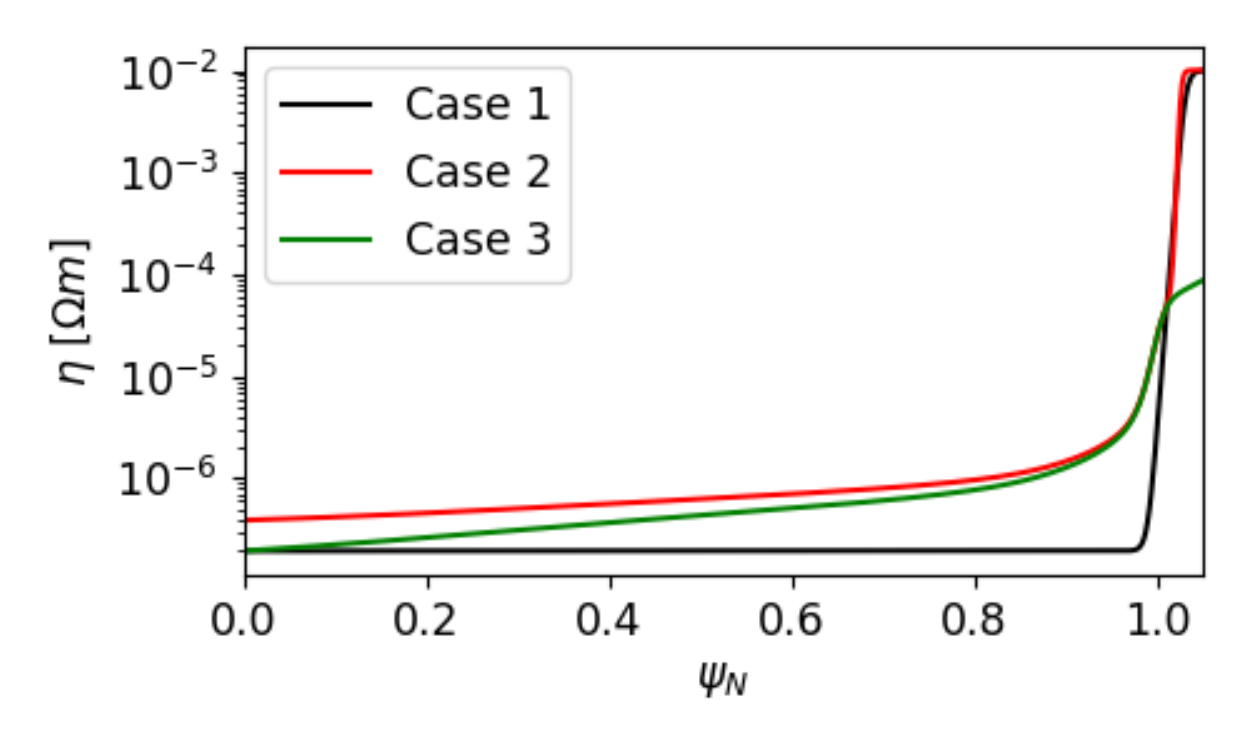}
    \centering
    
    \small{(a)}
    
    \includegraphics[width=0.41\textwidth]{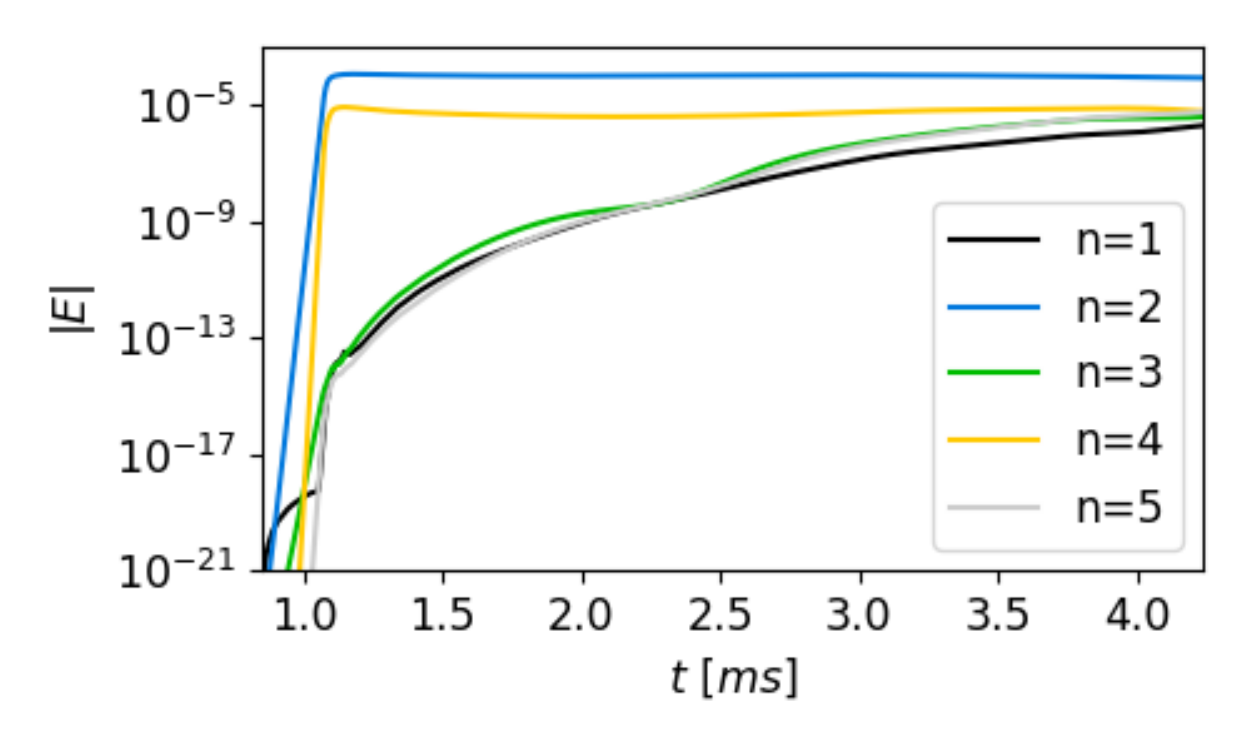}
    \centering
    
    \small{(b)}
    
    \includegraphics[width=0.41\textwidth]{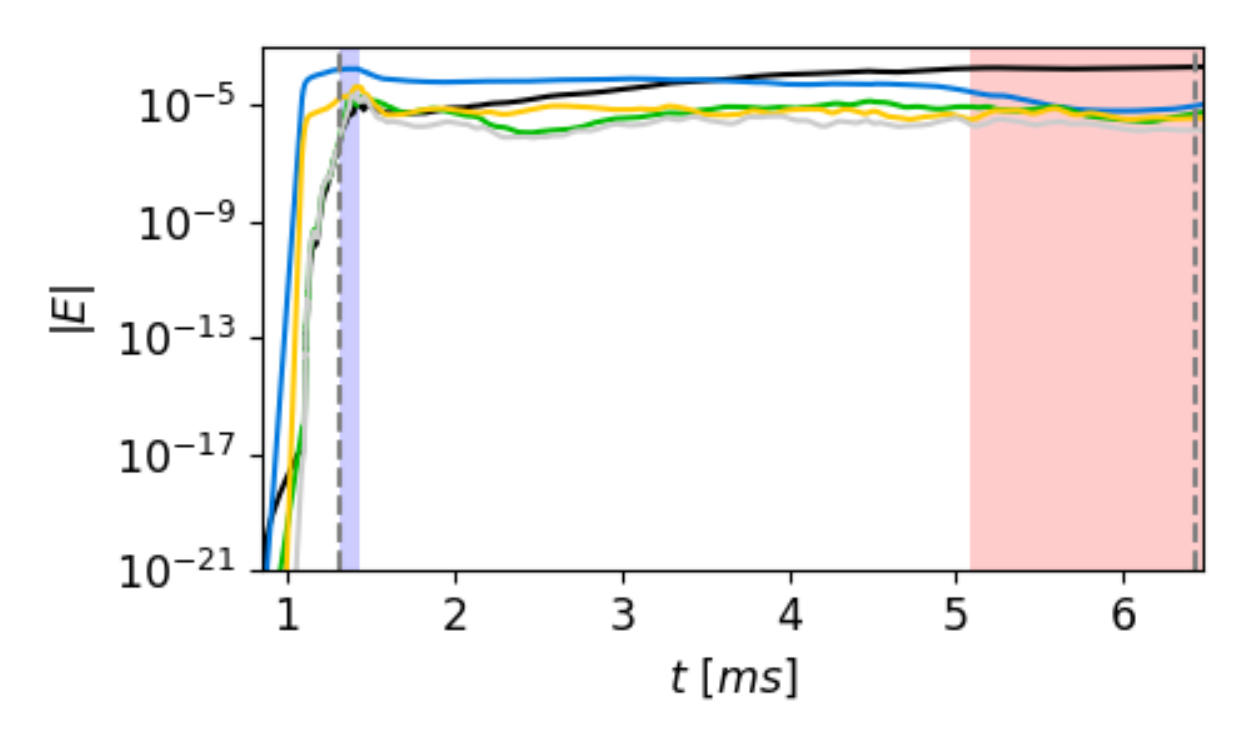}
    \centering
    
    \small{(c)}
    
    \includegraphics[width=0.41\textwidth]{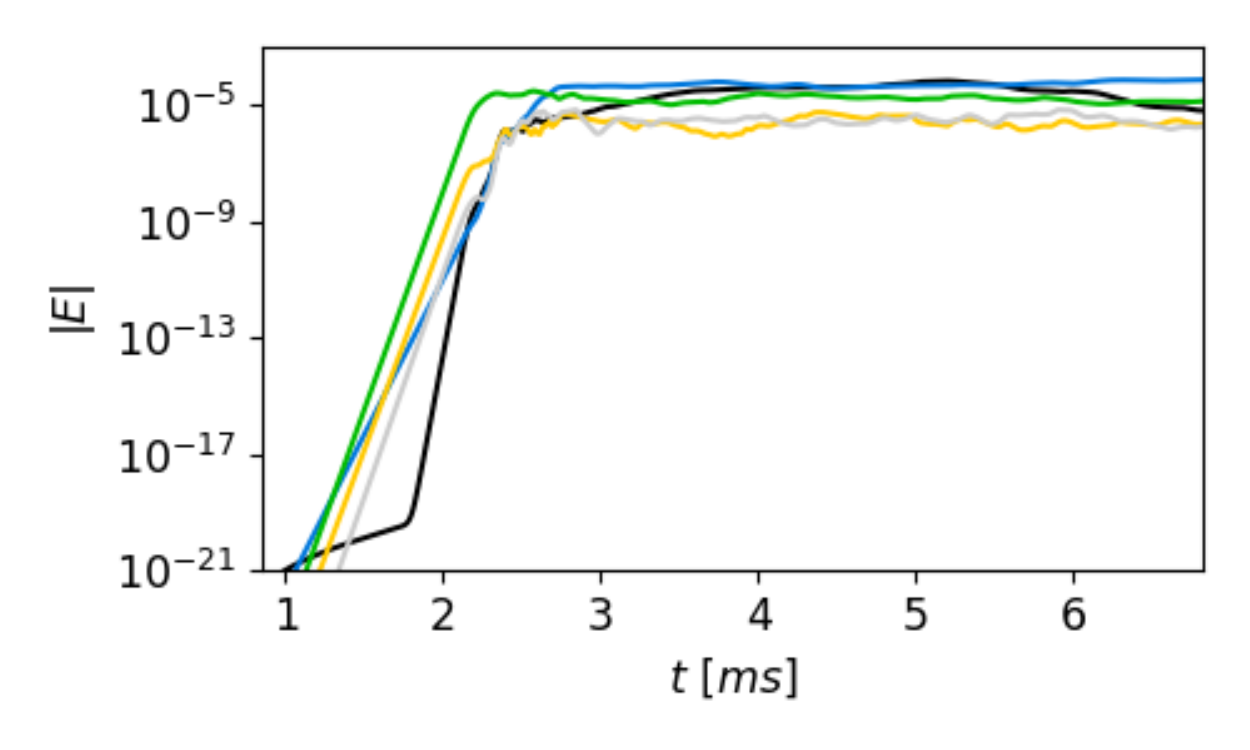}
    \centering
    
    \small{(d)}
    \caption{Toroidal magnetic energies {from JOREK} for the EHO test case simulated with different resistivity profiles (a).  {Note that the energies are normalised by a factor $\mu_\mathrm{0}$.} A larger core resistivity compared to case 1 (a) is used in case 2 (c) to accelerate the onset of resistive effects. The time points marked by grey dashed lines correspond to the Poincar\'e plots shown in Figure \ref{fig:jet_eho_poincare_comparison}, and pseudocolour plots in Figure \ref{fig:jet_eho_magnetic_energy_spectrum} (b) and (c). {The coloured regions correspond to the simulation time over which the energy spectrum is averaged over in Figure \ref{fig:jet_eho_magnetic_energy_spectrum} (a).} A Spitzer resistivity is used in case 3 (d).}
    \label{fig:jet_eho_E_comp}
\end{figure}

The evolution of the magnetic energies for the EHO test case is shown in Figure \ref{fig:jet_eho_E_comp}. Similar to Section \ref{sec:epfl_kink}, the test case was first simulated with the resistivity profile of case 1, shown in Figure \ref{fig:jet_eho_E_comp} (a), which approximates the ideal MHD conditions in VMEC. As expected from the linear analysis, the linear dynamics are led by the $n=2$ perturbation, resulting in a $n=2$ EHO, which grows and saturates on the fast ideal MHD timescale. The saturated state can be identified as an EHO by the nature  of the toroidal mode coupling, which includes only even toroidal mode numbers \cite{liu2015nonlinear}. 

After the initial saturation, the sub-dominant $n=1$, 3 and 5 modes begin to grow. The slow growth of these modes indicates that the nonlinear dynamics are resistive. To verify this, a second simulation was run with the resistivity inside the plasma {prescribed with a Spitzer-like dependence}. In such a way, the resistive dynamics are  accelerated, compared to case 1. As expected, the odd toroidal harmonics grow faster in this case, indicating resistivity is important in the nonlinear dynamics observed. To determine whether the $n=1$ mode could dominate later in the nonlinear phase, case 2 is extended over a longer timescale. It can be seen that after approximately $3.5\ ms$, the $n=1$ mode begins to suppress the initially prevailing $n=2$ structures. This $n=1$ saturated state is compared with the VMEC result in the following sections. 

At this point, it is important to note the timescale of the MHD dynamics that are observed in the JOREK simulations. In Section \ref{sec:epfl_kink_vmec}, the solution of the VMEC computation was interpreted as the saturated state immediately after the fast phase of the instability, before resistive effects can become important. In the JOREK simulation, this fast phase ends after the saturation of the $n=2$ mode, at $t\approx 1.3\ ms$ in case 1 and 2 of Figure \ref{fig:jet_eho_E_comp}. {The expectation from the VMEC result, is that the $n=1$ mode would dominate on this timescale, which is not observed in the JOREK simulations. This leads to two important questions --- why does the $n=1$ mode not dominate on the ideal timescale in JOREK simulations, and secondly, how much does the observed $n=1$ mode correspond to the expected ideal EHO mode? The second question is addressed in Section \ref{sec:jet_eho_flux_surface_comparison}.} 

{With respect to the first question, this is likely because the nonlinear influence of the $n=2$ mode reduces the drive for the competing $n=1$ mode for the simulated case.} {The rapid linear growth of the $n=2$ mode, stabilises the $n=1$ mode through quadratic coupling, as described in Ref. \onlinecite{krebs2013nonlinear}. The reason that the ideal $n=1$ mode must be stabilised by the $n=2$ mode is that any deformation of the flux surfaces related to the odd toroidal mode numbers, which overlaps with the dominant $n=2$ perturbation must compete with this nonlinearly saturated mode structure. It can be seen in Figure \ref{fig:jet_eho_eigenfunctions} (a) and (d) that the linear perturbation associated with the $n=1$ toroidal harmonic overlaps with the $n=2$ perturbation. This implies that the nonlinearly saturated $n=2$ EHO will have a stabilising influence on the linear $n=1$ instability during the its initial nonlinear saturation.}

{The sharp rise in the $n=1$ energy during the initial saturation of the $n=2$ mode implies that the $n=1$ mode structure is strongly influenced, as it is forced to accommodate the $n=2$ EHO.} After this point, the deformation of the plasma leads to thermal losses that increase the resistivity inside the plasma and reduce it in the vacuum region. The pressure gradient driving the mode in the plasma edge region also relaxes. The saturated $n=2$ mode {also} influences the evolution of the $n=1$ mode through these effects, leading to the slower timescale to saturation of the $n=1$ mode. While the observed dynamics in JOREK are physically reasonable, it means this study cannot be treated as a rigorous verification of the VMEC result, where viscoresistive and diffusive effects are neglected. {In this limit, the dynamics could be different.}

A final simulation, case 3, was run with a more realistic Spitzer resistivity inside, and outside the plasma. {This simulation serves to highlight the difficulty in predicting the nonlinear outcome of an EHO.} The lower resistivity in the vacuum region has a stabilising effect on the initial ideal MHD instabilities, which brings the ideal and resistive timescales closer together compared to case 1 and 2. The evolution of the magnetic energies is shown in Figure \ref{fig:jet_eho_E_comp} (d). It can be seen that for this case, all linearly unstable modes have been partially stabilised, and the $n=3$ mode leads the dynamics, before quickly being overtaken by the $n=2$ mode. After this point, there is a period of competition between the $n=1$ and $n=2$ modes, before the $n=2$ mode becomes the principal mode. The $n=1-3$ modes that dominate the perturbation at different points in the dynamics are all kink-peeling modes, such that the linear dynamics are similar to the simulation results for case 1 and 2. The nonlinearly saturated state at the end of the simulation time has changed due to the modification of the resistivity in the vacuum region, highlighting the sensitivity of the nonlinearly dominant mode on this parameter. 

\subsection{Comparison of flux surfaces and perturbed magnetic energies} \label{sec:jet_eho_flux_surface_comparison}

\begin{figure}
    \includegraphics[width=0.45\textwidth]{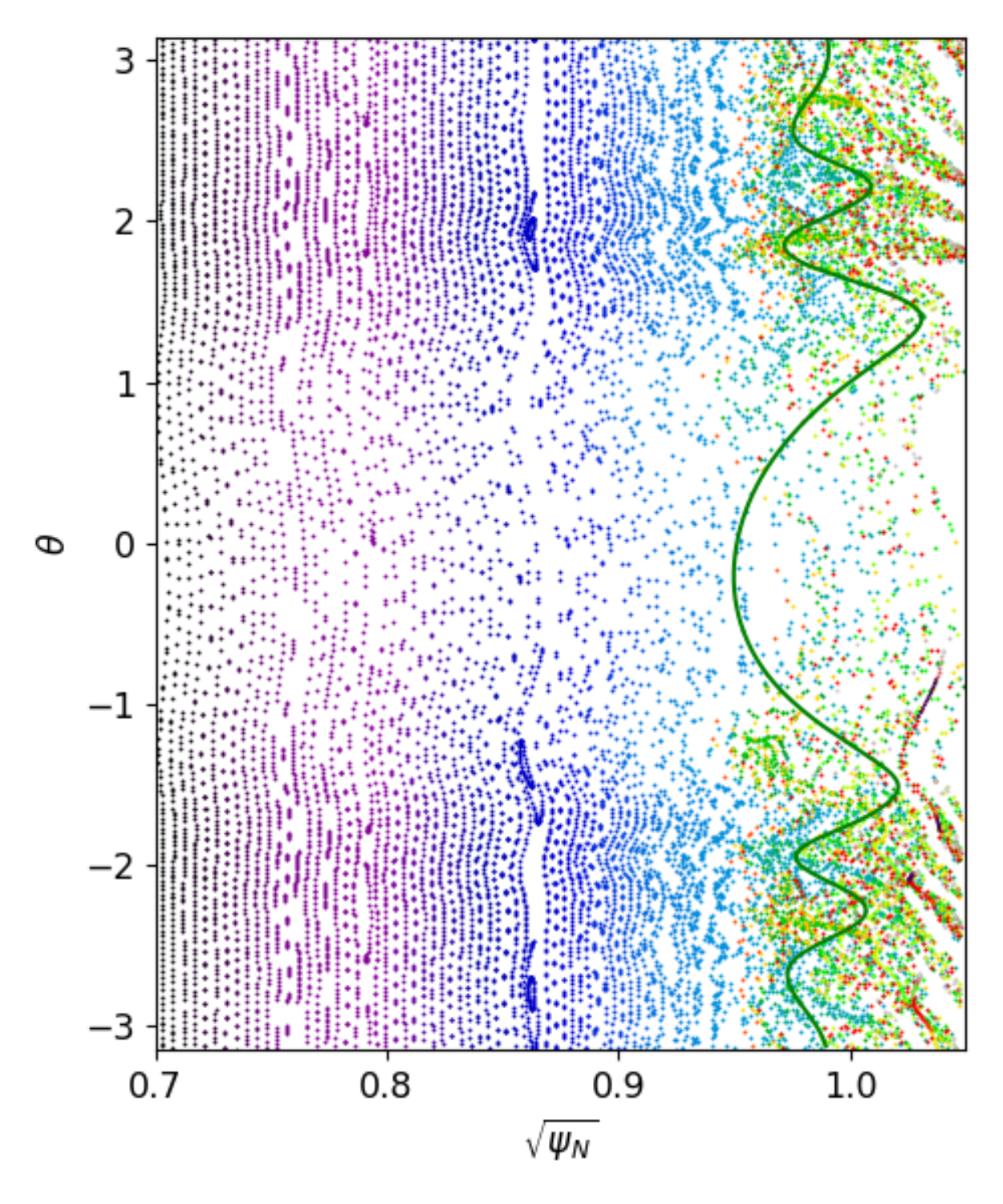}
    \centering
    \small{(a)}
    \includegraphics[width=0.45\textwidth]{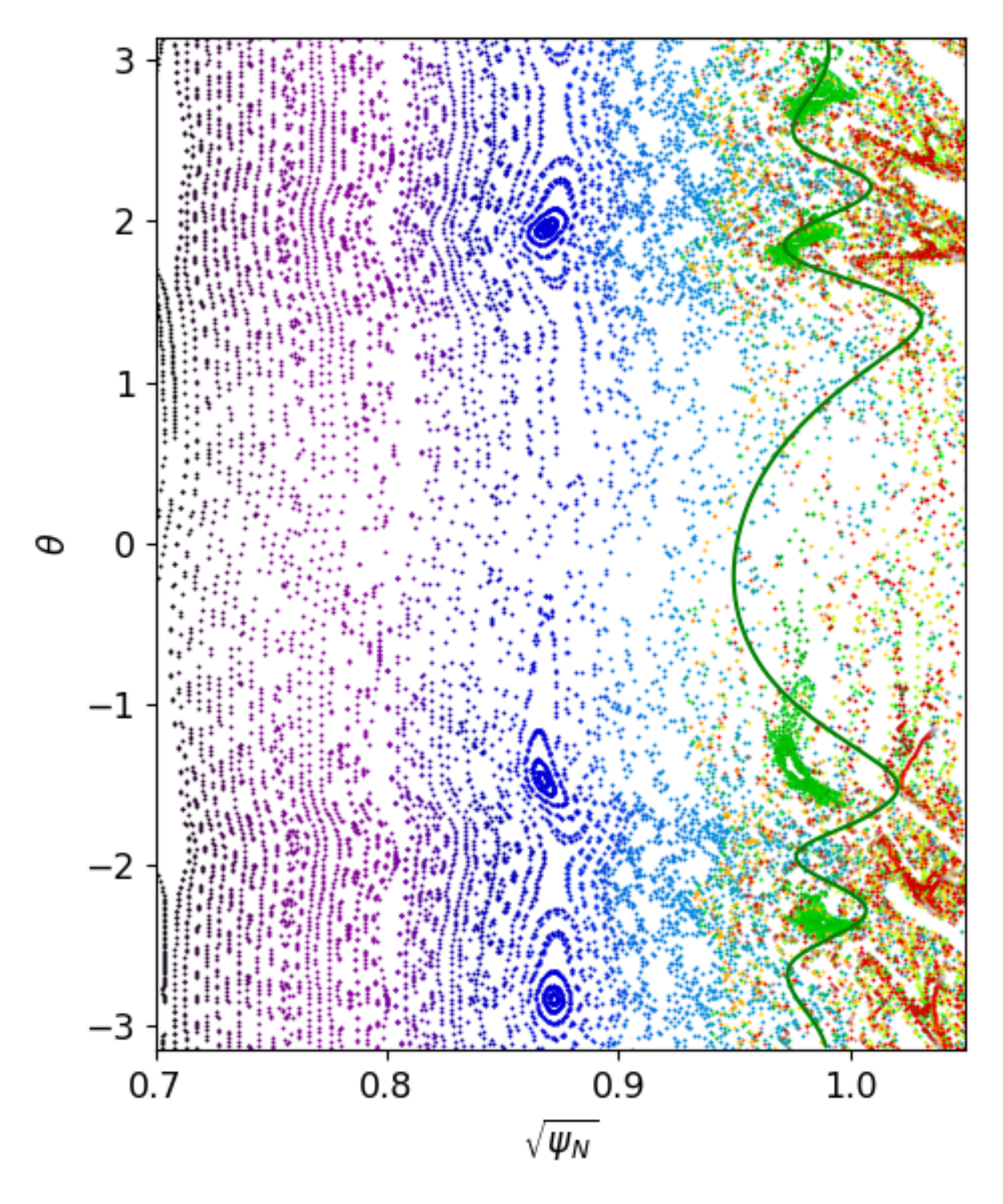}
    \centering
    \small{(b)}
    \caption{Poincar\'e comparisons are made with the perturbed VMEC equilibrium at two time points in the nonlinear evolution of case 2. The time points are marked by grey dashed lines in Figure \ref{fig:jet_eho_E_comp} (c). Each field line is given a different colour to identify the island structures more easily in ergodic regions. At the initial saturation of the $n=2$ harmonic, a (9, 2) island structure is identified at the plasma edge (a). At the end of the simulation, (5, 1) and (4, 1) island structures become dominant (b). The last closed flux surface from VMEC (green) is overlaid on the Poincar\'e plots.}
    \label{fig:jet_eho_poincare_comparison}
\end{figure}

To understand the differences between the nonlinearly perturbed state {found in case 2 and the ideal VMEC result}, the modification of the magnetic geometry and confinement is considered. The Poincar\'e plots in Figure \ref{fig:jet_eho_poincare_comparison} are taken at the saturation of the $n=2$ and $n=1$ mode. The $n=2$ mode initially saturates with a dominant (9, 2) perturbation of the poloidal flux. The corresponding island chain can be identified in Figure {\ref{fig:jet_eho_poincare_comparison} (a)}, but is faint due to the significant stochastisation in the edge region. It can be seen in Figure {\ref{fig:jet_eho_poincare_comparison} (b)} that later in the dynamics, when the $n=1$ mode is dominant, a (5, 1) island chain is observed outside the plasma{, as expected for the simulated EHO}.

The magnetic island structure in Figure {\ref{fig:jet_eho_poincare_comparison} (b) can be compared with} the approximate structure of the last closed flux surface from the equivalent VMEC computation, shown in green. The internal flux surfaces near the plasma boundary in VMEC are replaced by a strongly ergodic region with clearly visible magnetic islands in the region near the plasma bounday{, such that the agreement with the VMEC plasma boundary appears to be only marginal}. 

\begin{figure}
    \centering
    \includegraphics[width=0.475\textwidth]{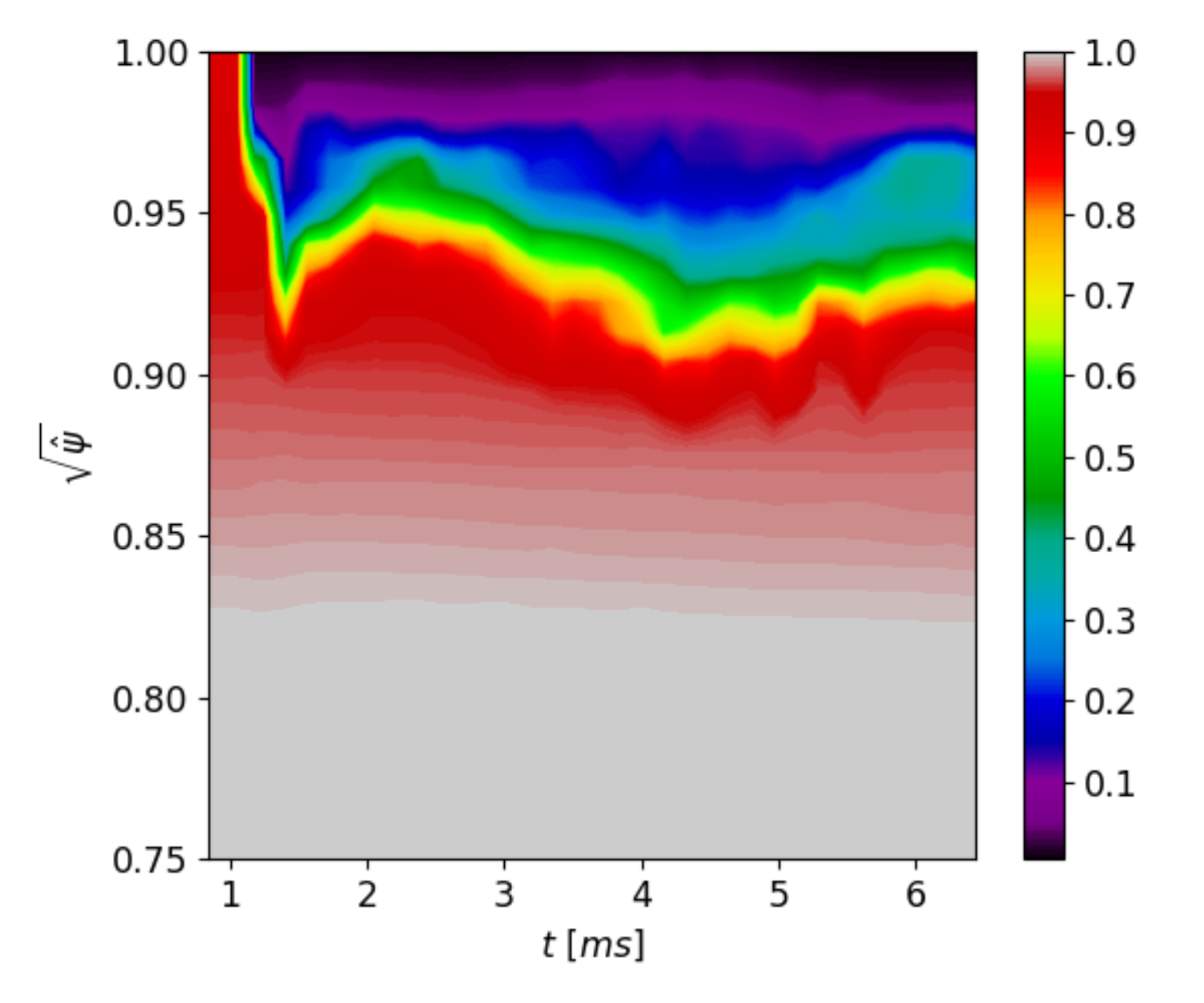}
    \caption{Average connection length of field lines to the {JOREK} simulation boundary as a function of time for case 2. The connection length is normalised by the maximum connection length, computed in the plasma core.}
    \label{fig:jet_eho_connection_length}
\end{figure}

To assess how these internal structures change the parallel transport, the connection length of magnetic field lines to the simulation boundary is shown in Figure \ref{fig:jet_eho_connection_length}. The connection length is calculated using the harmonic mean of 1000 sample field lines uniformly distributed along the poloidal angle of the $n=0$ flux surfaces. The radial coordinate is taken to be the square root of the $n=0$ component of the normalised poloidal flux. The results are normalised by the connection length in the plasma core. It can be seen that across the majority of the plasma region, field lines remain well confined within the plasma volume. There is only a significant loss of field lines near the plasma edge. Similar results have been observed in more realistic studies of EHOs in DIII-D using x-point geometry \cite{garofalo2015quiescent}. 

\begin{figure*}
    \centering
    \begin{minipage}{0.24\textwidth}
    \includegraphics[width=\textwidth]{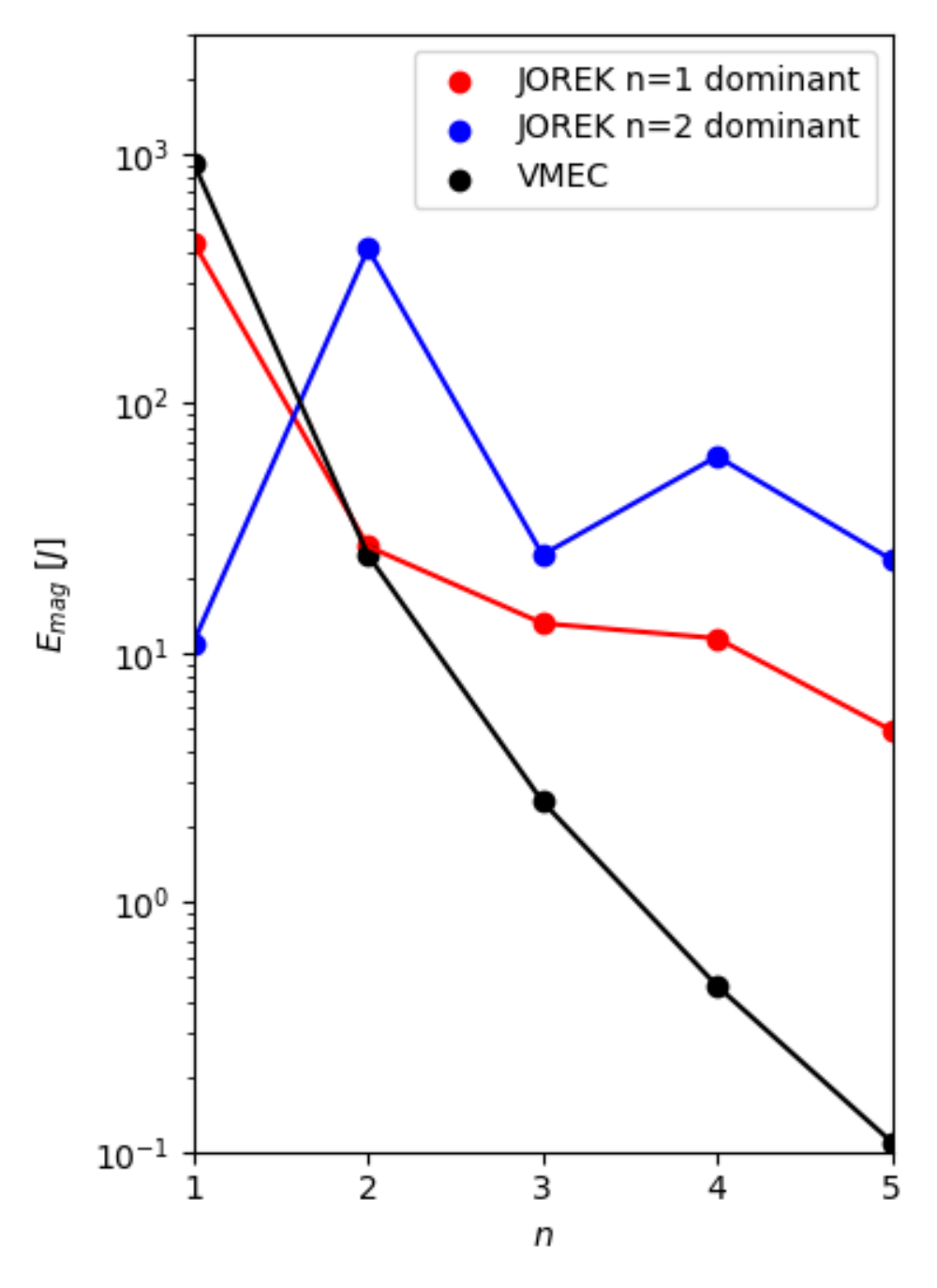}
    \centering
    \small{(a)}
    \end{minipage}
    \begin{minipage}{0.25\textwidth}
    \includegraphics[width=\textwidth]{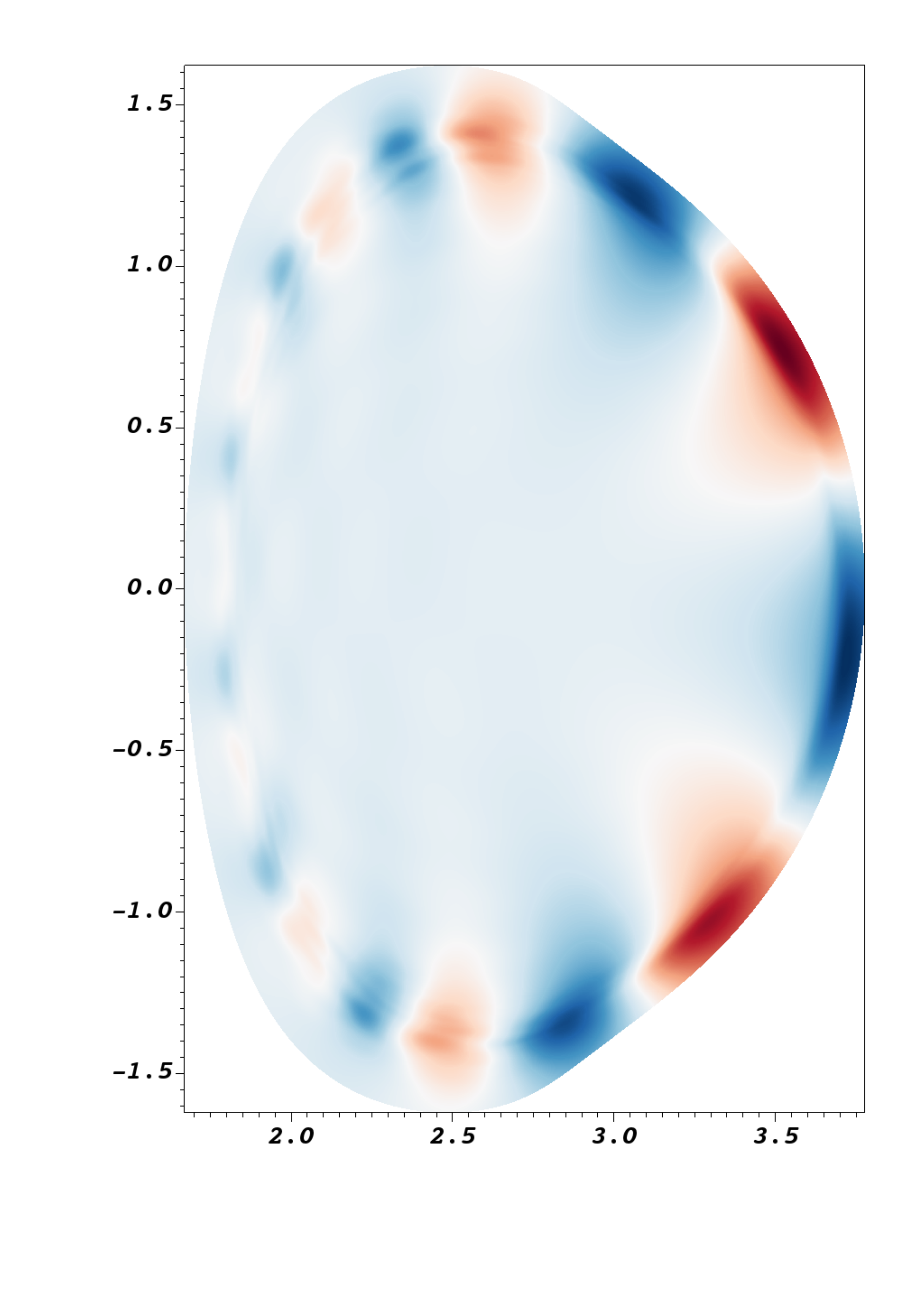}
    \centering
    \small{(b)}
    \end{minipage}
    \begin{minipage}{0.25\textwidth}
    \includegraphics[width=\textwidth]{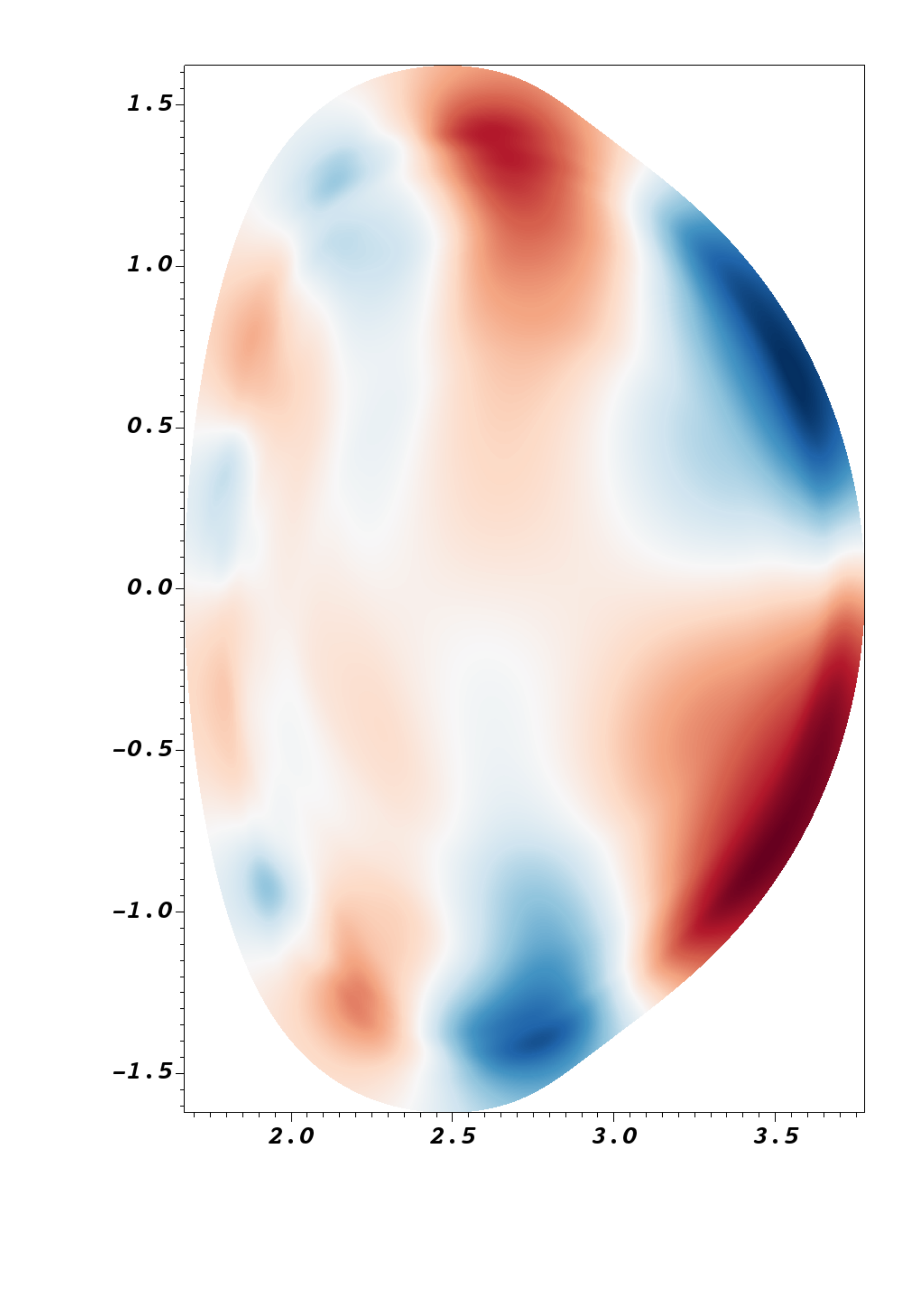}
    \centering
    \small{(c)}
    \end{minipage}
    \begin{minipage}{0.24\textwidth}
    \includegraphics[width=\textwidth]{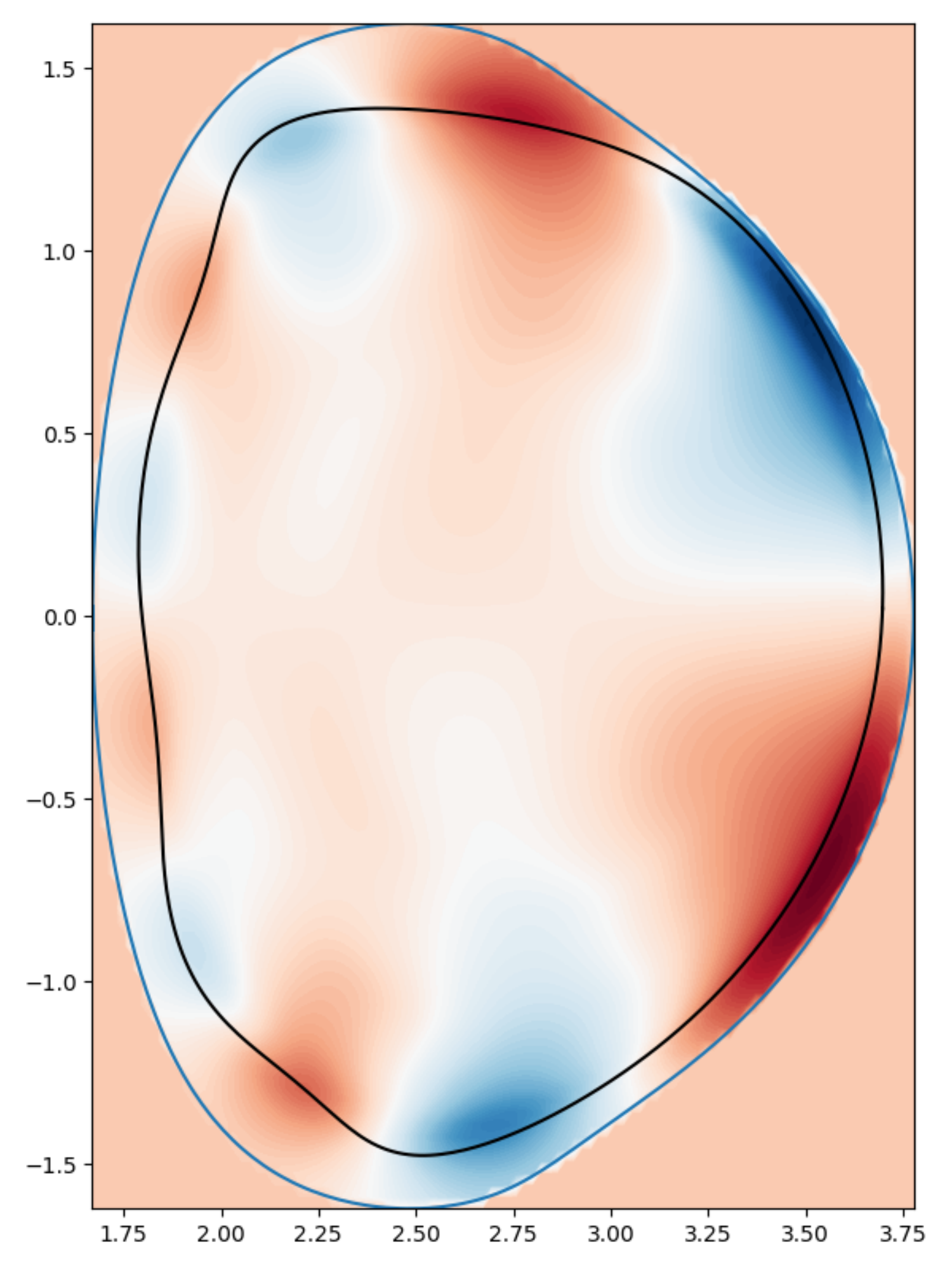}
    \centering
    \small{(d)}
    \end{minipage}
    \caption{Comparison of the perturbed magnetic energy spectrum in JOREK and VMEC solutions (a). {The energy spectrum during the $n=2$ dominant phase (blue), and the final $n=1$ dominant mode (red) are shown. The energy spectra are averaged over the corresponding blue and red regions shown in Figure \ref{fig:jet_eho_E_comp} (c).} The perturbed poloidal flux in JOREK is plotted for the time slices marked by grey dashed lines in Figure \ref{fig:jet_eho_E_comp} (c). These time points correspond to the initial saturation of the $n=2$ mode (b) and end of the simulation time (c). A comparable poloidal plane from the VMEC solution (d) is shown.  {The simulation boundary used in JOREK (blue), and the plasma boundary from the perturbed state in VMEC (black) are overlaid on the solution.} Note that the colour bars are not the same in the three pseudocolour plots, so that the poloidal structures can be more easily compared with one another.}
    \label{fig:jet_eho_magnetic_energy_spectrum}
\end{figure*}

Using the process outlined in Section \ref{sec:kink_jorek_comparison}, the magnetic energy spectrum has also been calculated and compared between the two codes, as shown in Figure \ref{fig:jet_eho_magnetic_energy_spectrum} (a). {The energy spectra in JOREK are taken during the initial saturation of the $n=2$ mode, and final saturation of the $n=1$ mode for comparison. For the final $n=1$ dominated state,} it can be seen that while the low n modes agree reasonably well, the high n modes are notably larger in JOREK. The flat spectra observed in the JOREK energies is typically seen when there is MHD activity in the higher n modes which is not driven by the lower toroidal harmonics, as shown in Figure 11 of Ref. \onlinecite{ramasamy2021nonlinear}. In other words, the higher toroidal harmonics are linearly unstable which is known to be the case. The decay of the energies seen in the VMEC spectrum is typical for an instability which is driven by the $n=1$ mode alone. 

The perturbation of the poloidal flux from JOREK is shown at the time of saturation for the $n=2$ mode in Figure \ref{fig:jet_eho_magnetic_energy_spectrum} (b). This shows that the instability is initially led by a (9, 2) external structure, as expected from the linear eigenfunction observed in Figure \ref{fig:jet_eho_eigenfunctions} (d). Near the end of the simulation time, shown in Figure \ref{fig:jet_eho_magnetic_energy_spectrum} (c), the poloidal flux perturbation looks very similar to the stationary perturbation in VMEC. {This indicates that the resistive dynamics have not changed the overall structure of the mode from the ideal result}. 

{The question remains, why does the energy spectrum differ in Figure \ref{fig:jet_eho_magnetic_energy_spectrum} (a)? The connection length diagnostic shows that the overall mode structure has strong similarities to previous observations of EHOs, and the poloidal flux perturbation is very similar to the VMEC result. It seems that the instability observed in the two codes has strong similarities, despite the discrepancy in the energies of higher toroidal harmonics. As a result, it appears that the VMEC solution does not capture the full toroidal mode coupling of the instability, represented by the sub-dominant modes in Figure \ref{fig:jet_eho_magnetic_energy_spectrum} (a).}

\subsection{Comparison of toroidal mode coupling in the pressure perturbation}

\begin{figure}
    \centering
    \includegraphics[width=0.475\textwidth]{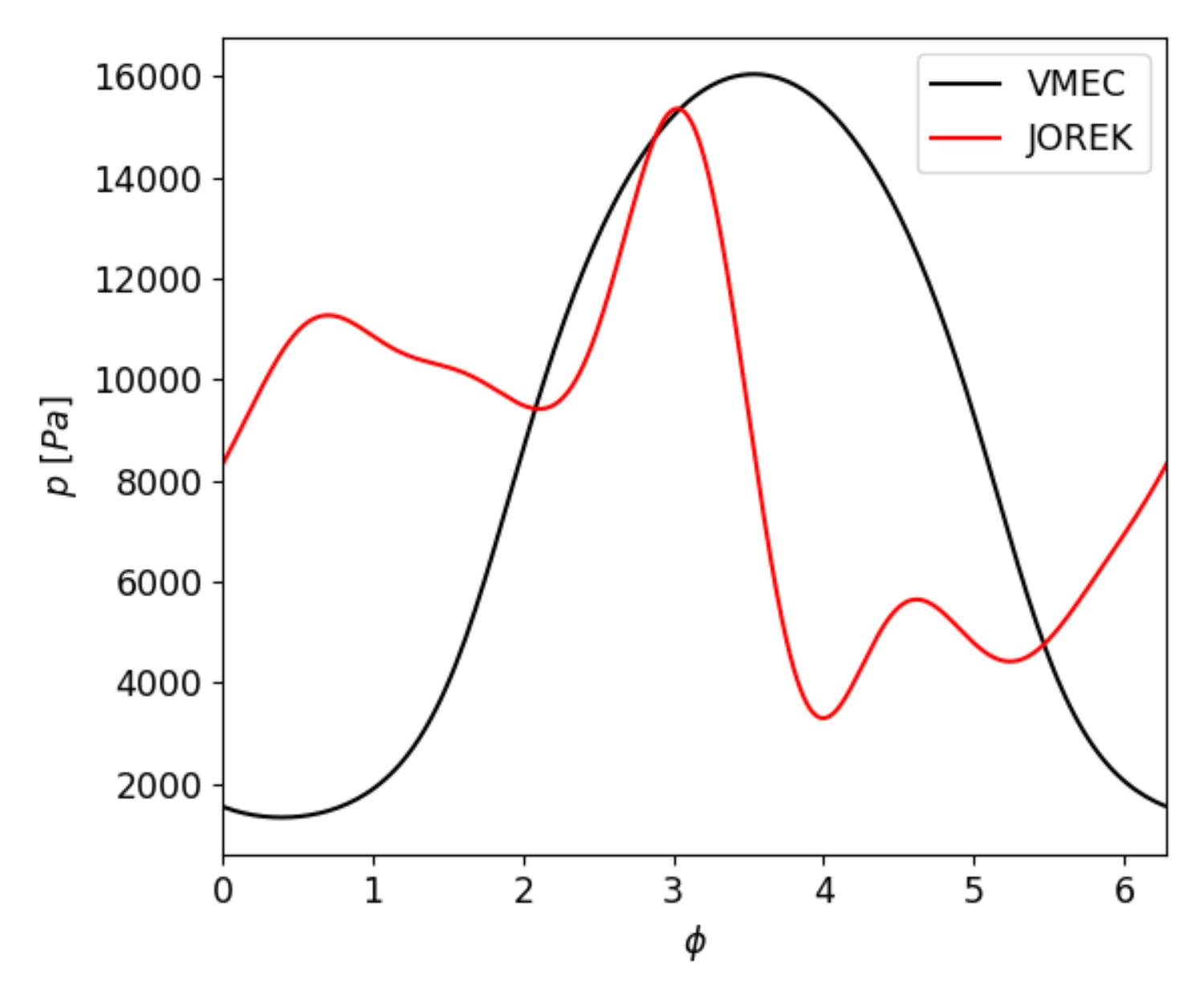}
    \caption{Pressure as a function of the toroidal angle at $R=3.667$, $Z=0.0$ in VMEC (black) and JOREK (red). The data from JOREK is taken at the end of the simulation time.}
    \label{fig:jet_eho_pressure_localisation}
\end{figure}

Experimental diagnostics of EHOs suggest that the saturated MHD structure consists of many coupled toroidal harmonics \cite{garofalo2015quiescent}. One such experimentally observed feature of EHOs is the toroidal localisation of the perturbed density and temperature in the nonlinearly saturated state, due to toroidal mode coupling. To observe this effect in JOREK, the pressure is sampled along the toroidal angle for a point on the low field side of the device along the midplane, at $R=3.667\ m$. The sampled pressure is shown in Figure \ref{fig:jet_eho_pressure_localisation} both in JOREK and VMEC. JOREK shows a more localised structure, with a steep gradient at $\phi\approx\pi$. In VMEC, the pressure profile follows the dominant $n=1$ perturbation alone. 

Toroidal mode coupling of the pressure perturbation cannot be observed {in the VMEC computation}, because the pressure profile is held fixed as a function of the radial coordinate during the computation. As such, the pressure is not allowed to relax naturally, and is constrained to follow the perturbation of the flux surface contours. As shown in Section \ref{sec:jet_eho_flux_surface_comparison}, the region near the plasma edge is ergodic, and so the flux surfaces in VMEC are not expected to capture the structure of the pressure correctly.

It is not clear what influence this error has on the nonlinear state that is found. In principle, ballooning stability is strongly dependent on the local pressure gradients in the device, and so the modification of the pressure in the edge region will modify the evolution of the pressure drive of the instability. The VMEC computation cannot follow the same relaxation of the pressure drive as in the nonlinear MHD simulation. As such, while VMEC computations have been shown to capture the correct initial pressure drive dependencies of the mode correctly, it is suspected that there must be some loss in accuracy due to the artificial constraint imposed on the pressure profile.

\subsection{Influence of flows} \label{sec:eho_flows}
{An important area of research in the context of EHOs is the influence of toroidal and poloidal flows on the dynamics. In particular, it is argued that flows play a crucial role in stabilising high-n ballooning modes that would otherwise modify the solution and potentially prevent low-n kink modes from becoming the dominant nonlinear mode structure. Previous nonlinear studies have observed the stabilisation of high-n modes, and the excitation of low-n external modes with increased toroidal and $\mathbf{E} \times \mathbf{B}$ flows \cite{chen2016rotational, liu2015nonlinear}.}

\begin{figure}
    \centering
    \begin{minipage}{0.45\textwidth}
    \centering
    \includegraphics[width=\textwidth]{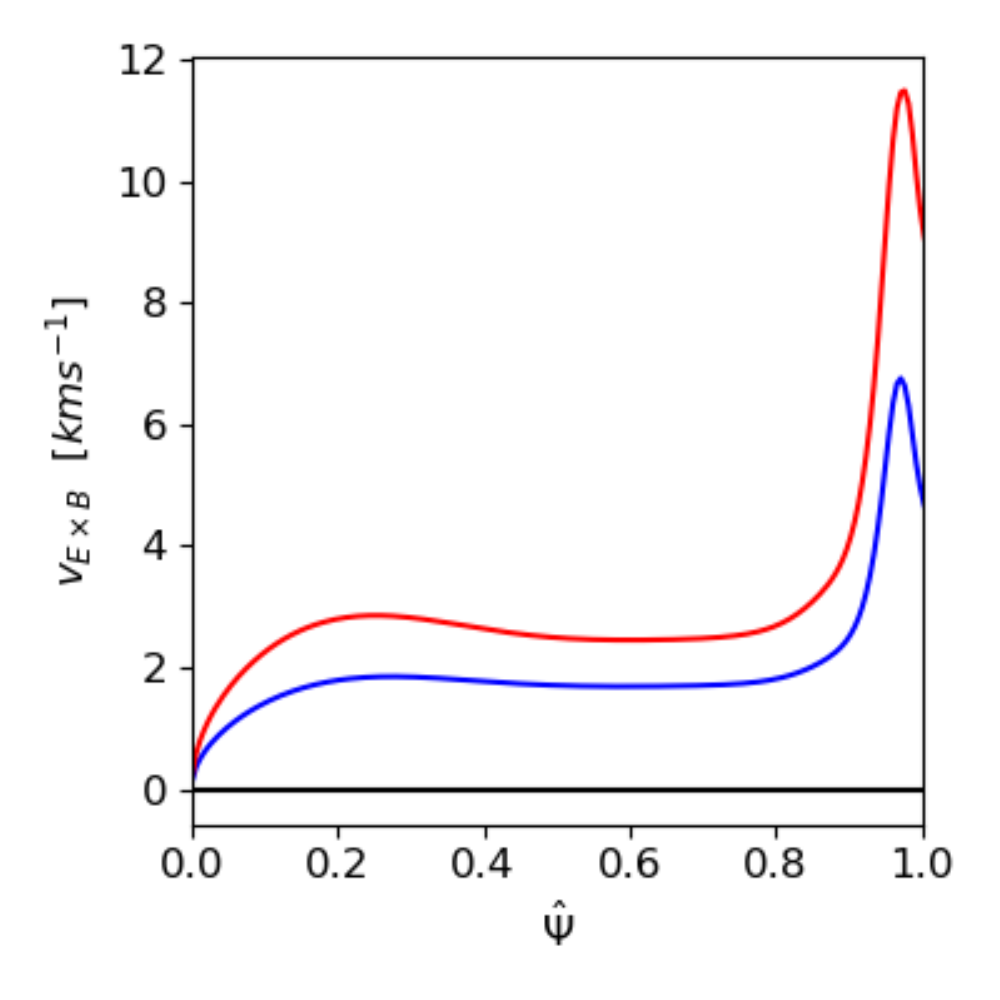}
    \small{(a)}
    \end{minipage}
    \begin{minipage}{0.45\textwidth}
    \centering
    \includegraphics[width=\textwidth]{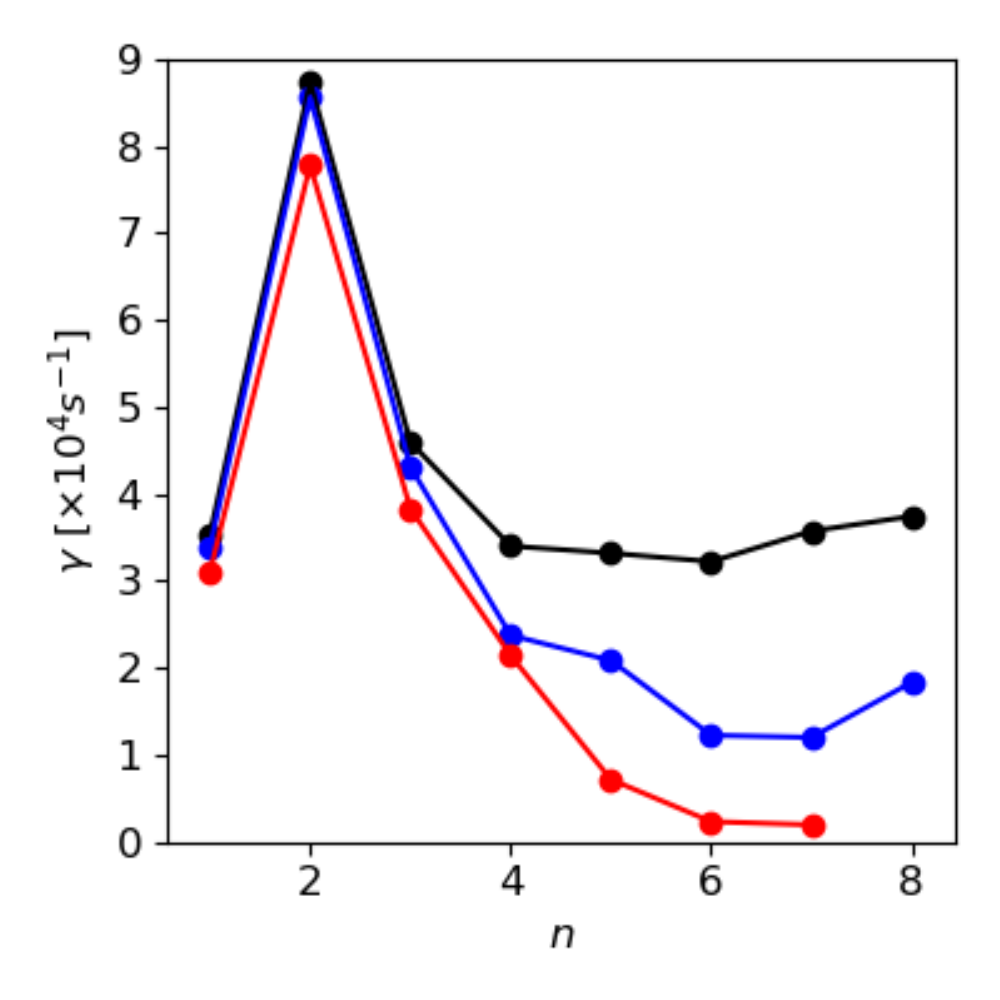}
    \small{(b)}
    \end{minipage}
    \caption{{Modification of the $\mathbf{E}\times\mathbf{B}$ velocity (a) with increasing diamagnetic source in JOREK. {The red curves correspond to a value for the diamagnetic source in equation \ref{eq:velocity_ansatz}, which is close to the expected value for the modelled equilibrium. Results are also shown without flows (black) and at an intermediate value (blue)} The linear growth rate of external modes up to $n=8$ (b) show that high n modes are stabilised significantly when moderate flows are introduced.}}
    \label{fig:eho_flows}
\end{figure}

{To understand the influence of diamagnetic flows on the simulated EHO test case, the diamagnetic source term in equation \ref{eq:velocity_ansatz} was introduced to the EHO equilibrium, using the resistivity profile from case 2 in Section \ref{sec:eho_resistivity_scan}. In JOREK, switching on the diamagnetic drift term implies that the experimentally observed $E_r$ well also forms, such that the diamagnetic and $\mathbf{E} \times \mathbf{B}$ velocities are varied together. For the ions, the diamagnetic and $\mathbf{E} \times \mathbf{B}$ velocities approximately cancel such that the ions are approximately at rest. This condition is similar to experimental observations, and the assumptions of analytical work on the influence of flows in Ref. \onlinecite{brunetti2019excitation}. It should be noted that the particle source, $S_\rho$, does not have an associated momentum source in equation \ref{eq:pol_momentum} and \ref{eq:par_momentum}, such that there is some damping of the flows in these JOREK simulations. For the moderate flows assumed in this study, it has been verified that this effect is not significant, however.}

{Three equilibrium conditions were considered with increasing values of the diamagnetic source term. The corresponding $\mathbf{E} \times \mathbf{B}$ velocity within the plasma is shown in Figure \ref{fig:eho_flows} (a). The red curve in Figure \ref{fig:eho_flows} (a) is considered close to the expected $\mathbf{E} \times \mathbf{B}$ velocity that would be driven in the given equilibrium, however it should be acknowledged that the assumed source has not been informed by experimental observations. The black curve corresponds to the case without a diamagnetic source, and therefore, also no corresponding $\mathbf{E} \times \mathbf{B}$ velocity. The blue curve is an intermediate value, between these two limits.}

{The modification of the linear growth rates in the no wall limit with increasing flows for $n<=8$ is shown in Figure \ref{fig:eho_flows} (b). The expected preferential stabilisation of high-n modes is observed, consistent with theoretical models of EHOs \cite{brunetti2019excitation}. This provides further evidence that the low-n modes can be expected to dominate the dynamics in an experiment for the given test equilibrium. The low-n modes were marginally stabilised by poloidal flows, unlike in previous studies of DIII-D. However, for the simulated JET-like discharge, it is not clear that this effect should be observed.}

It should be noted that the deviation of the energy spectra in Figure \ref{fig:jet_eho_magnetic_energy_spectrum} (a) could be influenced by flows. The main nonlinear JOREK simulations presented in this section were run without diamagnetic flows, which Figure \ref{fig:eho_flows} (b) shows would have a preferential stabilising effect on the linearly unstable high n modes. In such a way, including flows could improve the agreement in Figure \ref{fig:jet_eho_magnetic_energy_spectrum} (a) to some extent. These physical parameters have already been considered in detail in other JOREK studies, and so were neglected as part of this work \cite{liu2017nonlinear}.

\section{Conclusion} \label{sec:conclusion}
Two approaches for modeling saturated external MHD instabilities have been compared for a (5, 1) external kink mode, and edge harmonic oscillation. Simulations with JOREK are compared to VMEC, both {using a high Lundquist number within the plasma region to approach the ideal limit of the VMEC code}, and using typical parameters used in simulation studies for resistivity and diamagnetic flows. The results of the equilibrium approach using VMEC are in good agreement with the initial value problems that have been solved using JOREK, using equivalent physics that excludes equilibrium flows, and significant internal viscoresistive effects. The inclusion of diamagnetic flows suppresses the internal resistive dynamics that has been observed with more realistic resistivity profiles for the external kink case. Therefore, the results of nonlinear MHD simulations show good agreement with the VMEC result for this instability. 

For the EHO case, the fast growing higher toroidal harmonics influence the nonlinear dynamics{, making it harder to find agreement between the two approaches}. Using similar assumptions as in VMEC, the $n=2$ mode, rather than the $n=1$, dominates the JOREK simulation result on the ideal MHD timescale. {Eventually during the nonlinear phase, $n=1$ perturbations can develop with a similar magnetic field structure to the VMEC result}. Comparing the $n=1$ perturbations, deviations are observed due to the ergodisation of the plasma edge region and toroidal mode coupling in the pressure perturbation, both of which are observed in JOREK, and absent in VMEC, because closed magnetic flux surfaces are enforced. Even though the dynamics in the JOREK simulation extend beyond the fast MHD timescale, the solutions of the magnetic field structure calculated using the two approaches seem consistent. Using a Spitzer profile for the resistivity, the $n=2$ mode becomes the dominant mode by the end of the simulated time, indicating that the dynamics of the dominant low-n toroidal harmonics are sensitive to the assumed resistivity profile.

{It has therefore been shown that the results of a full nonlinear approach can disagree with the VMEC approach for complex test cases like the EHO. At the same time, even in this more advanced test case, there are clear similarities between the observed structures, when the result of the nonlinear code is dominated by the same toroidal harmonic as found in the equilibrium approach. This encourages} the use of VMEC as an efficient approach to understanding saturated external modes, at least for the instabilities with medium poloidal mode number that have been considered in this study{, provided such results are supported by experimental observations and the results of other numerical models}. 

How far both methods used in this study can be applied to understand experimentally relevant saturated modes depends on whether the assumptions of the two codes can be reasonably made for the dynamics. The validity is likely to depend on the particular experimentally observed mode of interest. For VMEC computations, the MHD activity is required to be dominated by ideal MHD. For nonlinear simulations, careful prescription of physically relevant viscoresistive and diffusive parameters, as well as source terms, is necessary.

There are a number of other directions for future work using these codes. JOREK has already been used for more advanced studies of EHOs in DIII-D and ITER. Similar to the results of this study, such simulations have started from an initially linearly unstable equilibrium. A natural extension would be to simulate EHOs from an initially stable equilibrium, transitioning across the stability boundary to understand the conditions for reaching an EHO.

The VMEC approach can be used to consider more experimentally relevant equilibria as well, by extending studies to consider x-point plasmas. An x-point cannot be modelled in VMEC, but, relaxing the up-down symmetry condition that is used in most studies, and truncating the equilibria close to the plasma edge, saturated states can be found which could be more experimentally relevant. This approach would be similar to that used in linear ideal MHD codes with a straight field line coordinate system, which cannot model x-points. 

{In addition, more generalised equilibrium codes, such as SIESTA \cite{peraza2017extension} or SPEC \cite{hudson2012computation}, could be used to further refine the results of the VMEC equilibrium approach. Such codes would allow the formation of magnetic islands and stochastic regions, relaxing some of the assumptions that have been made using VMEC. In particular, it would be interesting to know how these effects could modify the EHO result, as this would help identify whether the discrepancies we have observed are due to closed flux surfaces being enforced.}

Lastly, some of the authors are interested in considering external modes in stellarators using VMEC. Such studies are intended to complement the recent development of stellarator-capable nonlinear MHD codes, to inform intuition for how ideal MHD instabilities saturate in such devices. An initial study on this topic is currently in progress.

\section*{Acknowledgements}
The authors would like to thank Florian Hindenlang for his contribution to this work through fruitful discussions{, and Michael Drevlak for providing access to codes, which were used in the construction and analysis of free boundary VMEC calculations}. 

This work has been supported in part by the Max-Planck/Princeton Center for Plasma Physics and the Swiss National Science Foundation. This work has been carried out within the framework of the EUROfusion Consortium, funded by the European Union via the Euratom Research and Training Programme (Grant Agreement No 101052200 — EUROfusion). Views and opinions expressed are however those of the author(s) only and do not necessarily reflect those of the European Union or the European Commission. Neither the European Union nor the European Commission can be held responsible for them. Some of this work was carried out on the high performance computing architectures COBRA operated by MPCDF in Germany, and JFRS-1 operated by IFERC-CSC in Japan.

\appendix
\section{Conservation of helicity in VMEC} \label{app:helicity_conservation}
\setcounter{section}{1}

The global helicity, $K$, of an equilibrium is

\begin{equation} \label{eq:helicity_def_i}
    K = \int \mathbf{A} \cdot \mathbf{B}\ dV,
\end{equation}

where, in VMEC, the magnetic field, $\mathbf{B}$, is defined by the vector potential, $\mathbf{A}$, such that

\begin{eqnarray}
    \begin{split}
    \mathbf{B} &= \nabla \times \mathbf{A} \nonumber\\
    &= \nabla \times \left( \Phi(s) \nabla \theta^* - \Psi(s) \nabla \phi\right) \nonumber\\
    &= \Phi'(s) \nabla s \times \nabla \theta^* - \Psi'(s) \nabla s \times \nabla \phi,
    \end{split}
\end{eqnarray}

where $\Phi$ and $\Psi$ are the toroidal and poloidal flux respectively, and $s$, $\theta$, and $\zeta$ have their usual meanings as the radial, poloidal, and toroidal coordinates. $\theta^*$ is the transformed poloidal angle in straight field line coordinates. The local helicity can therefore be written as

\begin{eqnarray}
    \begin{split}
    \mathbf{A} \cdot \mathbf{B} &= \left(\Psi' \Phi - \Phi' \Psi \right) \left[\nabla s \cdot \left( \nabla \theta^* \times \nabla \phi \right) \right] \nonumber\\
    &= \left(\Psi' \Phi - \Phi' \Psi \right) \frac{1}{J^*},
    \end{split}
\end{eqnarray}

{where $J^*$ is the Jacobian of the transformation from laboratory to plasma coordinates.} Finally, it can be shown that equation \ref{eq:helicity_def_i} can be simplified to

\begin{equation} \label{eq:helicity_vmec}
    K = 4 \pi^2 \int_\mathrm{0}^\mathrm{1}  \left(\Psi' \Phi - \Phi' \Psi \right) ds.
\end{equation}

As the $\iota$ profile, toroidal flux profile, $\Phi(s)$, and total toroidal flux are fixed during VMEC computations, the profile for the poloidal flux is also fixed, because

\begin{equation}
    \Psi(s) = \Phi(s=1) \int_\mathrm{0}^s \iota(s) ds.
\end{equation}

As such, the profiles in equation \ref{eq:helicity_vmec} are all fixed, such that the helicity is kept constant during the VMEC computation. This provides a physically meaningful link between the initial and final states, such that the trajectories can be reasonably compared against linear and nonlinear MHD codes. It should be noted that when computing the perturbed state, a small initial perturbation is applied to the magnetic axis{, such that the initial condition is not the exact unperturbed equilibrium.}

At this point, it is worth providing an interpretation for computations where the current profile is fixed. This constraint is not found in ideal MHD, and so the perturbed equilibria that are computed must have some form of external control, such as a modification of the loop voltage in the device, in order to link them with the initial axisymmetric state. As such, the constraint could be used for studies of RMPs where external control is implied, but for dynamics on the ideal MHD timescale, constraining the current profile is not justified.

\bibliography{references}

\end{document}